\documentclass[twocolumn]{aastex631}

\usepackage[super]{nth}
\usepackage{savesym}
\savesymbol{tablenum}
\usepackage[T1]{fontenc}
\usepackage{pifont}
\usepackage{inputenc}
\usepackage[range-units=brackets,product-units=single,multi-part-units=single,range-phrase={-},separate-uncertainty,print-unity-mantissa=false]{siunitx}
\usepackage{multirow}
\DeclareSIUnit\angstrom{\text {Å}}
\restoresymbol{SIX}{tablenum}
\newcommand{\cmark}{\ding{51}}
\newcommand{\xmark}{\ding{55}}

\graphicspath{{./}{figures/}}

\accepted{April 28, 2024}

\submitjournal{ApJ}

\shorttitle{Mapping feedback signatures in the CGM with microcalorimeters}
\shortauthors{Schellenberger et al.}

\begin{document}

\title{Mapping the imprints of stellar and AGN feedback in the circumgalactic medium with X-ray~microcalorimeters}

\correspondingauthor{Gerrit Schellenberger}
\email{gerrit.schellenberger@cfa.harvard.edu}

\author[0000-0002-4962-0740]{Gerrit Schellenberger}
\affiliation{Center for Astrophysics $|$ Harvard \& Smithsonian, 60 Garden St., Cambridge, MA 02138, USA}

\author[0000-0003-0573-7733]{\'Akos Bogd\'an}
\affiliation{Center for Astrophysics $|$ Harvard \& Smithsonian, 60 Garden St., Cambridge, MA 02138, USA}

\author[0000-0003-3175-2347]{John A. ZuHone}
\affiliation{Center for Astrophysics $|$ Harvard \& Smithsonian, 60 Garden St., Cambridge, MA 02138, USA}

\author[0000-0002-3391-2116]{Benjamin D. Oppenheimer}
\affiliation{University of Colorado, Boulder, 2000 Colorado Avenue, Boulder, CO 80305, USA}

\author[0000-0003-4983-0462]{Nhut Truong}
\affiliation{NASA Goddard Space Flight Center, Code 662, Greenbelt, MD 20771}
\affiliation{Center for Space Sciences and Technology, University of Maryland, Baltimore County, MD 21250, USA}
\affiliation{Max-Planck-Institut für Astronomie, Königstuhl 17, D-69117 Heidelberg, Germany}

\author[0000-0003-3701-5882]{Ildar Khabibullin}
\affiliation{Universitäts-Sternwarte, Fakultät für Physik, Ludwig-Maximilians-Universität München, Scheinerstr.1, 81679 München, Germany}
\affiliation{Max-Planck-Institut für Astrophysik, Karl-Schwarzschild-Straße 1, 85748 Garching, Germany}
\affiliation{Space Research Institute (IKI), Profsoyuznaya 84/32, Moscow 117997, Russia}

\author[0009-0000-0152-9983]{Fred Jennings}
\affiliation{Institute for Astronomy, University of Edinburgh, Royal Observatory, Blackford Hill, Edinburgh, EH9 3HJ, United Kingdom}

\author[0000-0003-1065-9274]{Annalisa Pillepich}
\affiliation{Max-Planck-Institut für Astronomie, Königstuhl 17, D-69117 Heidelberg, Germany}

\author[0000-0002-1979-2197]{Joseph Burchett}
\affiliation{Department of Astronomy, New Mexico State University, Las Cruces, NM 88001, USA}

\author[0000-0002-5840-0424]{Christopher Carr}
\affiliation{Department of Astronomy, Columbia University, 550 West 120th Street, New York, NY, 10027, USA}

\author[0000-0002-4469-2518]{Priyanka Chakraborty}
\affiliation{Center for Astrophysics $|$ Harvard \& Smithsonian, 60 Garden St., Cambridge, MA 02138, USA}

\author[0000-0001-6258-0344]{Robert Crain}
\affiliation{Astrophysics Research Institute, Liverpool John Moores University, 146 Brownlow Hill, Liverpool L3 5RF, UK}

\author[0000-0002-9478-1682]{William Forman}
\affiliation{Center for Astrophysics $|$ Harvard \& Smithsonian, 60 Garden St., Cambridge, MA 02138, USA}

\author[0000-0003-2206-4243]{Christine Jones}
\affiliation{Center for Astrophysics $|$ Harvard \& Smithsonian, 60 Garden St., Cambridge, MA 02138, USA}

\author[0000-0001-9464-4103]{Caroline A. Kilbourne}
\affiliation{NASA Goddard Space Flight Center, Code 662, Greenbelt, MD 20771}

\author[0000-0002-0765-0511]{Ralph P. Kraft}
\affiliation{Center for Astrophysics $|$ Harvard \& Smithsonian, 60 Garden St., Cambridge, MA 02138, USA}

\author[0000-0003-0144-4052]{Maxim Markevitch}
\affiliation{NASA Goddard Space Flight Center, Code 662, Greenbelt, MD 20771}

\author[0000-0002-6766-5942]{Daisuke Nagai}
\affiliation{Department of Physics, Yale University, New Haven, CT 06520, USA}

\author[0000-0001-8421-5890]{Dylan Nelson}
\affiliation{Universität Heidelberg, Zentrum für Astronomie, Institut für theoretische Astrophysik, Albert-Ueberle-Str. 2, 69120 Heidelberg, Germany}

\author[0000-0003-4504-2557]{Anna Ogorzalek}
\affiliation{NASA Goddard Space Flight Center, Code 662, Greenbelt, MD 20771}
\affiliation{Department of Astronomy, University of Maryland, College Park MD, USA}

\author[0000-0002-3984-4337]{Scott Randall}
\affiliation{Center for Astrophysics $|$ Harvard \& Smithsonian, 60 Garden St., Cambridge, MA 02138, USA}

\author[0000-0002-5222-1337]{Arnab Sarkar}
\affiliation{Department of Physics, Kavli Institute for Astrophysics and Space Research, Massachusetts Institute of Technology, Cambridge, MA 02139, USA}

\author[0000-0002-0668-5560]{Joop Schaye}
\affiliation{Leiden Observatory, Leiden University, Niels Bohrweg 2, NL-2333 CA Leiden, the Netherlands}

\author[0000-0002-3158-6820]{Sylvain Veilleux}
\affiliation{Department of Astronomy and Joint Space-Science Institute, University of Maryland, College Park, MD 20742, USA}

\author[0000-0001-8593-7692]{Mark Vogelsberger}
\affiliation{Department of Physics, Kavli Institute for Astrophysics and Space Research, Massachusetts Institute of Technology, Cambridge, MA 02139, USA}

\author[0000-0002-9279-4041]{Q. Daniel Wang}
\affiliation{Astronomy Department, University of Massachusetts, Amherst, MA 01003, USA}

\author[0000-0001-7630-8085]{Irina Zhuravleva}
\affiliation{Department of Astronomy and Astrophysics, University of Chicago, Chicago IL 60637, USA}

\begin{abstract}
The Astro2020 Decadal Survey has identified the mapping of the circumgalactic medium (CGM, gaseous plasma around galaxies) as a key objective. 
We explore the prospects for characterizing the CGM in and around nearby galaxy halos with a future, large grasp X-ray microcalorimeter. 
We create realistic mock observations from hydrodynamical simulations (EAGLE, IllustrisTNG, and Simba) that demonstrate a wide range of potential measurements, which will address the open questions in galaxy formation and evolution. 
By including all background and foreground components in our mock observations, we show why it is impossible to perform these measurements with current instruments, such as X-ray CCDs, and only microcalorimeters will allow us to distinguish the faint CGM emission from the bright Milky Way (MW) foreground emission lines. 

We find that individual halos of MW mass can, on average and depending on star formation rate, be traced out to large radii, around R$_{500}$, and for larger galaxies even out to R$_{200}$, using prominent emission lines, such as O\,VII, or O\,VIII. 
Furthermore, we show that emission line ratios for individual halos can reveal the radial temperature structure. Substructure measurements show that it will be possible to relate azimuthal variations to the feedback mode of the galaxy. 
We demonstrate the ability to construct temperature, velocity, and abundance ratio maps from spectral fitting for individual galaxy halos, which reveal rotation features, AGN outbursts, and enrichment. 

\end{abstract}


\section{Introduction} \label{ch:intro} 
Structure formation models predict that galaxies reside in massive dark matter halos and are embedded in large-scale gaseous halos, the circumgalactic medium (CGM). The CGM plays a crucial role in the evolution of galaxies as gas flows through the CGM and regulates galaxy growth over cosmic time. To establish a comprehensive picture of the formation and evolution of galaxies, it is essential to probe the interplay between the stellar body, the supermassive black hole (SMBH), and the large-scale CGM. 
However, our understanding of the CGM, especially its hot X-ray-emitting component that is critical to the mass budget of galaxies, is still limited. This lack of knowledge poses major gaps in our understanding of galaxy formation and evolution.
The importance of the CGM is highlighted by the fact that it plays key roles on a wide range of spatial scales from small-scale processes (e.g.\ galactic winds driven by supernovae) to the largest scales of galaxies (e.g.\ accretion of gas from large-scale structure filaments). 

Theoretical studies hint that the CGM has a complex and multi-phase structure. In Milky Way-type and more massive galaxies, the dominant phases of the CGM have characteristic temperatures of millions of degrees and are predominantly observable at X-ray wavelengths (e.g., \citealp{Crain2010-lw,Van_de_Voort2013-lj,Nelson2018-an,Oppenheimer2019-ub,Wijers2022-cf}). Indeed, in the well-established picture of structure formation, dark matter halos accrete baryonic matter, which is thermalized in an accretion shock (\citealp{White1978-vm,White1991-ht}). The characteristic temperature is determined by the gravitational potential of the galaxies and reaches X-ray temperatures ($\gtrsim10^6$~K) for Milky Way-type galaxies.  Since the cooling time of the hot gas is much longer than the dynamical time, the CGM is expected to be quasi-static and should be observable around galaxies in the present-day universe.

Theoretical studies suggest that the CGM is multiphase and, for Milky Way mass and more massive galaxies, the bulk of the CGM resides in the hot ($\SI{e6}{K} < T < \SI{e7}{K}$) phases \citep{Nelson2018-an}. Absorption studies carried out with the Cosmic Origin Spectrograph on the Hubble Space Telescope probed the cooler phases ($\SI{e4}{K} < T < \SI{e6}{K}$) of the CGM (e.g., \citealp{Xavier_Prochaska2011-it,Tumlinson2011-eh,Putman2012-qx,Werk2014-yq,Werk2016-dr,Prochaska2017-mn}). Stacking analyses of galaxies using Sunyaev-Zeldovich measurements indicated the presence of hot phases in the CGM \cite{Wu2020-ws,Bregman2022-tb,Moser2022-uc,Das2023-ha}. However, X-ray observations are best suited to map and explore the physical characteristics of the hot phases of the CGM.

Because the importance of studying the X-ray emitting large-scale gaseous component of galaxies was recognized decades ago, all major X-ray observatories attempted to explore this component. Studies of elliptical (or quiescent) galaxies achieved significant success in the early days of X-ray astronomy (\citealp{Nulsen1984-fk,OSullivan2001-cq}). Observations of massive ellipticals with the \textit{Einstein} and \textit{ROSAT} observatories revealed the presence of gaseous X-ray halos that extend beyond the optical extent of galaxies 
out to ${\sim\SI{100}{kpc}}$ \citep{Forman1985-ki,Trinchieri1985-xl,Trinchieri1986-kp,Canizares1987-gw,Mathews1990-nk,Mathews2003-fg}.
These observations not only revealed the ubiquity of the gaseous halos, but allowed characterizations of the physical properties of the X-ray gas, and measurements of its mass. Follow-up observations with \textit{XMM-Newton} and \textit{Chandra} played a major role in further probing the gaseous emission around a larger sample of nearby massive elliptical galaxies \citep{Anderson2011-ne,Bogdan2013-vw,Kim2013-hr,Bogdan2015-bc,Goulding2016-gm,Bogdan2017-mm,Li2018-ac,Li2020-vy,Nicastro2023-tr,Mathur2023-sn}.
Thanks to the sub-arcsecond angular resolution of \textit{Chandra}, it became possible to clearly resolve and separate point-like sources, such as low-mass X-ray binaries and AGN, from the truly diffuse gaseous emission \citep{Revnivtsev2009-em,Bogdan2011-sv}. This allowed more detailed studies of the X-ray-emitting interstellar medium and the larger-scale CGM. However, a major hindrance to studying elliptical galaxies is that the dominant fraction of galaxies explored by \textit{Chandra} and \textit{XMM-Newton} reside in rich environments, such as galaxy groups or galaxy clusters. The presence of these group and cluster atmospheres makes it virtually impossible to differentiate the CGM component of the galaxy from the large-scale group or cluster emission. Because the group or cluster atmosphere will dominate the overall emission beyond the optical radius, it becomes impossible to separate these components from each other and determine their relative contributions. Additionally, the gaseous component around quiescent galaxies is likely a mix of the infalling primordial gas onto the dark matter halos and the ejected gas from evolved stars, which was shock heated to the kinetic temperature of the galaxy. 
Due to quenching mechanisms, most quiescent galaxies reside in galaxy groups and clusters, which are not ideal targets to probe the primordial gas. 

As opposed to their quiescent counterparts, star-forming galaxies provide the ideal framework to probe the gas originating from primordial infall. The main advantage of disk (or star-forming) galaxies is their environment. While quiescent galaxies form through mergers, which happen at a higher likelihood in rich environments, due to the higher galaxy density, a substantial fraction of star-forming galaxies preferentially reside in relatively isolated environments. 
The CGM around disk galaxies has been probed in a wide range of observations. Using \textit{ROSAT} observations, the X-ray gas around disk galaxies remained undetected \citep{Benson2000-us}. 
However, this posed a challenge to galaxy formation models that predicted bright enough gaseous halos to be observed around nearby disk galaxies (\citealp{White1991-ht,Crain2010-lw,Vogelsberger2020-hc}). Revising these models and involving efficient feedback from supernova, and later AGN, feedback drastically decreased the predicted X-ray luminosity of the X-ray CGM, implying that non-detection by \textit{ROSAT} was consistent with theoretical models (\citealp[e.g.][]{Crain2010-lw,Oppenheimer2020-ix})
More sensitive observations with \textit{Chandra} and \textit{XMM-Newton} led to the detection of the CGM around isolated massive disk galaxies (e.g., \citealp{Wang2001-bg,Li2006-pr,Li2007-hd,Crain2010-lw}). Most notably, the CGMs of two massive galaxies, NGC\,1961 and NGC\,6753, were detected and characterized out to about $50-60$~kpc radius, which corresponds to about $0.15r_{\rm 200}$, where $r_{\rm 200}$ is the radius within which the density is 200 times the critical density of the Universe, and we consider it to be the virial radius of the galaxies. These studies not only detected the gas, but also measured the basic properties of the CGM, such as 
temperature and abundance, and established simple thermodynamic profiles beyond the optical radii of the galaxies \citep{Anderson2011-ne,Bogdan2013-vw,Anderson2015-go,Bogdan2017-mm}. 
Following these detections, the CGM of other disk galaxies was also explored albeit to a much lesser extent due to the lower signal-to-noise ratios of these galaxies \citep[e.g.][]{Anderson2012-ph,Anderson2011-ne,Dai2012-dp,Bogdan2013-ca,Bogdan2015-bc,Li2017-bk}.
Despite these successes, however, it is important to realize that all these detections explored massive galaxies (few $\times10^{11} \ \rm{M_{\odot}}$ in stellar mass), while X-ray observations could not detect the CGM emission around Milky Way-type galaxies, with exception of our own Galaxy (\citealp{Das2019-uz,Das2019-zo,Das2021-gz,Bhattacharyya2022-nk}).

The main challenge in detecting the extended CGM of external galaxies is due to the hot gas residing in our own Milky Way. Specifically, our Solar system is surrounded by the local hot bubble (LHB), which has a characteristic temperature of $kT\approx\SI{0.1}{keV}$ \citep{McCammon2002-bm,Das2019-zo}. On larger scales, the Milky Way also hosts an extended hot component with a characteristic temperature of $\sim\SI{0.2}{keV}$ \citep{McCammon2002-bm,Das2019-zo}. These gas temperatures are comparable to those expected from other external galaxies and, since both the Milky Way and the other galaxies exhibit the same thermal emission component, the emission signal from the low-surface brightness CGM of external galaxies can be orders of magnitude below the Milky Way foreground emission. Because the X-ray emission from the Milky Way foreground is present in every sightline and this component cannot be differentiated at CCD resolution ($\sim50-100$~eV), its contribution cannot be easily removed from the CGM component of other galaxies. A direct consequence of this is that even future telescopes with larger collecting areas with  CCD-like instruments will be limited by the foreground emission and thus cannot probe the large-scale CGM. To achieve a transformative result in exploring the extended CGM, we must utilize high spectral resolution spectroscopy to spectroscopically differentiate the emission lines of the Milky Way foreground from those emitted by the external galaxies. We emphasize that mapping the CGM around individual galaxies is essential to learn about its 2D distribution, enrichment, and thermodynamic structure, which is more challenging with dispersive (grating) spectroscopy due to line broadening  (\citealp{Li2020-vy}) and is pursued by concepts such as ARCUS (\citealp{Smith2022-dh}).

Recent advances in technology allow us to take this transformative step. The development of high spectral resolution X-ray Integral Field Units (IFUs) provides the much-needed edge over traditional CCD-like instruments. Notably, X-ray IFUs can simultaneously provide traditional images with good spatial resolution and very high, $1-2$~eV, spectral resolution across the array. In this work, we explore how utilizing the new technology of X-ray IFUs can drastically change our understanding of galaxy formation. We assume capabilities similar to the Line Emission Mapper (LEM) Probe mission concept \citep{Kraft2022-vi}. The LEM concept is designed to have a large field-of-view ($\gtrsim \SI{900}{arcmin^2}$), state-of-the-art X-ray microcalorimeter with 1~eV spectral resolution in the central array, and 2~eV spectral resolution across the field-of-view. The single-instrument telescope is planned to have $\SI{2500}{cm^2}$ collecting area at $\SI{1}{keV}$ energy. Overall, the spectral resolution of LEM surpasses that of CCD-like instruments by $50-100$ times, allowing us to spectrally separate the Milky Way foreground lines and the emission lines from the CGM of external galaxies. 

Modern cosmological hydrodynamical simulations are able to model the detailed distribution of the hot CGM (e.g., \citealp{Cen1999-eb,Cen2000-cw}).
The divergence among these simulations is chiefly driven by the difference in modelling baryonic physics, most notably the modelling of feedback processes, such as those driven by star formation activities or accretion onto SMBHs (see e.g. \citealp{Vogelsberger2020-hc} for a recent review). 
Intrinsic limitations due to the finite resolution of these simulations require sub-resolution models, which add uncertainties to the results. Sub-resolution feedback effects are implemented to mimic net effects of AGN feedback, but depend on the numerical implementation. The simulations of the IllustrisTNG project, for example, switch from thermal to kinetic feedback mode, depending on the chosen thresold of the AGN accretion rate (\citealp{Weinberger2017-pj,Vogelsberger2020-hc}). 
Other simulations, such as EAGLE, use only the thermal AGN feedback channel to reheat the gas (\citealp{Schaye2015-po}). 
Thus, simulations are significantly diverse in predicting the CGM properties (e.g., X-ray line emission profiles, see \citealp{Van_de_Voort2013-lj,Wijers2022-cf,Truong2023-am}). 
Therefore, probing the hot phases of the CGM is essential to understand how feedback processes operate on galactic scales, and future observations will constrain models by comparison of observations with simulations. 

In this work, we utilize three modern hydrodynamical structure formation simulations, IllustrisTNG, EAGLE, and Simba, to demonstrate that a large grasp imaging microcalorimeter will provide an unprecedented view into the formation and evolution of galaxies. 
In section \ref{ch:analysis} we describe the hydrodynamical simulations and the setup of the mock observations. We show the surface brightness profiles of four bright emission lines for galaxy subsamples selected by halo mass and star formation rate in section \ref{ch:results}. We also quantify the level of substructure and present 2D maps of the temperature and element abundance ratios inferred from a spectral analysis. Section \ref{ch:discussion} discusses the results.

\section{Methods} \label{ch:analysis} 
Here we describe our methodology for the analysis of microcalorimeter mock observations. 

\subsection{X-ray microcalorimeter}\label{ch:xrayprobe} 
There are currently several X-ray missions and mission concepts with a microcalorimeter, such as Athena X-IFU (\citealp{Barret2013-hy,Barret2018-hb,Barret2023-ow}), the X-Ray Imaging and Spectroscopy Mission (XRISM, \citealp{Terada2021-dr}), the Hot Universe  Baryon  Surveyor (HUBS, \citealp{Zhang2022-ze}), the Explorer of Diffuse Emission and Gamma Ray Burst Explosions (Edge, \citealp{Piro2007-ou}), Super DIOS (\citealp{Sato2022-en}), and the Line Emission Mapper (LEM, \citealp{Kraft2022-vi}). 
\begin{deluxetable}{ccccc}
	\tablecaption{FoM for CGM detection of selected past, current and future X-ray missions and concepts, adapted from \cite{Li2020-vy}.\label{tab:missions}}
    \tablehead{
    Mission & $R = E/\Delta E$ & $A_{\rm eff}$   &  $\Omega_{\rm FoV}$ & FoM\\
            & at 1\,keV        & $\si{cm^2}$     &  $\si{deg^2}$       & $\si{cm^2\,deg^2}$
    }
    \startdata
    \multicolumn{5}{c}{Currently operating missions} \\
    XMM-Newton & 20 & 2000 & 0.20 & 8000 \\
    eROSITA & 20 & 2000 & 0.79 & $\num{31000}$ \\
    XRISM & 200 & 250 & $\num{2e-3}$ & 125 \\
    \hline
    \multicolumn{5}{c}{Previously proposed missions} \\
    Super DIOS & 300 & 1000 & 0.25 & $\num{75000}$\\
    EDGE & 330 & 1000 & 0.5 & $\num{165000}$\\
    \hline
    \multicolumn{5}{c}{Currently proposed missions} \\
    Athena/XIFU & 250 & 5800 & $\num{4e-3}$ & 6400 \\
    HUBS & 300 & 500 & 1 & $\num{150000}$\\
    LEM & 500 & 2500 & 0.25 & $\num{312000}$ \\
    \enddata
\end{deluxetable}
However, the CGM can be best probed by a large area, large field of view instrument (i.e., large grasp, which is the product of the field of view and collecting area),  designed for the X-ray detection of the faint emission lines of the CGM. 
We follow the definition of \cite{Li2020-vy} for the figure of merit to map the hot phase of the CGM, ${\rm FoM} = R \,A_{\rm eff} \,\Omega_{\rm FoV}$, where the spectral resolving power $R=E/\Delta E$ is measured at 1\,keV, the effective area $A_{\rm eff}$ at 1\,keV measured in cm$^2$, and the solid angle $\Omega_{\rm FoV}$ in deg$^2$. 
Using this metric it can be seen in Tab. \ref{tab:missions} that a LEM-like mission is ideally suited to detect and measure the properties of the CGM. We use the LEM mission design to illustrate the power of an large grasp microcalorimeter to probe the CGM. 
We note that the spectral resolution must allow distinguishing bright Milky Way foreground lines from the typically fainter and redshifted CGM emission lines, namely C\,VI, O\,VII, O\,VIII, Fe\,XVII. 
Therefore, lower spectral resolution can in principle be compensated by selecting more distant galaxies, which will be fainter and require deeper exposures. For comparison we also include two current CCD instruments in Tab. \ref{tab:missions}, XMM-Newton (\citealp{Jansen2001-cw,Lumb2012-pa}) and eROSITA (\citealp{Predehl2021-rj}).

Spatial resolution is not a key parameter as the point source contribution (e.g., from AGNs) to X-ray flux from nearby galaxies can be modeled in narrow bands. However, a smaller PSF is beneficial as it allows spatial masking of the brightest point sources in the field from structure in the CGM. 
We note that in the case of LEM, the spatial resolution (half-power diameter, HPD) of $\SI{10}{\arcsec}$ is also superior to most other missions listed in Tab. \ref{tab:missions}.  

Although the LEM concept includes a high-resolution ($\sim \SI{1}{eV}$) central array ($5\times 5\,\si{arcmin}$), we only consider the energy resolution of the main array (2\,eV).  
A galaxy with similar stellar mass as the Milky Way extends out to  $R_{500}=\SI{165}{kpc}$. 
At redshift $z=0.01$, this galaxy will span $\SI{26}{arcmin}$ and fit well within the ${32}{\arcmin}\times \SI{32}{\arcmin}$ field of view.
Therefore, we conservatively only consider a 2\,eV resolution throughout the field of view. 
Larger and brighter galaxies at a redshift $z=0.035$, which contain about twice the Milky Way stellar mass, can be tested for CGM emission even beyond $R_{500}$. Therefore, we investigate the CGM emission for galaxies at these two redshifts assuming a LEM-like instrument. 

\subsection{Mock observations of hydro-dynamical cosmological simulations}\label{ch:mocks}
\begin{figure*}[htbp]
    \centering
    \includegraphics[width=0.9\textwidth]{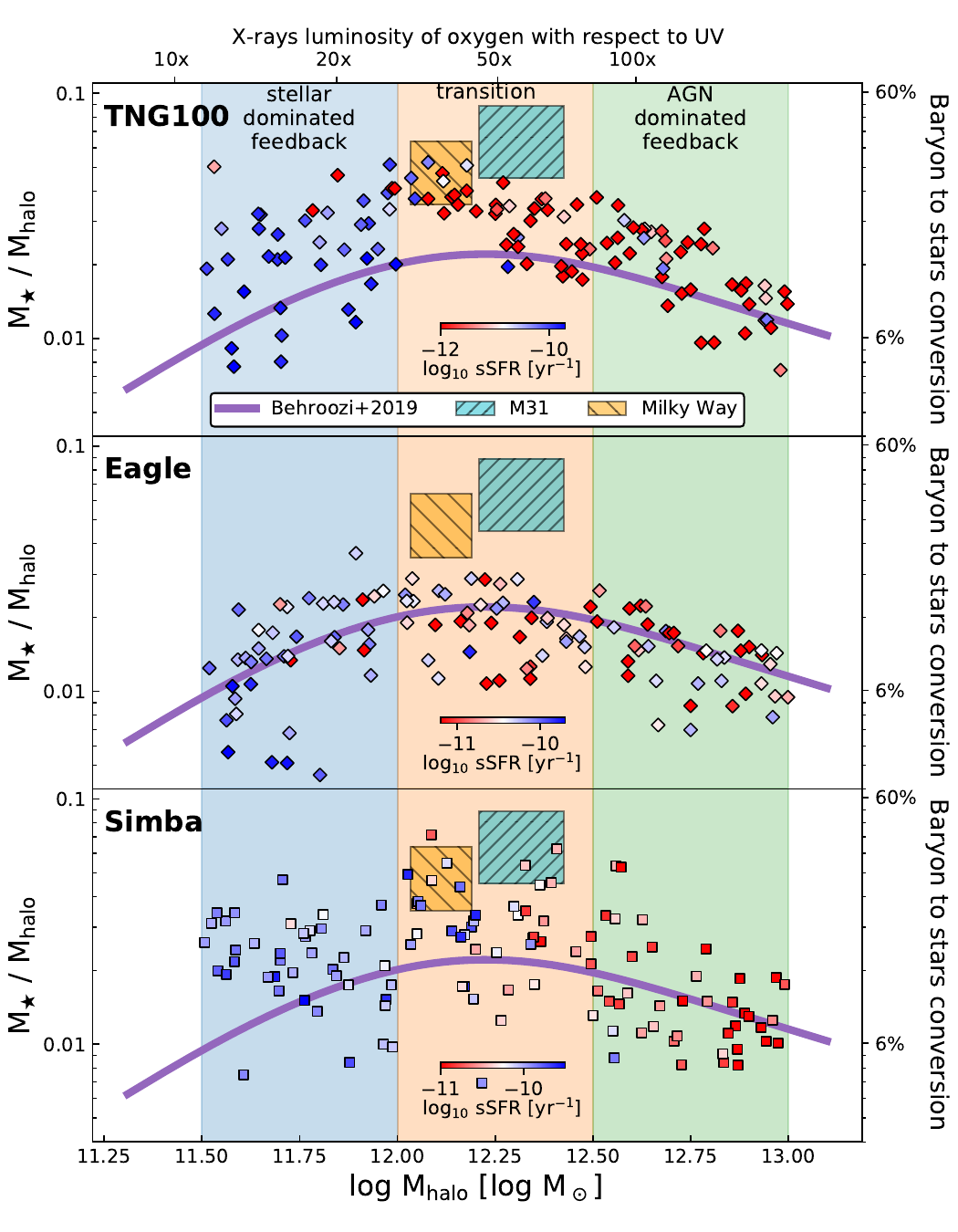}
    \caption{Stellar mass fraction as a function of the halo mass of the galaxies. 
    The three panels (top, middle, bottom) show the selected galaxies from TNG100, Eagle, and Simba, respectively. 
    The blue, orange and green shaded halo mass bins represent the three feedback regimes that we explore. The baryon to star conversion efficiency on the right y-axis assumes a cosmic baryon fraction of $f_{\rm b} = 0.17$. 
    The top axis indicates the fraction the X-ray emitting oxygen ions (O\,VII, O\,VIII) with respect to O\,VI, which is dominant in the UV (as predicted by \citealp{Nelson2018-an}. The purple line shows the predictions from \cite{Behroozi2019-fk} for $z=0$ and excludes satellite galaxies within larger halos.}
    \label{fig:stellar_mass}
\end{figure*}

\subsubsection{Simulations}\label{ch:sims}
Our galaxy  selection from the cosmological simulations follows clear criteria. 
We focus here on  three state-of-the-art simulations, TNG100 of the IllustrisTNG project (\citealp{Vogelsberger2014-br,Vogelsberger2014-bw,Pillepich2017-ge,Weinberger2017-pj,Naiman2018-gi,Nelson2018-br,Pillepich2017-bz,Pillepich2018-ur,Springel2018-wz,Marinacci2018-lz,Nelson2019-ud,Nelson2019-vh,Pillepich2019-sw}, with $\sim \SI{110.7}{cMpc}$ box size (comoving coordinates), and a baryon mass $m_{\rm baryon} = \SI{1.4e6}{M_\odot}$), EAGLE-Ref-L100N1504 (EAGLE, \citealp{Crain2015-jo,McAlpine2016-la,Schaller2015-iw,Schaye2015-po,Wijers2022-cf}, with $\sim \SI{100}{Mpc}$ box size, and a baryon mass $m_{\rm baryon} = \SI{1.81e6}{M_\odot}$), and Simba 100Mpc (Simba, \citealp{Hopkins2015-uo,Dave2019-cc}, with $\sim \SI{147}{Mpc}$ box size, and a baryon mass $m_{\rm baryon} = \SI{1.82e7}{M_\odot}$), which all have similar volumes and cosmological parameters, are tuned to reproduce observed stellar properties, but vary in hydrodynamic codes and modules for galaxy formation, including AGN feedback prescriptions (e.g., \citealp{Davies2019-yu,Davies2019-nz,Zinger2020-qu,Truong2021-ds,Byrohl2022-xn,Donnari2021-on}). All three simulations trace the evolution of gas, stellar, dark matter, and SMBH particles over a large redshift range. Gas heating and cooling processes are included, as well as star formation and stellar evolution, and  various feedback processes (AGN, SNe, stellar winds). The implementation of the AGN feedback is significantly different in the three simulations, emphasizing the need for a detailed comparison of observables that can trace the feedback mechanisms. 
For example, EAGLE uses a thermal AGN feedback model with a single efficiency of mass-to-energy (\citealp{Booth2009-me}), while TNG and Simba have multiple energy injection modes (\citealp{Oppenheimer2020-ix,Truong2023-am}). 
In order to understand the impact of stellar-driven and AGN feedback on the CGM, and how it can be traced with a large grasp microcalorimeter, we subdivide simulated galaxies into halo mass bins based on the $M_{200}$ (the total mass within 200 times the critical density of the Universe at the redshift of the galaxy). 

\subsubsection{Samples}\label{ch:samples}
\begin{figure*}[htbp]
    \centering
    \includegraphics[width=0.99\textwidth]{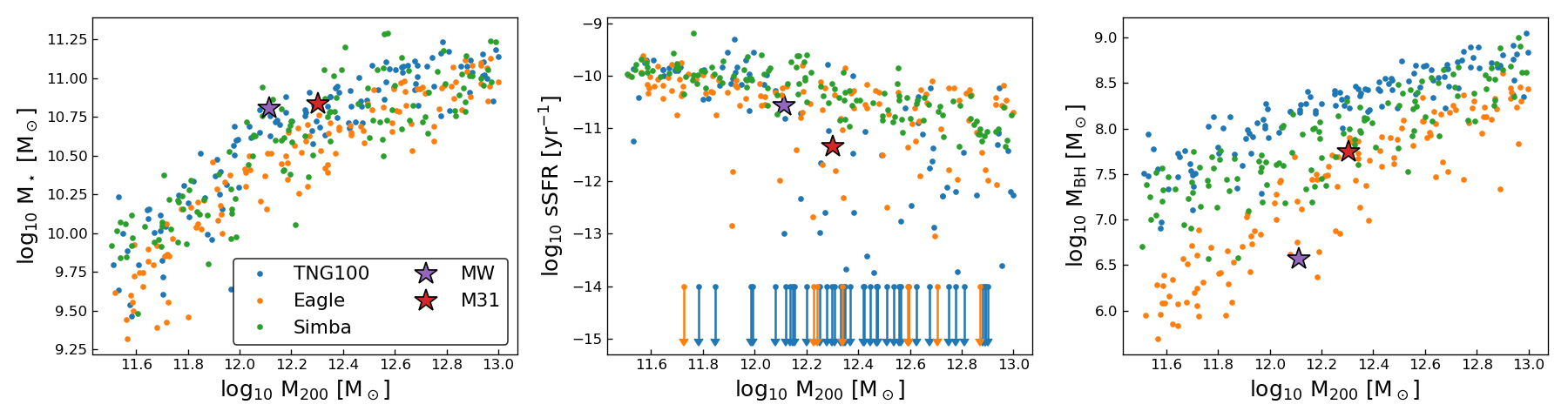}
    \caption{Properties of the galaxies in the three samples (low/medium/high mass) for each of the three simulations (TNG100 in blue, EAGLE in orange, Simba in green). The x-axis of each of the three panels is the halo mass M$_{200}$ in log$_{10}$. The left panel shows the stellar mass dependence M$_\star$, similar to what is shown in \ref{fig:stellar_mass}. We also highlight the location of the Milky Way (purple star), and M31 (red star). The middle panel illustrates the specific star formation rate, sSFR in units of $\si{yr^{-1}}$, clearly decreasing with increasing halo mass. The right panel shows the SMBH mass dependence, clearly indicating a different trend between the simulations, where EAGLE produces lower mass black holes than TNG100.}
    \label{fig:samples}
\end{figure*}

In simulations (\citealp{Donnari2021-on,Sorini2022-op,Mitchell2022-pv,Ayromlou2022-aj}) and observations (\citealp{Mandelbaum2006-fl,Zheng2007-hv,Gavazzi2007-wd,Hansen2009-bm,Guo2010-vt,Wang2010-mp}) a peak of the stellar mass fraction of non-satellite galaxies is observed around the Milky Way and M31  halo mass and stellar mass fraction (see \citealp{Schodel2002-xz,Licquia2015-qs,Posti2019-es} for MW properties, and \citealp{Tamm2012-fp,Rahmani2016-os,Al-Baidhany2020-ee,Carlesi2022-zc} for M31, both indicated in Figs. \ref{fig:stellar_mass} and \ref{fig:samples}). 
MW and M31 analogs have been found by \cite{Ramesh2022-vg} in TNG50 simulations with realistic multiphase gas properties, e.g., warm and cold gas reservoirs. 
The peak of the stellar mass fraction, and growing mass of the SMBHs implies a transition from a supernovae dominated mode of feedback to AGN dominated feedback. This transition is poorly understood and only a detailed analysis of the CGM  will  reveal the driving mechanism of this feedback mode change. 
Therefore, we select three galaxy samples, based on halo mass, to encapsulate the different forms of the dominant feedback (see blue, orange, and green regions in Fig. \ref{fig:stellar_mass}). 

We select galaxies with M$_{200}$ from $10^{11.5}$ to $\SI{1e12}{M_\odot}$ as our \textit{low mass sample} (Fig. \ref{fig:samples} and Tab. \ref{tab:samples}). In this mass range, the stellar mass fraction increases with halo mass, which means more and more gas is converted into stars indicating strong stellar feedback (see e.g., \citealp{Behroozi2010-vq,Moster2010-ut,Harrison2017-sv,Behroozi2019-fk} and references therein). Galaxies in the low mass sample are also generally have a lower central black hole masses (around $\SI{7e7}{M_\odot}$ in TNG100), and a high specific star formation rate (${\rm sSFR} \sim \SI{6e-11}{yr^{-1}}$ for instance in the case of EAGLE), which are not representative of the typical galaxy across the three samples. 

Galaxies in the mass range from M$_{200} = 10^{12}$ to $10^{12.5}\,\si{M_\odot}$ form our \textit{medium mass sample} (Fig. \ref{fig:samples} and Tab. \ref{tab:samples}), and generally are not clearly dominated by either one feedback mechanism. 
The impact of both stellar and AGN feedback on the CGM should be visible in these galaxies, which have significantly more massive central AGNs (median at least three times as massive as in the low mass sample, Tab. \ref{tab:samples}), and, depending on the simulation, also  lower specific star formation rates. 

Our \textit{high mass sample} consists of halo masses from M$_{200} = 10^{12.5}$ to $10^{13}\,\si{M_\odot}$ (Fig. \ref{fig:samples} and Tab. \ref{tab:samples}), and is generally dominated by AGN feedback, while star formation and stellar feedback become less and less important. The central black hole masses are on average a factor of 2.3 larger than in the medium mass sample, while the specific star formation rate decreases by a factor of 2 (see Fig. \ref{fig:samples}).

From each simulation, TNG100, EAGLE, and Simba, we select 40 galaxies for each of the three samples. 
We also exclude any galaxy that is a non-central galaxy of the dark matter halo, e.g., a member galaxy of a cluster that evolves very differently due to an early onset of quenching by the surrounding ICM. 
We note that the stated halo masses refer to the parent dark matter halo of each galaxy. 
By restricting our sample to halo masses below $\SI{e13}{M_\odot}$ we exclude all galaxies within clusters or groups. 
The galaxies have been selected to be uniformly distributed in $\log M_{200}$. 
Our samples are summarized in Table \ref{tab:samples}, and we describe all the individual galaxies in more detail in Tables \ref{tab:TNG}, \ref{tab:Eagle}, and \ref{tab:Simba}, and show the galaxy properties, including halo and stellar mass, star formation rate, and black hole mass in Fig. \ref{fig:samples}.

\begin{deluxetable*}{cc|ccc|ccc|ccc}
	\tablecaption{Galaxy halos samples for tracing and mapping the CGM emission.\label{tab:samples}}
    \tablehead{
    \multicolumn{2}{c}{Simulation} & \multicolumn{3}{c}{\textbf{Illustris TNG}}  & \multicolumn{3}{c}{\textbf{EAGLE}}   & \multicolumn{3}{c}{\textbf{Simba}} \\
    \multicolumn{2}{c}{box} & \multicolumn{3}{c}{TNG100}  & \multicolumn{3}{c}{Ref-L0100N1504}   &  \multicolumn{3}{c}{100 Mpc/h} \\
    \colhead{}  & \colhead{}  &  \colhead{low}  & \colhead{medium}  & \colhead{high}  & \colhead{low}  & \colhead{medium}  & \colhead{high} & \colhead{low}  &\colhead{medium}  & \colhead{high}
    }
    \startdata
Sample size &  & 40    & 40    & 40    & 40    & 40    & 40    & 40    & 40    & 40     \\\hline
\bf log$\mathbf{_{10}}$ M$\mathbf{_{200}}$ &\bf median  & \bf11.79   & \bf12.29   & \bf12.73   & \bf11.72   & \bf12.26   & \bf12.76   & \bf11.76   & \bf12.21   & \bf12.73    \\
$[\log_{10} {\rm M_\odot}]$ & 25\%  & 11.65    & 12.17    & 12.63    & 11.63    & 12.15    & 12.64    & 11.63    & 12.15    & 12.58     \\
& 75\%  & 11.92   & 12.39   & 12.89   & 11.84   & 12.35   & 12.88   & 11.87   & 12.35   & 12.87    \\\hline
\bf log$\mathbf{_{10}}$ M$\mathbf{_\star}$ &\bf median & \bf10.15   & \bf10.77   & \bf11.06   & \bf9.91   & \bf10.52   & \bf10.93   & \bf10.07   & \bf10.72   & \bf10.95    \\
$[\log_{10} {\rm M_\odot}]$ & 25\% & 9.94   & 10.71   & 10.95   & 9.64   & 10.41   & 10.80   & 9.96   & 10.60   & 10.74    \\
 & 75\% & 10.32   & 10.85   & 11.11   & 10.09   & 10.65   & 10.97   & 10.21   & 10.84   & 11.05    \\\hline
\bf log$\mathbf{_{10}}$ SFR &\bf median  & \bf-0.01   & \bf-3.95   & \bf-1.36   & \bf-0.32   & \bf0.03   & \bf0.15   & \bf0.22   & \bf0.61   & \bf0.33    \\
$[{\rm M_\odot}\,{\rm yr^{-1}}]$ & 25\%  & -0.34   & -5.00   & -5.00   & -0.51   & -0.61   & -0.93   & 0.11   & 0.34   & 0.18    \\
 & 75\%  & 0.13   & -0.71   & -0.41   & -0.12   & 0.32   & 0.50   & 0.33   & 0.78   & 0.66    \\\hline
\bf log$\mathbf{_{10}}$ M$\mathbf{_{\rm BH}}$ &\bf median  & \bf7.73   & \bf8.33   & \bf8.68   & \bf6.33   & \bf7.46   & \bf8.14   & \bf7.38   & \bf7.94   & \bf8.43    \\
$[\log_{10} {\rm M_\odot}]$ & 25\%  & 7.50   & 8.21   & 8.53   & 6.09   & 7.17   & 7.96   & 7.18   & 7.62   & 8.25    \\
 & 75\%  & 7.99   & 8.43   & 8.75   & 6.67   & 7.69   & 8.31   & 7.57   & 8.07   & 8.62    \\\hline
\bf R$\mathbf{_{500}}$ &\bf  median & \bf123   & \bf176   & \bf242   & \bf116   & \bf170   & \bf250   & \bf117   & \bf166   & \bf238    \\
$[{\rm kpc}]$ & 25\%  & 110   & 163   & 222   & 105   & 158   & 225   & 107   & 154   & 218    \\
 & 75\% & 135   & 193   & 274   & 125   & 184   & 272   & 126   & 183   & 264    \\
$[{\rm arcmin}]$ & median  & 9.96   & 14.32   & 5.78   & 9.42   & 13.84   & 5.97   & 9.54   & 13.47   & 5.68   
    \enddata
    \tablecomments{For each quantity, the halo mass M$_{200}$, the stellar mass within 30\,kpc M$_\star$, the halo star formation rate (SFR), the SMBH mass M$_{\rm BH}$, and the characteristic radius R$_{500}$, we show the median value of the sample, as well as the 25\textsuperscript{th} and 75\textsuperscript{th} percentiles. To calculate R$_{500}$ in arcmin, we  place the high mass samples at $z=0.035$, while the low and medium mass samples are at $z=0.01$.}
\end{deluxetable*}

\subsubsection{Mock X-ray observations}\label{ch:mocks_xray}
We produce mock observations from the hydro simulations using pyXSIM\footnote{\url{http://hea-www.cfa.harvard.edu/~jzuhone/pyxsim/}} (see \citealp{ZuHone2016-vy}), which creates large photon samples from the 3D simulations. 
We create mock observations from a high spectral resolution imaging instrument assuming a Gaussian PSF with 10~arcsec FWHM using SOXS\footnote{\url{https://hea-www.cfa.harvard.edu/soxs/}}. We use a detector size of $128\times 128$ pixels with $\SI{15}{\arcsec}$ per pixel, yielding a FoV of $\SI{32}{\arcmin}\times\SI{32}{\arcmin}$. The spectral bandpass covered is 0.2 to 2\,keV, with 2\,eV FWHM resolution. 
The X-ray emission is modeled from each emitting, non-star-forming gas cell with  $T > \SI{3e5}{K}$ and $\rho < \SI{5e-25}{g\,cm^{-3}}$ (a value of the density close to the star formation density threshold in the TNG simulations). 
The electron and hydrogen number densities in each gas cell are determined from the underlying simulations (TNG100, EAGLE, ans Simba, see section \ref{ch:sims}), which track the conversion of gas into stars, starting from primordial initial abundance. The electron number density depends on the density, temperature and abundance of the gas. The emission measure of each gas cell is calculated based on electron and hydrogen number densities. 
In each galaxy, there is also a small set of isolated gas cells which are abnormally bright in X-rays--these typically have extreme values of cooling time and/or thermal pressure, and on this basis are excluded from the analysis to improve visualizations, but we
do not find that leaving them in changes any of our conclusions (see \citealp{ZuHone2023-le}, for more details). 

The plasma emission of the hot gas surrounding the galaxy is based on the Cloudy emission code (\citealp{Ferland2017-cq}), and includes the effect of resonant scattering from the CXB, which enhances the O\,VIIr line (\citealp{Chakraborty2020-po,Chakraborty2022-lp}). An extensive description is provided in \cite{Churazov2001-ap,Khabibullin2019-ls}. 
In contrast to other emission models, such as APEC/AtomDB (\citealp{Foster2012-op,Foster2018-jj}), we utilize a density-dependent model, that is sensitive to the photo-ionization state of the gas at densities $\lesssim \SI{1e-4}{cm^{-3}}$ (see, e.g., \citealp{Bogdan2023-qg}). We updated the code with respect to \cite{Khabibullin2019-ls} to include the latest version of Cloudy, and ensured that the intrinsic resolution matches the sub-eV requirements of a microcalorimeter. The various metal species are independent of each other, allowing a consistent modelling of gas with arbitrary metal abundance patterns. 

However, we neglect the effects of resonant scattering from  the hot ISM gas in the galaxy (\citealp{Nelson2023-my}).
We place the low and medium mass samples at a nearby redshift of $z=0.01$, which separates the forbidden O\,VII line, the C\,VI line, the O\,VIII line and the 725 and 739\,eV Fe\,XVII lines from the MW foreground. However, the O\,VII resonance line is blended with the MW foreground forbidden line. The extent of the low and medium mass halos (R$_{500}$) fits well within the assumed field of view of the detector. However, since galaxies in the high mass sample are too large to fit in the $32\times\SI{32}{arcmin^2}$ field of view,  we chose to place these galaxies at $z=0.035$. This allows us to include the resonance line of O\,VII, but blends the 739\,eV Fe\,XVII line with MW foreground (see Fig. \ref{fig:SBR_windows}). 

The Galactic foreground emission is assumed to consist of a thermal component for the local hot bubble (LHB, temperature $\SI{0.1}{keV}$, \citealp{McCammon2002-bm}), an absorbed thermal model to account for Galactic halo emission (GHE, temperature of $\SI{0.23}{keV}$, \citealp{McCammon2002-bm}, and a velocity broadening of $\SI{100}{km\,s^{-1}}$), and the North Polar Spur or hotter halo component (NPS, temperature $\SI{0.7}{keV}$, see \cite{Das2019-uz,Bluem2022-ow}, also broadened by $\SI{100}{km\,s^{-1}}$). Each thermal component is implemented with the APEC model  (\citealp{Foster2018-jj}) with solar abundances (\citealp{Anders1989-ou}), and the absorption with the tbabs model (\citealp{Wilms2000-uu}) with a hydrogen column density of $\SI{1.8e20}{cm^{-2}}$. The normalizations are $\mathcal{N}_{\rm LHB} = \SI{1.7e-6}{cm^{-5}\,arcmin^{-2}}$ for the LHB, and $\mathcal{N}_{\rm GHE} = 0.43 \,\mathcal{N}_{\rm LHB}$ for the GHE, and $\mathcal{N}_{\rm NPS} = 0.05\,\mathcal{N}_{\rm LHB}$ for the NPS. The spatial distributions are flat in the mock event files. 

The astrophysical background contains unresolved X-ray point sources, mostly distant AGNs (cosmic X-ray background, CXB). On average, the flux distribution of a source follows a powerlaw $S_\nu \propto \nu^{-2}$ (\citealp{De_Luca2004-le,Hickox2006-fo}), and the $\log N - \log S$ distribution from \cite{Lehmer2012-th}.  The average powerlaw normalization is $\SI{4.1e-7}{pht\,s^{-1}\,keV^{-1}\,cm^{-2}\,arcmin^{-2}}$ after excising the brightest 50 point sources from the event file, which make up half of the total CXB flux (and about 3\% of the FoV area).  

Considerations on the particle background based on Athena X-IFU (\citealp{Barret2013-hy,Barret2018-hb}) studies showed that a spectral component due to Galactic cosmic rays will be a factor of 30 to 60 lower than the second lowest component, the CXB. We included a conservative estimate on the residual particle background, after anti-coincidence filtering, of $\SI{1}{cts~s^{-1}\,keV^{-1}}$ for the field of view. The particle background is assumed to have a flat spectrum and no spatial features. 
Our mock event files include all the above-mentioned components and simulate a 1\,Ms observation.

\subsection{Analysis}
The analysis of the mock event files relies partly on existing software, such as CIAO, but most routines are re-implemented in python using \textit{astropy}  (\citealp{The_Astropy_Collaboration2013-lw,The_Astropy_Collaboration2018-gx}), and the \textit{scipy} packages. 

\subsubsection{Preparation}
While the pixel size of a LEM-like detector array is $\SI{15}{arcsec}$, the optics and mirror assembly reach a spatial resolution of $\SI{10}{arcsec}$, which will be utilized through Lissajous-dithering.  Therefore, we oversample the detector pixels by a factor of 2 for all images that are produced, e.g. for point source detection. 
To start the analysis of the simulated event files, we visually inspect the images and spectra around the O\,VII(f) emission line, 
determine the redshift by locating the peak of the emission line, and extract the line surface brightness $\pm \SI{2}{eV}$ around the O\,VII(f) line. We use the surface brightness distribution to calculate the emission weighted centroid, which is used as the center of our profiles.

\begin{deluxetable}{lccccc}
    \tablecaption{CGM emission lines used throughout this work.\label{tab:lines}}
    \tablehead{
    Element & Energy & Peak $kT$  & \multicolumn{2}{c}{Line Photometry} & Spectral\\
            &  [eV]  & [keV]      &  $z=0.01$ & $0.035$ & analysis
    }
    \startdata
    C\,VI & 367.5 & 0.11 & \cmark & \cmark & \xmark \\ 
    O\,VII (f) & 561.0 & 0.17 & \cmark & \cmark & \cmark \\ 
    O\,VII (i) & 568.6 & 0.17 & \xmark & \xmark & \cmark  \\
    O\,VII (r) & 574.0 & 0.17 & \xmark & \cmark & \cmark \\ 
    O\,VIII & 653.7 & 0.27 & \cmark & \cmark & \cmark \\ 
    Fe\,XVII & 725.1 & 0.43 & \cmark & \cmark & \cmark \\ 
    Fe\,XVII & 727.0 & 0.43 & \cmark & \cmark & \cmark \\
    Fe\,XVII & 739.0 & 0.43 & \cmark & \xmark & \cmark \\ 
    Fe\,XVII & 825.8 & 0.54  & \xmark & \cmark & \cmark \\
    \enddata
    \tablecomments{The peak temperature for each transition is the plasma temperature at which the excitation rate is maximum assuming collisional ionization equilibrium. However, this number is for information only, since we include effects such as photoionization in our simulations. }
\end{deluxetable}

\subsubsection{Point sources}
Following these initial tasks, we use the \verb|wavdetect| algorithm included in the CIAO 4.14 package (\citealp{Fruscione2006-wt}) to detect point sources from the CXB in the observation, and in the corresponding background file. Since the point sources are expected to have a continuum powerlaw spectrum, we use a broad band image, $\SIrange{250}{950}{eV}$ for the detection. While several hundred  point sources are typically detected, we select the 50 most significant and brightest sources, which contribute about 50\% to the total CXB flux, but only cover about 2-3\,\% of the detector area. The least significant of the top 50 sources is still detected at $100\sigma$. Examples of the detected sources can be seen in the top left panel of Fig. \ref{fig:spectrum_panels}. 
The area of these 50 point sources is masked out in the observation and background event file for the following analysis steps. More details on the point source contribution is given in section \ref{ch:appendix_ps}. 

\subsubsection{Surface brightness profiles}\label{ch:method_sbr}
\begin{figure*}[htbp]
    \centering
    \includegraphics[width=0.99\textwidth]{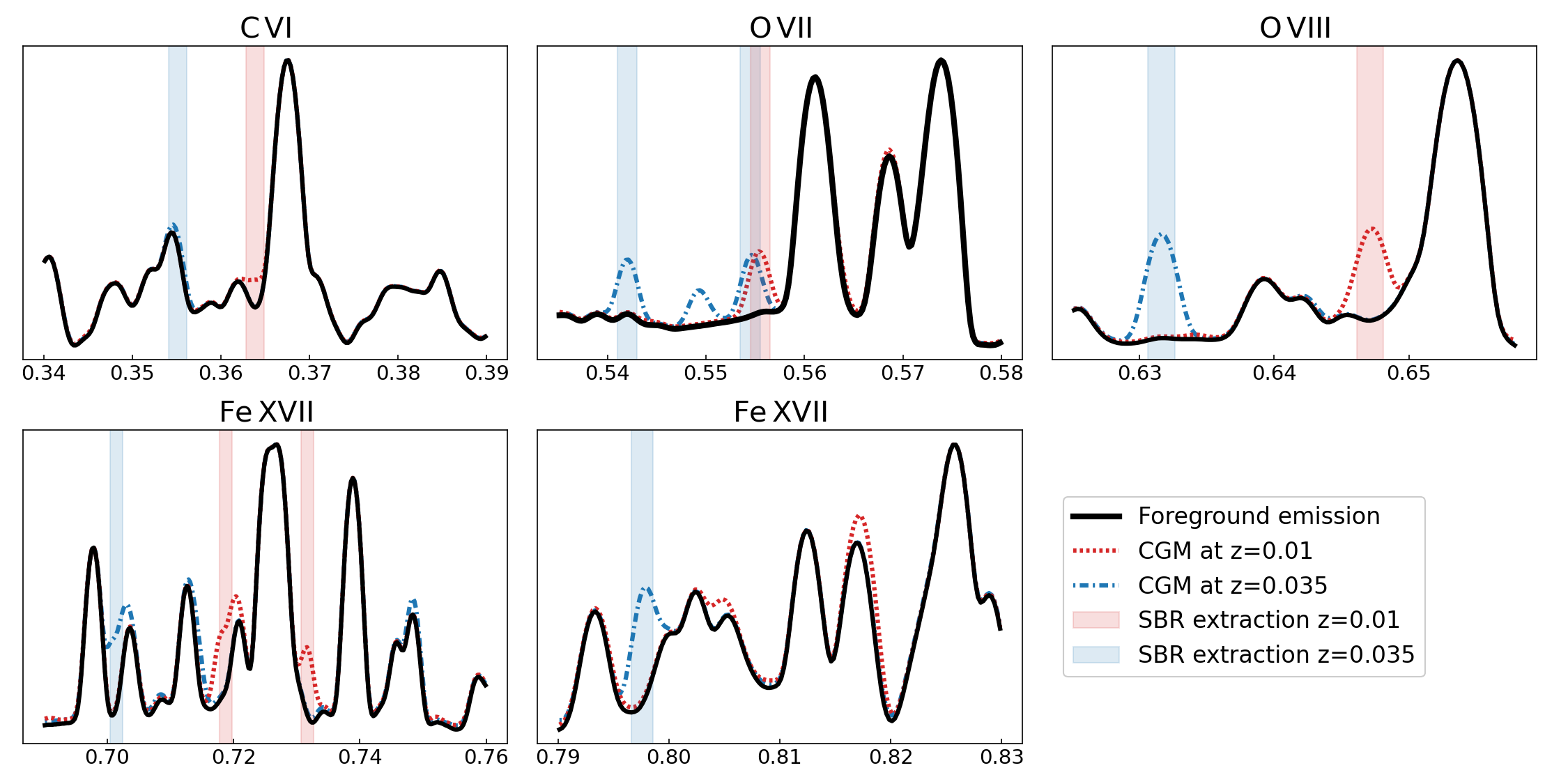}
    \caption{Illustration of the spectral windows used for the surface brightness extraction (see Tab. \ref{tab:lines}). The x-axes of the panels are in keV, and the flux on the y-axes in arbitrary units. The black lines show a spectral model of the total background, i.e. Milky Way foreground, and CXB emission. 
    We show the thermal emission with arbitrary normalization (to highlight the location of lines) with a thermal model at $\SI{0.2}{keV}$, marked by the blue and red lines for $z=0.035$ and $z=0.01$, respectively. The small windows marked by the colored bars are used to distinguish the CGM from the foreground emission.}
    \label{fig:SBR_windows}
\end{figure*}
Our goal is to quantify the CGM emission out to large radii for galaxies in our three subsamples (see section \ref{ch:samples}). Surface brightness profiles of bright emission lines (see \textit{Line Photometry} column in Tab. \ref{tab:lines}) allow us to understand the structure of the CGM as a function of radius. Therefore, we extract for each line the counts from the mock event files around the previously defined emission weighted center. We use narrow 2\,eV spectral windows centered at the redshifted line energies.  
The 2\,eV spectral window is the optimal width in terms of signal-to-noise, and it includes 76\% of the line counts. 
Figure \ref{fig:SBR_windows} shows the redshifted spectral extraction with respect to the foreground emission for several interesting line windows considered here. 
The red shaded regions correspond to the extraction for the nearby ($z=0.01$) galaxies, the blue shaded regions to the more distant ($z=0.035$) galaxies. 
In the case of O\,VII, we can use the forbidden and resonant lines at $z=0.035$, while for the nearby galaxies at $z=0.01$, we only use the forbidden line. For Fe\,XVII, we coadd the three lines, 725\,eV, 727\,eV and 739\,eV, at lower redshift ($z=0.01$), while at $z=0.035$, we use the 725\,eV, 727\,eV and the 826\,eV lines.

For the surface brightness profiles we determine the width of the radial bins (annuli) to achieve, a) a minimum signal-to-noise ratio of 3, and b) a minimum source-to-background ratio of 10\%.  While a) limits our statistical uncertainties, b) prevents systematics in the background from biasing our result.  
We estimate the background counts based on a simulated blank field observation with only foreground and CXB background components, where we repeat the same steps that were performed for the CGM observation, in particular the point source detection, as this field has a different realization of the CXB point sources. 
The background counts are estimated in the same narrow bands, but from the whole field of view (minus the excised point sources) to reduce the statistical uncertainty in the background. 
Since this background estimate assumes a field of view averaged residual CXB contribution (point sources that are fainter than the 50 brightest that are excluded), we introduce a scaling factor to the background counts in each annulus, based on the broad band emission ($\pm \SI{200}{eV}$ around the line) of the continuum CXB sources (see appendix \ref{ch:appendix_sbr} for details).

To combine the results from all individual galaxy profiles of the various subsamples, we build a median (not stacked) profile, where we take the median surface brightness at each radius and use the 68\% scatter among the profiles of our subsample to represent the galaxy-to-galaxy variation.

\subsubsection{Structural clumping in the gas}
\label{ch:clumping}
While the radial surface brightness profiles demonstrate our ability to detect the CGM to large distances, they do not quantify the level of substructure present in the gas at a given radius, 
nor do they show how well an X-ray microcalorimeter can detect/characterize the substructure.
It is expected that different feedback mechanisms leave imprints in the CGM. 
Stellar feedback is able to expel gas from the inner region near the disk to larger radii and cause  not only metal enrichment, but also an observable anisotropic distribution of structure within the gas (\citealp{Peroux2020-hp,Truong2021-ds,Nica2022-gr}), especially within intermediate radii ($\sim 0.5 R_{500}$). A very dominant central AGN will have a major impact on the halo gas distribution. After several feedback cycles, it is expected that the gas distribution is smoothed by the impact of the AGN.

To capture this information from our mock observations, we divide each galaxy azimuthally into sectors, for which we compute the surface brightness of emission lines. The ratio between mean and median surface brightness of all the sectors traces the asymmetry and clumpiness of the X-ray gas (approximated by the ratio of the average squared density to the square of the average density, e.g., \citealp{Nagai2011-la}),
\begin{equation}
    \mathcal{C} (r) = \frac{\langle \rho^2 \rangle}{\langle\rho\rangle^2} \approx \frac{S_{\rm mean}(r)}{S_{\rm median}(r)}~,
\end{equation}
where $\rho$ is the gas density. We note that $\mathcal{C}(r)$ is also known as the emissivity bias in the literature (\citealp{Eckert2015-mx}). Typical values from observations range from 1, meaning no azimuthal asymmetry or substructure, to about 2 at large radii. 
Emission from a single line is more prone to vary, due to temperature variations in the gas. We calculate $\mathcal{C}$ in annuli of width $0.25R_{500}$ from stacked images of the O\,VII(f), the O\,VIII, C\,VI, and the Fe\,XVII 725\,eV line. 
This combines forbidden and resonant lines, and the latter are more sensitive to geometry. As \citealp{Nelson2023-my} points out, the O\,VIII line is only very mildly sensitive to resonant scattering. 
We combine the signal from these emission lines, which also increases the signal to noise. We trace the substructure out to a radius of $R_{500}$ (4 radial bins), while using 8 sectors (45\,deg each). We then stack the $\mathcal{C}$ profiles for galaxies to derive the median profile. We notice that, especially for lower mass halos, the scatter in $\mathcal{C}$ is substantial. Since we are only interested in the type of galaxy where $\mathcal{C}$ is significantly larger than 1, we use the range of $\mathcal{C}$ values in the 50\textsuperscript{th} to 75\textsuperscript{th} percentile as a diagnostic.  

\subsubsection{Spectral analysis}
\label{ch:spec_analysis}
Besides the line photometry to measure the extent and distribution of the CGM, a large grasp microcalorimeter can also map the dynamical, thermal, and chemical structure of the CGM by analyzing the spectrum. This provides insights on gas motion, outflows, enrichment history, and the dominant re-heating process. For example, we can use the signal of the various emissions lines to constrain the CGM temperature, abundance ratios, and line-of-sight velocity. 
To avoid fitting all spectral features within the instrument bandpass, we decide to constrain the main properties by focusing on the most important emission lines. We fit a model, that includes background and foreground components, in small (8\,eV) spectral windows around the emission lines (Tab. \ref{tab:lines}) 

For the spectral mapping, we choose the region size based  on the brightness distribution of the three lines, O\,VII(f), O\,VIII, and Fe\,XVII, through an adaptive binning technique (see \citealp{OSullivan2013-ef,Kim2019-fa}), which we briefly describe here: 
At every pixel of the combined line image, we derive a radius at which we reach a threshold signal to noise. This radius can be different for neighboring pixels. The spectral extraction region for each pixel is given by the determined radius. As a consequence, the spectra of neighboring pixels are not independent, and we will oversample the map. 
We use a signal-to-noise of 10 as a threshold parameter, and we do not include pixels if the radius has to be larger than $\SI{7}{\arcmin}$. 

We assume the foreground model and CXB models are not known a priori, and constrain their parameters through spectral fitting of a separate background spectrum. 
This background spectrum is extracted from the same observation, using a region outside the central $\SI{15}{\arcmin}$ radius, and remove bright regions from the galaxy CGM and exclude point sources.
This leaves about 30\% of the detector area for the background spectrum, which is enough to measure all parameters (temperatures and normalizations of LHB, GHE, NPS, and CXB) with high precision. 
The spectral extraction is done using the CIAO tools dmextract, and dmgroup to have a grouped spectrum file with at least one count per bin.

For the spectral fitting, we use Sherpa (\citealp{Freeman2001-hz,Burke2020-vd}), which is distributed  with CIAO, and also provides the Xspec models, such as APEC and tbabs \citep{Arnaud1996-uy,Foster2012-op}.
We model the background/foreground emission with two absorbed APEC models (GHE and NPS, including thermal and velocity broadening with a velocity of $\SI{100}{km\,s^{-1}}$), and one unabsorbed APEC model (LHB) plus one absorbed power low (CXB) to fit the background components in the mock observations (see section \ref{ch:mocks}). 
Contamination of the background spectrum with CGM emission of the targeted galaxy is small, since we use a large radius for spectral extraction. However, to account for residual contamination, we add a CGM component to the background spectral fitting as well. The parameters of this CGM component are not used later on. 
We are able to reproduce the input background parameters within 2\% relative accuracy. 

For spectral fitting of the actual CGM regions (determined through our adaptive binning) we apply the previously determined background parameters and freeze these values. 
We only leave the CXB powerlaw normalization free to vary, since each spectral region can contain a slightly different population of CXB sources, while the CXB normalization from the background spectrum is just the average over a larger area. 

Our CGM emission model consists of 19 absorbed components: The individual elements C, N, O, Ne, Mg, Si, S, Fe, plus a model for all other elements, one component for each element that includes the effect of resonant scatter of the cosmic microwave background  (\citealp{Ferland2017-cq,Khabibullin2019-ls}), and one component for the emission from plasma without metals.
The normalization of the resonant scattered components is frozen to half the normalization of the corresponding element, since we removed about half the flux from CXB emission through the point source masking. 
Each CGM component has a temperature, redshift, and intrinsic hydrogen  density (for the photo-ionization). We link the temperature, redshift, and density between the elements, as each chemical element has its own emission model component. In order to account for velocity broadening of the CGM model components we apply a Gaussian smoothing kernel (xspec gsmooth, with $\alpha=1$) to the CGM model before convolution with the instrumental response matrix. 
Furthermore, we note that photoionization is mostly unimportant for the regions (within ${\sim 0.8 R_{500}}$) and masses considered here (see appendix \ref{ch:deviation_cie}), and therefore we do not leave the density free to vary.
We use Cash statistics (\citealp{Cash1979-vm}) for the fitting of the input spectrum. 

As mentioned before, we do not use a broad band of the spectrum, but rather select 8\,eV narrow spectral windows centered on the 11 interesting CGM lines listed in Tab. \ref{tab:lines}. 
The reduced $\chi^2$ values are typically very close to 1.  
After the best-fit parameters are found, we use the MCMC sampling method integrated into Sherpa to find the parameter uncertainties using the Metropolis Hastings algorithm. 

\begin{figure*}[t]
    \centering
    \includegraphics[width=0.99\textwidth]{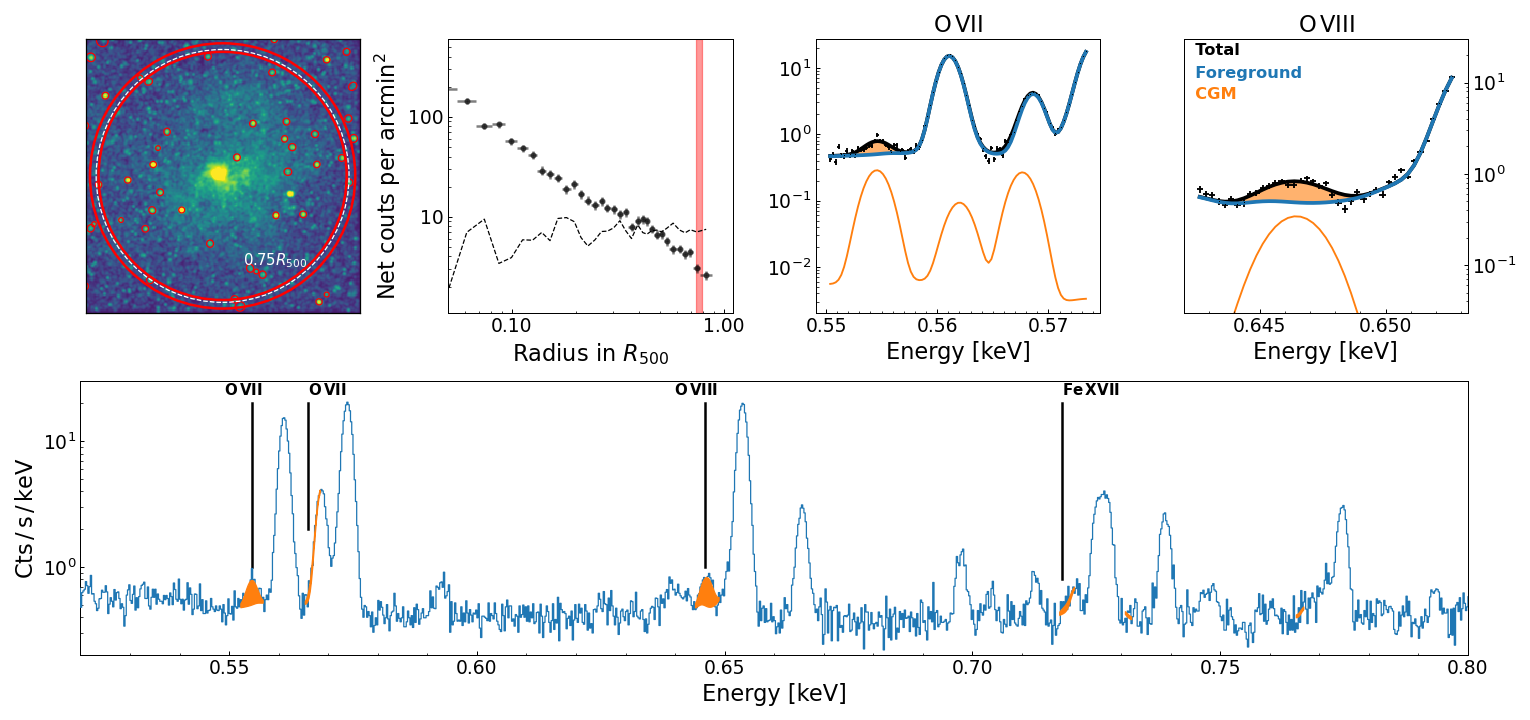}
    \caption{Example for a $10^{11.18}\,\si{M_\odot}$ stellar mass galaxy with $R_{500}=\SI{243}{kpc}$ (TNG50, ID 358608, at $z=0.01$). A map of the narrow band O\,VIII emission (including background/foreground) is shown in the top left panel, where all the detected point sources are marked by small red circles, and the extraction region used for the other panels is shown as a large red annulus around the $0.75\,\si{R_{500}}$ marker. A surface brightness profile of the O\,VIII line is shown in second to right top panel, and the spectral region of both, the O\,VII and \,VIII emission including the Milky Way foreground, in the two top left panels. A wider spectrum is shown in the bottom panel, with the CGM emission in excess of the foreground/background emission as the orange shaded area. }
    \label{fig:spectrum_panels}
\end{figure*}
\section{Results} \label{ch:results} 

The results of our analysis demonstrate the extraordinary capabilities of a large grasp X-ray microcalorimeter for detecting and characterizing the CGM. 
As an example, we show in Figure \ref{fig:spectrum_panels} the line emission image, surface brightness profile and spectra for a TNG50 galaxy at the upper mass end ($\log_{10} M_\star / M_\odot = 11.18$, and $R_{500}=\SI{243}{kpc}$)  at $z=0.01$. The same galaxy is also analyzed in detail in section \ref{ch:spectral_analysis} using temperature, velocity and abundance ratio maps. However, we note that typical galaxies of the low and medium mass samples will be fainter. 

Figure \ref{fig:spectrum_panels} illustrates the power of a LEM-like mission to probe the CGM properties to large radial distances.  We illustrate this capability by focusing on a narrow annulus at $0.75R_{500}$. The top left panel shows thecombined image of O\,VII, O\,VIII, and Fe\,XVII, which fills the field of view. The red annulus shows the extraction region that has a 1\,arcmin width. The panel to the right shows the surface brightness profile, and the background level (dashed line). At the red shaded extraction region, which is the same as in the first panel, the CGM emission still reaches 50\% of the background. The other panels show spectra extracted from the same region, and highlight the CGM model (orange), over the foreground emission (blue). The two top right panels show close-ups of the spectral region around the O\,VII triplet and the O\,VIII line, while the bottom panel shows the spectrum from 544 to $\SI{800}{eV}$ with the prominent CGM emission lines indicated.

\begin{figure*}[htbp]
    \centering
    \fbox{\includegraphics[width=0.45\textwidth]{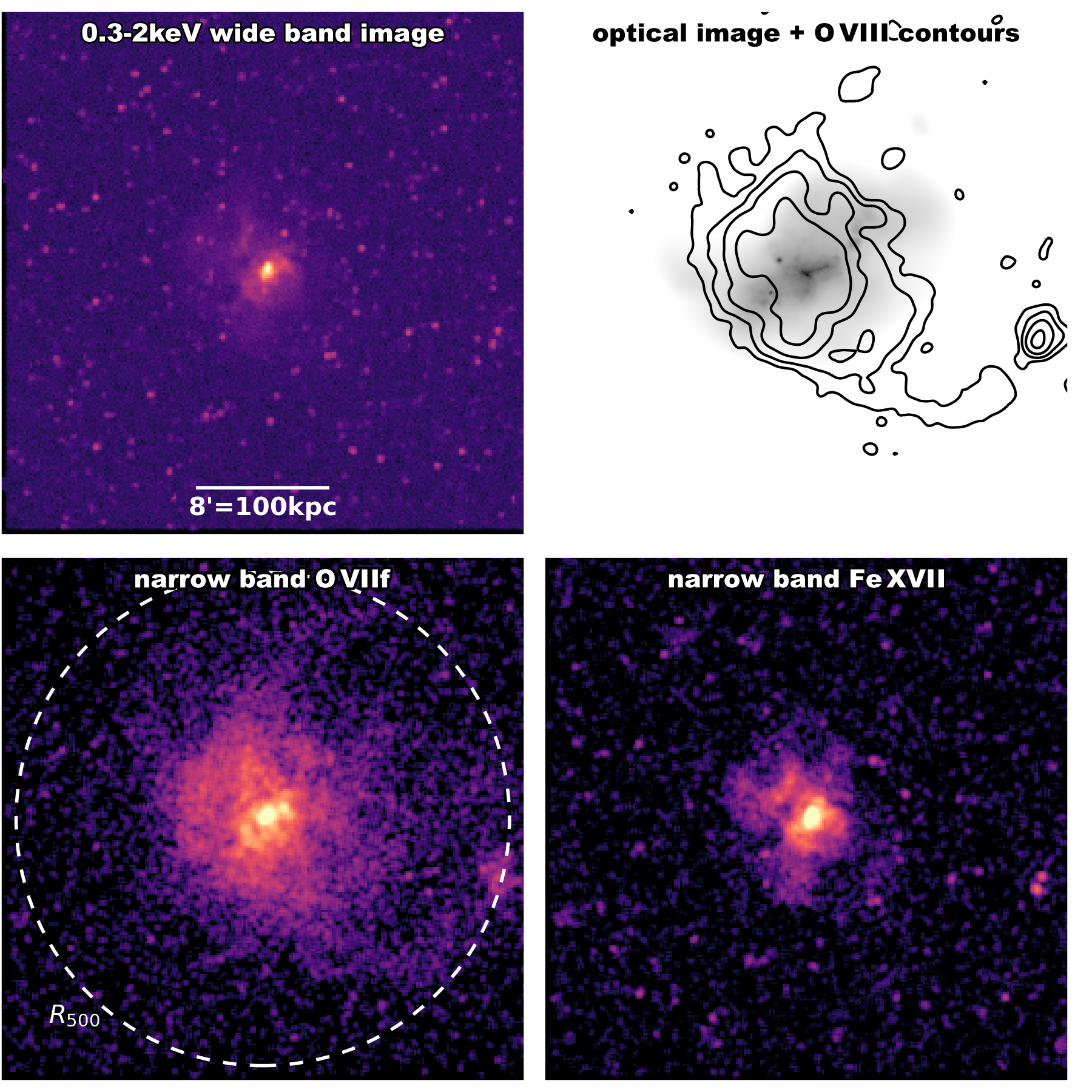}}~~
    \fbox{\includegraphics[width=0.45\textwidth]{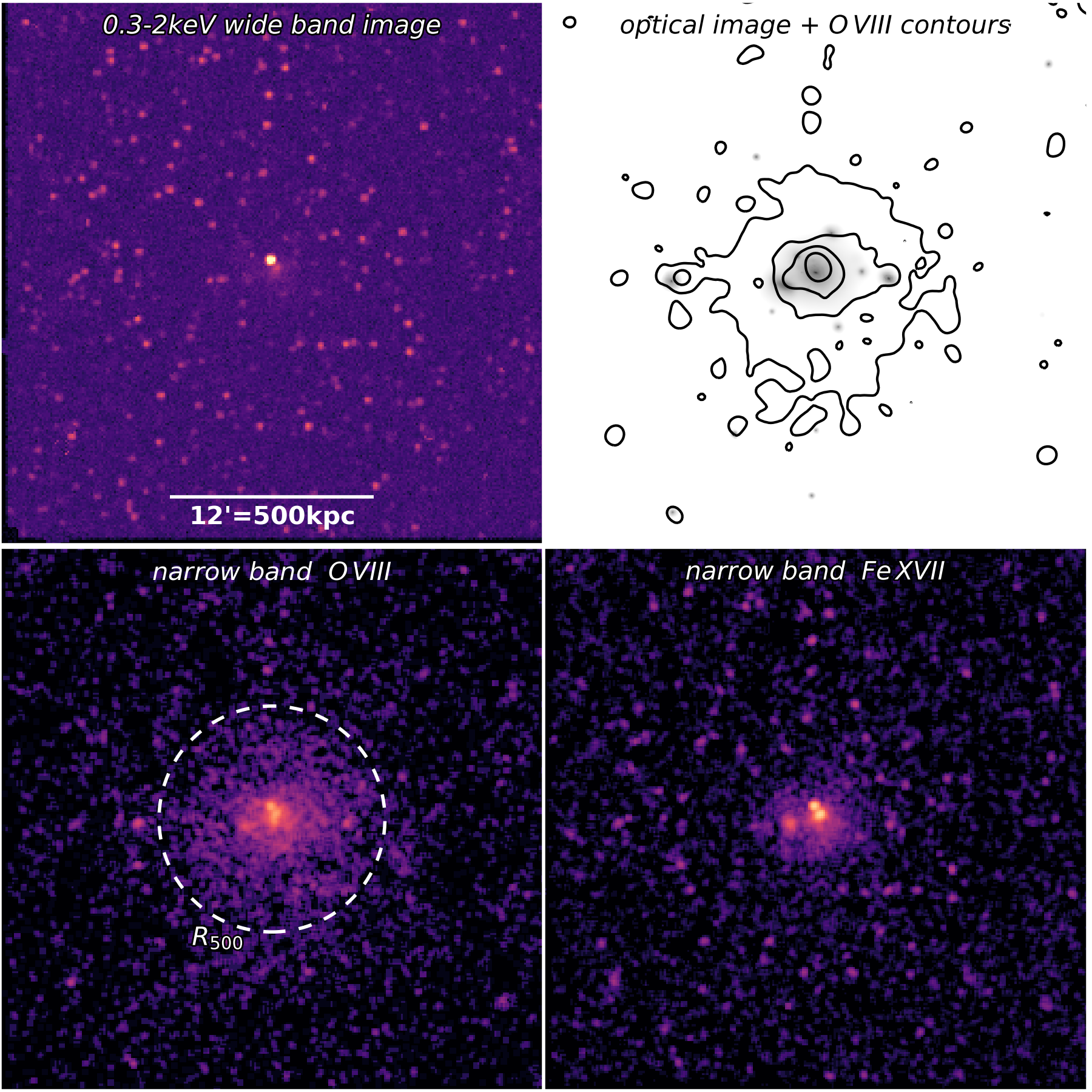}}
    \caption{Illustrating the capabilities of a LEM-like mission for mapping the CGM: \textit{Left:} Galaxy halo of the size of the Milky Way at z=0.01 (TNG100, ID 417281). \textit{Right:} High mass galaxy at z=0.035 (TNG100, ID 337444). The 4 panels in each image show the broad band image similar to a CCD resolution instrument (top left), the optical r-band image with O\,VIII contours (top right), the O\,VII(f) or O\,VIII image with R$_{500}$ indicated (bottom left), and the Fe\,XVII image (bottom right).}
    \label{fig:4panels_others}
\end{figure*}

In figure \ref{fig:4panels_others} we exhibit the imaging capabilities and highlight the superiority of a high spectral resolution microcalorimeter over an otherwise equivalent CCD imaging X-ray instrument for CGM science. For the two TNG100 galaxies shown in the panels (left is a MW-like galaxy at $z=0.01$, right is a high mass galaxy at $z=0.035$) we select narrow spectral bands for imaging (O\,VII or O\,VIII and Fe\,XVII, bottom panels). 
The top left quadrant of each panel shows a broad-band image ($0.3-2$\,keV) where only the core of the galaxy is visible. An optical r-band image (top right quadrant in each panel) shows the distribution of stellar light, which is much smaller than the X-ray CGM emission (black contours). To quantitatively understand how far the emission can be traced, we analyze median line surface brightness profiles.

\subsection{Line surface brightness profiles} \label{ch:profiles} 
We extract surface brightness profiles of four important emission line complexes, C\,VI, O\,VII, O\,VIII, and Fe\,XVII (see also Tab. \ref{tab:lines}). 
The profiles are extracted in a $\SI{2}{eV}$ window around the redshifted line energy, which minimizes the contamination with foreground lines from the MW. We extract a profile of a simulated background/foreground observation in the same way and subtract this from the observation of the CGM. 

Each of the low, medium, and high mass samples contains 40 galaxies, while the former two have the galaxies at $z=0.01$, and the latter at $z=0.035$. 
Details on the individual galaxies, such as stellar mas, gas mass, and black hole mass, are given in Tables \ref{tab:samples}, \ref{tab:TNG}, \ref{tab:Eagle}, and \ref{tab:Simba}. 

\begin{figure*}[htpb]
    \centering
    \includegraphics[width=0.97\textwidth]{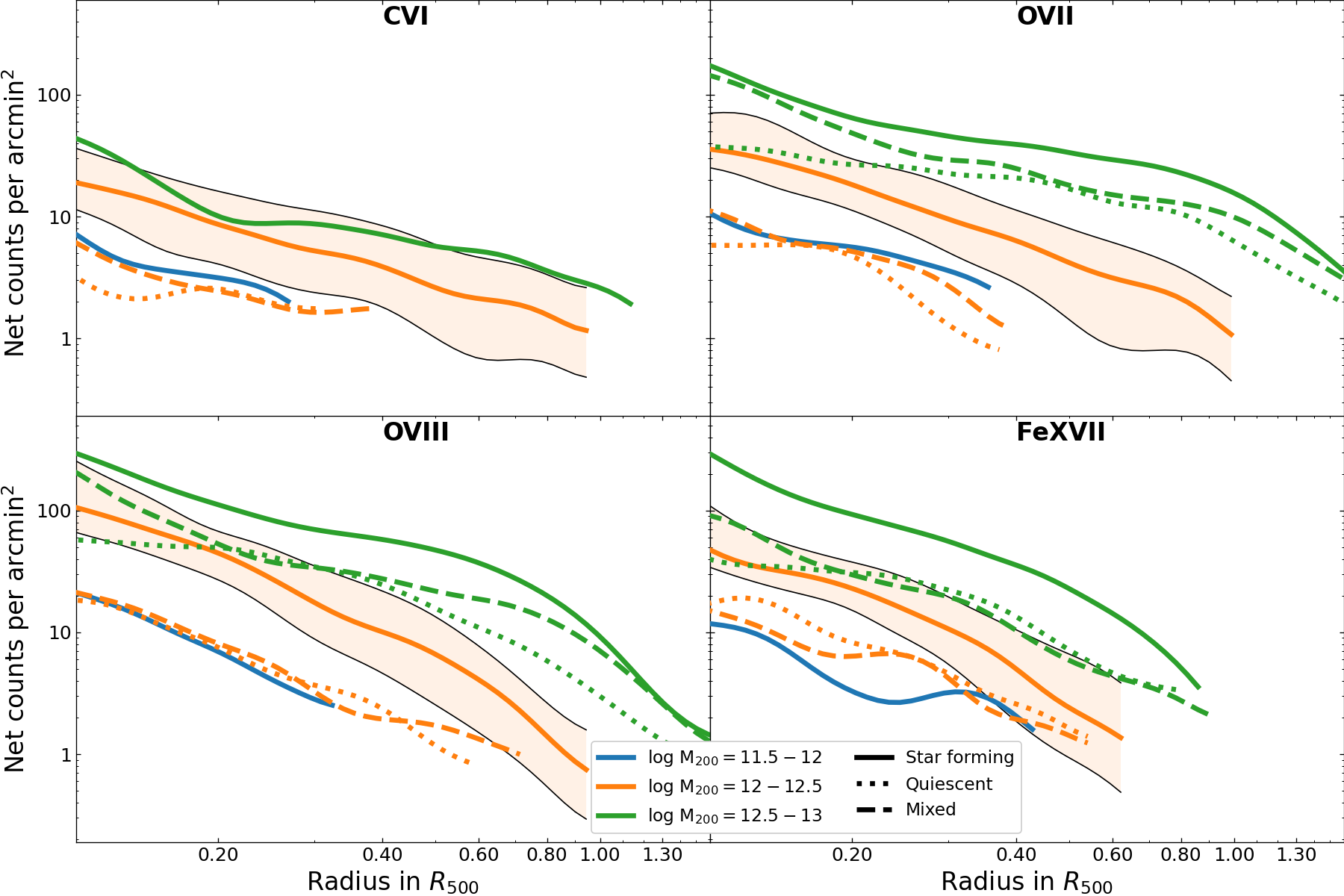}
    \caption{Median profiles of the CGM emission from the EAGLE galaxies, binned by halo mass ($M_{200}$). The solid, dotted, and dashed lines correspond to star-forming, quiescent, and mixed subsamples, respectively. Blue, orange, and green curves show the low, medium, and high mass galaxy halos, respectively. The orange shaded region represents the 68\% scatter, shown only for star-forming, medium mass galaxies for visibility purposes.}
    \label{fig:profiles_Eagle}
\end{figure*}

\begin{figure*}[htpb]
    \centering
    \includegraphics[width=0.97\textwidth]{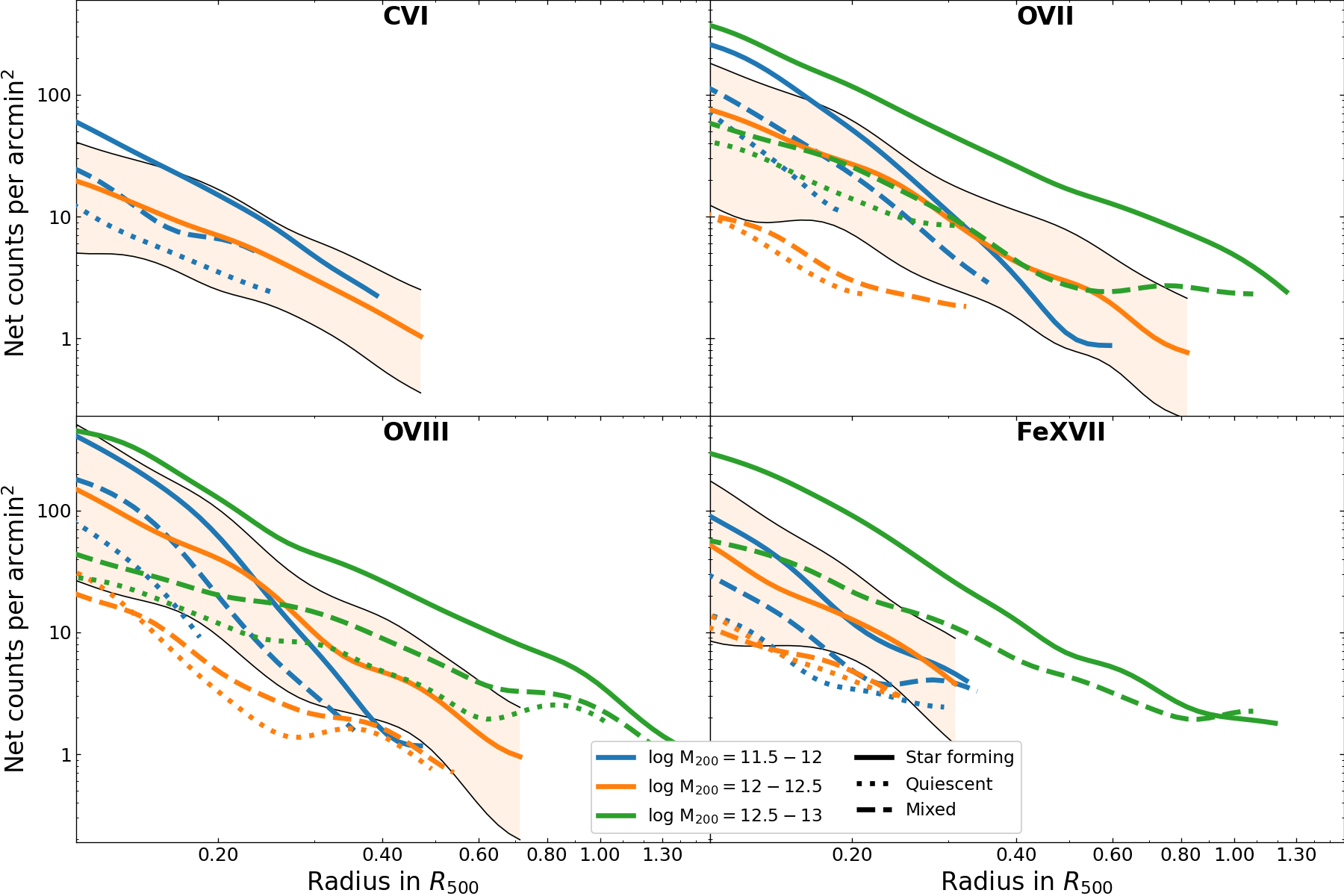}
    \caption{As Fig. \ref{fig:profiles_Eagle}, but for TNG100 galaxy halos.}
    \label{fig:profiles_TNG}
\end{figure*}

\begin{figure*}[htpb]
    \centering
    \includegraphics[width=0.97\textwidth]{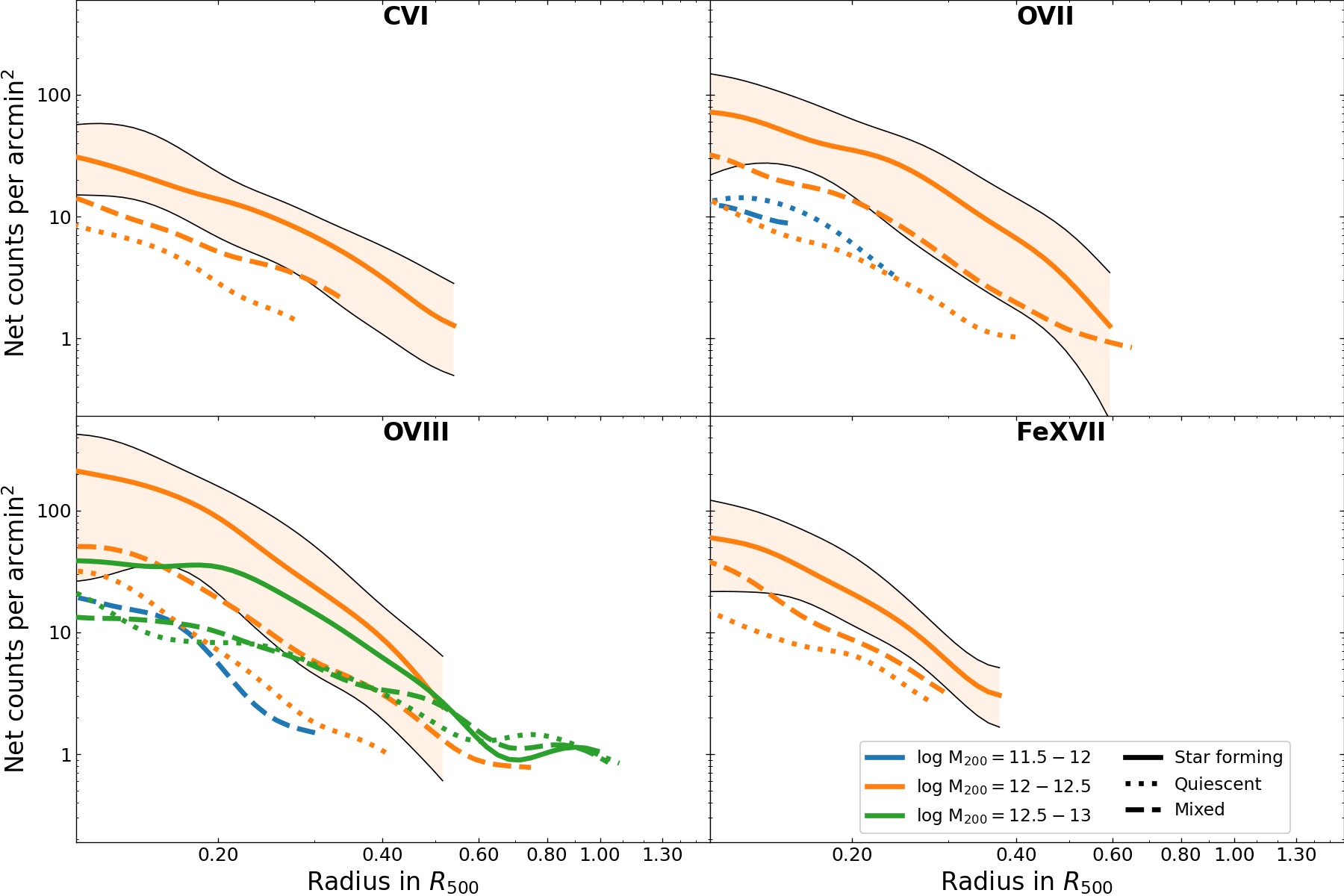}
    \caption{As Fig. \ref{fig:profiles_Eagle} but for Simba galaxy halos.}
    \label{fig:profiles_Simba}
\end{figure*}

We analyze the median profile with the radius scaled by R$_{500}$. 
Within each sample (low/medium/high mass) we use the 33 and 66 percentiles of the star formation rates as thresholds to subdivide further. We therefore have 9 (sub-)samples, three star formation samples per mass, which all comprise either 13 or 14 galaxies.

We conservatively consider CGM detection if the measured signal is at least 10\% of the background, and the signal-to-noise of each extracted radial bin is at least 3. 
We show profiles in Figs. \ref{fig:profiles_Eagle}, \ref{fig:profiles_TNG}, and \ref{fig:profiles_Simba} for EAGLE, TNG100, and Simba, respectively. 
The low-mass, medium-mass, and high-mass samples are shown in blue, orange, and green, respectively. The subsamples are shown in solid, dotted, and dashed lines for the top, lowest, and intermediate thirds, respectively. Therefore, each line is the median profile of 13 or 14 galaxies. 
The typical 68\% scatter is shown for the star-forming, medium mass sample as the orange shaded region. 

We detect the C\,VI, O\,VII, O\,VIII, and Fe\,XVII lines in emission in all simulations. Based on the galaxy mass, all simulations detect the CGM emission out to R$_{500}$. 
Simba is clearly the faintest. However, there is significant scatter between the galaxies of a single simulation, and between the different simulations. 
O\,VIII can be detected out $R_{200} \approx 1.5 R_{500}$ ($M_{200} \approx 1.35 M_{500}$) for the more massive galaxies in TNG100 and EAGLE, and out to $R_{500}$ in Simba. Other lines that we tested are basically undetected in Simba for the high and low mass samples. 

The O\,VIII is the brightest line, with the highest number of counts and a relatively low background, followed by O\,VII and Fe\,XVII. While C\,VI can be detected in most galaxies, it is very weak in Simba. For the oxygen and carbon, TNG100 and EAGLE are comparable, but TNG100  typically has a steeper shape, leading to a smaller detection radius.
Especially for galaxies in the low and medium mass sample (blue, orange), TNG100 has a clear trend that galaxies with higher star formation rates are also brighter (solid line above dashed line, and the dotted line is lowest). This has also been pointed out, e.g., by \cite{Oppenheimer2020-ix}. 
The lowest mass galaxies with little star formation appear very faint in EAGLE. However, for the medium mass sample, we only partially find the same trend with star formation rate (solid line highest, but dotted and dashed lines comparable). 
For the high mass sample, we find in both, TNG100 and EAGLE, that at larger radii close to $R_{500}$ the brightness is independent of star formation rate (see also \citealp{Oppenheimer2020-ix}). 

The medium mass sample shows oxygen and iron emission at about 0.6 to $1\,R_{500}$ in both simulations, EAGLE and TNG. For C\,VI we find a big difference between EAGLE and TNG100, where in EAGLE carbon is detected out to $0.8\,R_{500}$, and in TNG100 only to about $0.3\,R_{500}$. 
The difference in the visibility of C\,VI between EAGLE and TNG100, at higher halo mass, is not explained by the higher EAGLE CGM  temperatures, but possibly by a different metal composition, as carbon is produced also by AGB stars.
The visibility of the high mass galaxies placed at $z=0.035$ is not limited by the field of view, and can therefore be traced far beyond R$_{500}$ as in the case of O\,VII (both resonant and forbidden line combined, see Fig. \ref{fig:SBR_windows}) and O\,VIII. We also can detect Fe\,XVII in both, EAGLE and TNG100, nearly to R$_{500}$, depending on the star formation rate. C\,VI was not detected in the high mass TNG100 galaxies, but it is very clearly visible in EAGLE.

Clearly, Fe\,XVII is detected best in EAGLE, likely because some of the galaxy halos are hotter. 
Comparing the scatter between the galaxy halos of a given sample, we notice a slightly larger scatter among TNG100 galaxies. The scatter of the CGM profiles between the Simba galaxies can only be measured at smaller radii. 

\subsection{Emission line ratios} \label{ch:ratios} 
\begin{figure*}
    \centering
    \includegraphics[width=0.95\textwidth]{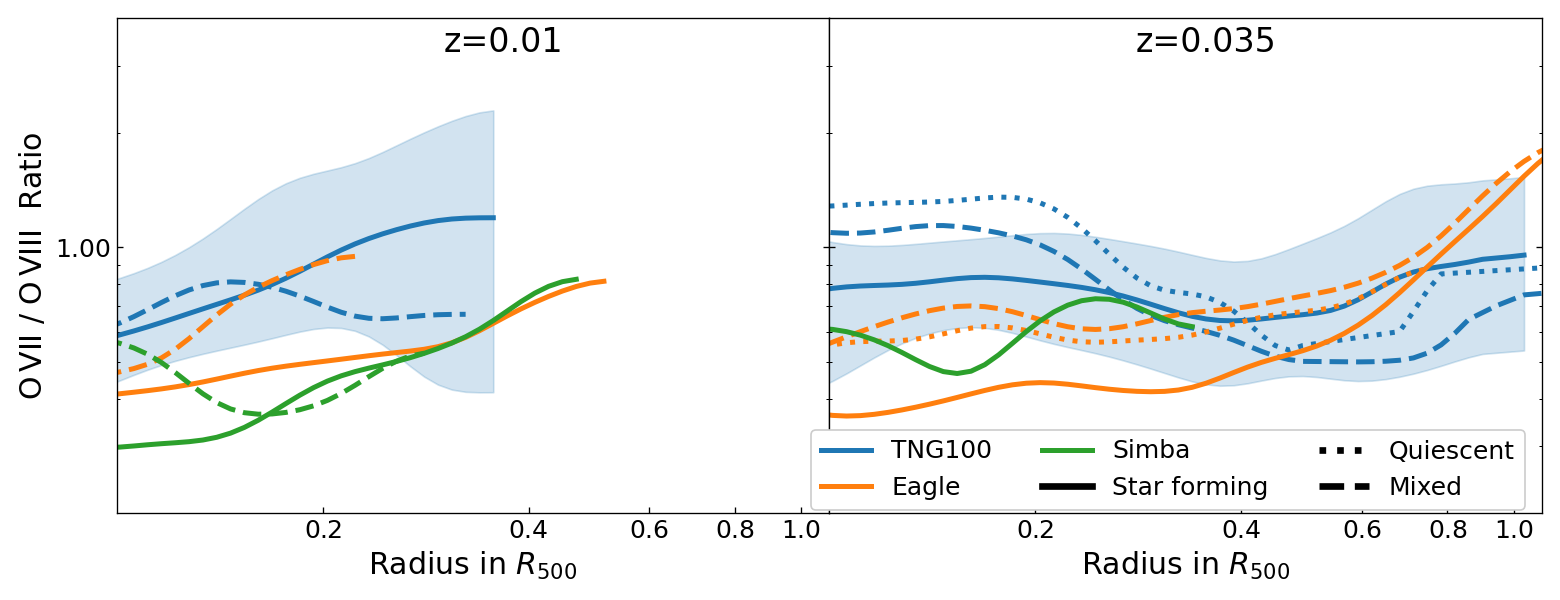}
    \caption{The O\,VII to O\,VIII line ratios for the low and medium mass samples (left), and high mass samples at larger redshift (right). As in the profile plots, the solid, dotted, and dashed lines correspond to star-forming, quiescent, and mixed subsamples, respectively, while the orange curves show EAGLE galaxies, the blue curves TNG100, and the green curves Simba. The blue shaded region represents the 68\% scatter, shown only for star-forming TNG100 galaxies for visibility purposes.}
    \label{fig:ratios}
\end{figure*}
If the hot gas in and around galaxies is not isothermal, we expect the emission line ratio profiles to reflect any deviations from isothermalities of the gas because of the sensitivity of the emission to gas temperature. In the simplest case of a constant (with radius) abundance of heavy elements, and in the limit of collisional ionization equilibrium (CIE), we expect the Fe\,XVII lines to become stronger relative to the O\,VII lines, if the temperature increases with radius (see Tab. \ref{tab:lines}).  
In the case of the same element (e.g., oxygen) the line ratio, e.g., O\,VII and O\,VIII, is proportional to the gas temperature. 
At large radii with low gas densities, photo-ionization becomes important and changes the line ratios. We quantified this transition in appendix \ref{ch:deviation_cie}. In our default analysis we co-add the O\,VII(f) and O\,VII(r) for the high mass samples (i.e. at $z=0.035$), to increase the signal. We also show in appendix \ref{ch:deviation_cie} that the typical densities of these higher mass halos do not show biasing effects of photoionization. Therefore, using both O\,VII lines for higher mass halos to infer temperature does not create a bias. However, the line ratios cannot be compared to the $z=0.01$ samples, where only O\,VII(f) is used. When comparing the emission lines of different elements (e.g., C\,VI and Fe\,XVII), the conclusions are less clear, since not only the temperature changes in the gas, but the enrichment mechanism differ: SNIa contribute significantly to the abundance of iron, but not to that of carbon or oxygen. 

Figure \ref{fig:ratios} shows the O\,VII to O\,VIII line ratio for the same samples that were shown in section \ref{ch:profiles}. 
Looking at the nearby galaxies in Fig. \ref{fig:ratios} (left), we notice a systematic offset between star-forming TNG100 and EAGLE galaxies (solid blue and orange lines), where the EAGLE ratios are always below the TNG100 ones. This indicates that the EAGLE galaxy halos are systematically hotter within the covered radius $< 0.5R_{500}$ (see also \citealp{Truong2023-am}). 
For the mixed star forming galaxies (33rd to 66th percentile of star formation rates), we find comparable line ratios within the scatter among TNG100 and EAGLE, while Simba halos have lower line ratios (see also \citealp{Truong2023-am}).
 Unfortunately, statistics only allow us to derive line ratios to about $0.4\,R_{500}$, until which we see a rising line ratio, indicating a hotter core and cooler outer regions (Fig. \ref{fig:ratios}). 
For the high mass galaxy halos (Fig. \ref{fig:ratios}, right), we see a similar trend of star-forming galaxies being hotter in EAGLE with respect to TNG100 and Simba (lower O\,VII to O\,VIII ratio). TNG100 halos appear almost isothermal (constant line ratio), while EAGLE galaxies have an increasing line ratio toward the outer regions, and become even steeper beyond $0.6\,R_{500}$. At these large radii, the difference between star-forming and quiescent galaxies appears to vanish: the impact of star formation is most prominent in the core. 
In principle, we need to include the effects of photo-ionization in our interpretation of line ratios at large radii, since at very low plasma densities the assumption of CIE is no longer applicable. However, simulating several line ratios with fixed plasma temperature and only changing the density showed that above a density of $\SI{3e-5}{cm^{-3}}$ the changes in the ratio are less than 5-6\%. Densities within $R_{500}$ are expected to be larger than that (e.g., \citealp{Bogdan2013-ca}). We note that for regions with column densities above $N_{\rm H} = \SI{e21}{cm^{-2}}$, the O\,VII to O\,VIII ratio will be affected by electron scattering escape (\citealp{Chakraborty2020-ld}). However, for a typical high mass galaxy ($\log M_{500} \sim 12.5$) column densities are generally below this value.

\subsection{Substructure in the CGM emission} \label{ch:substructure} 
In order to analyze the substructure that can be detected in the mock observations, we apply the  method introduced in section \ref{ch:clumping}. $\mathcal{C}$ is calculated for the TNG100, EAGLE, and Simba halos in all three mass samples.

As pointed out before, $\mathcal{C}\approx 1$  means that there was no clumping detected, and observations have shown that it can rise  up to 2 at $R_{500}$. This can be associated with the substructure in the outskirts being accreted. 
For clusters \citep{Eckert2013-tn,Eckert2015-mx,Zhuravleva2015-zs} have observed very low clumping factor values, even at $R_{500}$, while \cite{Simionescu2011-yq} found higher values for the Perseus cluster. 

\begin{figure}[t]
    \centering
    \includegraphics[width=0.4\textwidth]{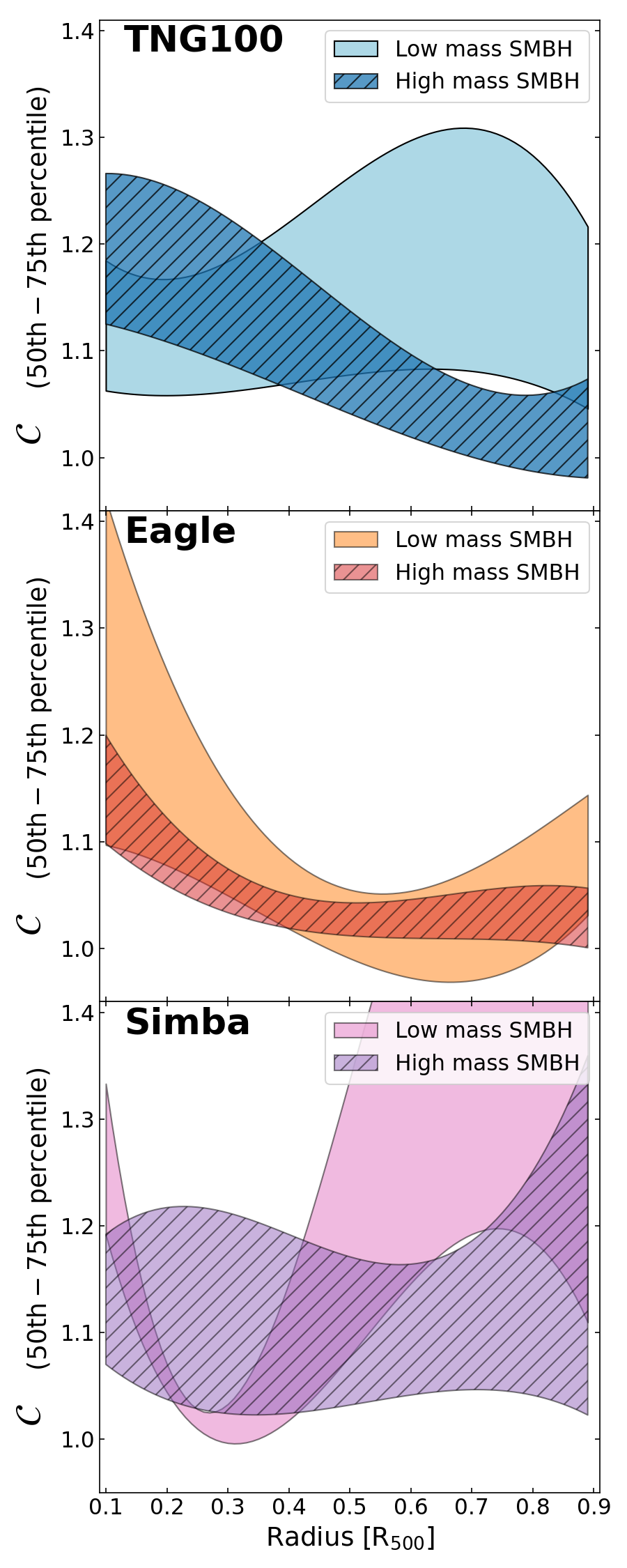}
    \caption{Azimuthal asymmetry as a function of radius, parameterized by the clumping factor $\mathcal{C}$ (defined in section \ref{ch:clumping}). We show the  50\textsuperscript{th} to  75\textsuperscript{th} percentile of the sample to illustrate the galaxy halos among the sample with high substructure, which are different in TNG100 (left) and EAGLE (right). The darker, shaded area is for higher mass black holes in the galaxy sample, while the lighter shaded area shows low mass black holes. }
    \label{fig:clumping}
\end{figure}

Since the calculation of the clumping factor of a single emission line, like the O\,VII, could bias the results due to the sensitivity to temperature changes in the CGM, we use the stacked signal of the O\,VII, O\,VIII, and Fe\,XVII lines, which are all sensitive to different temperatures (Tab. \ref{tab:lines}). 
The combined samples cover a wide range of galaxy halos in terms of mass, star formation rate, or temperature. Therefore, we divided the sample into galaxies with a central SMBH below the median SMBH mass, and the ones above, shown in Fig. \ref{fig:clumping} as dark shaded and light shaded regions. 

We find an interesting trend for the black hole mass distinction: 
We do not see any difference in EAGLE for different SMBH masses (all values outside the core are within  1 to 1.1, see Fig. \ref{fig:clumping} middle panel). However, galaxies in TNG100 show a dichotomy (Fig. \ref{fig:clumping} top panel), as we find systematically higher $\mathcal{C}$ for lower SMBH masses outside $0.4 R_{500}$, while massive SMBHs show a very similar trend between EAGLE and TNG100. 
Simba galaxy halos (\ref{fig:clumping} bottom panel) are generally fainter and statistical uncertainties are larger, but lower mass SMBH appear similar to the trend in TNG100, as they have higher $\mathcal{C}$, but less significant. 
Recent results from simulations by \cite{Rasia2014-qh,Planelles2014-mh,Planelles2017-ls} indicate that SNe, and especially AGN feedback, smooth out the gas distribution and suppress a higher clumping factor. While TNG100 and Simba have efficient AGN feedback prescriptions that increase the gas kinematics, AGN feedback only becomes dominant with higher SMBH masses. 
We can see this trend in our results (Fig. \ref{fig:clumping} top panel, and less significant in the bottom panel), where higher mass black holes have more clumping near the core and a smoother gas distribution at larger radii, compared to low mass black holes. EAGLE instead pressurizes the gas more efficiently, which lowers impact of the SMBHs on the gas clumping. 
It should be noted that the numerical schemes, hydrodynamic solvers, and sub-grid physics, especially of feedback in the simulations suites are different, which will impact the gas distributions. A Smoothed Particle Hydrodynamics (SPH) simulation will also produce a smoother gas distribution (e.g., \citealp{Rasia2014-qh}). 

\subsection{Spectral analysis} 
\label{ch:spectral_analysis} 
The detectability of CGM emission lines with a LEM-like instrument out to large radii offers the opportunity to not only derive one-dimensional profiles, but, to map the emission out to R$_{500}$, and measure temperatures, line-of-sight velocities, and abundance ratios. 
We described our spectral fitting approach in section \ref{ch:spec_analysis}, and apply it here to two galaxies selected from TNG50 of the IllustrisTNG project (\citealp{Nelson2019-ud,Pillepich2019-sw}). TNG50 provides a higher particle resolution than TNG100 (baryon particles about 16 times smaller in mass, and simulation volume about 10 times smaller), enabling in-depths studies of the velocity structure of individual galaxies. However, we verified that our previous results are not biased by properties of TNG100 (which provides a larger volume and larger galaxy samples to select galaxies from). 
Since we do not want to select a large number of galaxies, but rather study two examples in more detail, a larger simulation box won't provide any advantage.  

We explore two galaxy halos in more detail: The first galaxy (358608) has a halo mass of $10^{12.7}\,\si{M_\odot}$, which would place it in our high mass sample. It has a relatively high stellar mass of $10^{11.18}\,\si{M_\odot}$, and a high star formation rate, $\SI{3.87}{M_\odot\,yr^{-1}}$. With a mass of $10^{8.64}\,\si{M_\odot}$ its central SMBH is relatively dominant, and we expect both AGN and stellar feedback to be present. 
Placing our galaxy at $z=0.1$, the galaxy slightly exceeds our FOV ($R_{500} = \SI{243}{kpc}$, corresponds to $\SI{19.7}{\arcmin}$).
\begin{figure*}
    \centering
    \includegraphics[width=0.99\textwidth]{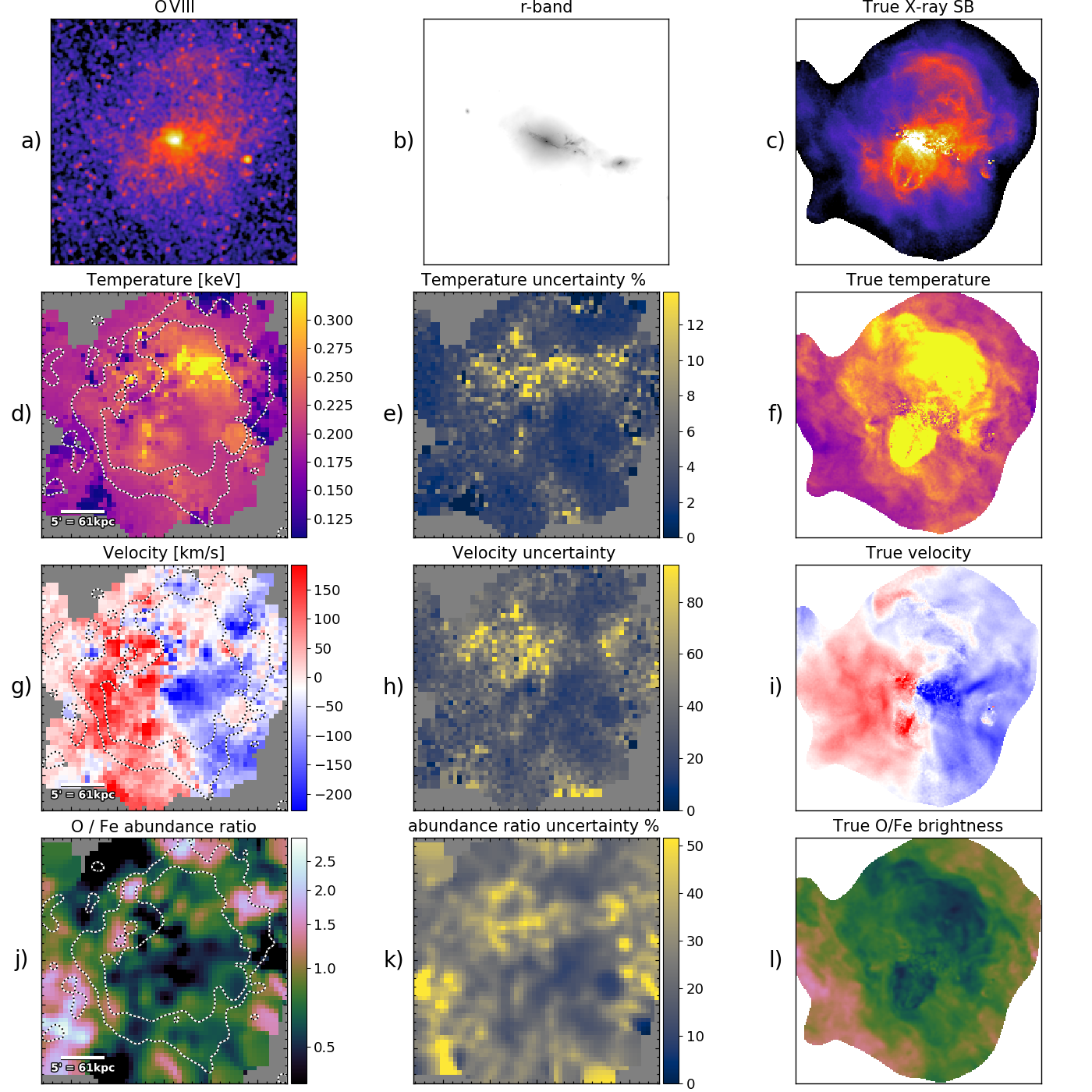}
    \caption{Spectral map of galaxy 358608 from TNG50 (for details see section \ref{ch:spectral_analysis}), showing observed O\,VIII surface brightness (a), and the predicted optical r-band signal (b), the simulated X-ray brightness (c), the observed temperature in keV (c, and error map d) from a simultaneous fit to O\,VII, O\,VIII, and Fe\,XVII lines), the emission weighted temperature from the simulation (f),    
    the observed average line velocity shift in $\si{km\,s^{-1}}$ (g, and error map h), the predicted, emission weighted LOS velocity in the simulation (i), the observed O/Fe abundance ratio (j, and error map k), and the predicted O/Fe brightness ratio in the simulation (l). }
    \label{fig:specmap358608}
\end{figure*}
Figure \ref{fig:specmap358608}~a) shows the observed (i.e., including background and foreground) O\,VIII emission: We see a bright core and extended CGM emission out to the edge of the field of view ($\SI{16}{\arcmin}$ radius or $\SI{195}{kpc}$). Brighter filaments extend to the north (forming a rim around a lower surface brightness region), and south-east, perpendicular to the galactic disk, which is edge-on and oriented south-west to north-east (see the optical r-band image tracing the stellar population in Fig. \ref{fig:specmap358608}~b ). 
The optical image also shows a smaller structure, about 100\,kpc to the west, which is a smaller galaxy. It also has an X-ray counterpart in the O\,VIII image. The  X-ray brightness in the simulation (Fig. \ref{fig:specmap358608}~c) shows the ``true'' distribution of the hot gas, which is very filamentary.

We derive the spectral maps following our strategy laid out in section \ref{ch:spec_analysis}. We simultaneously fit the spectrum within 8\,eV narrow bands around emission lines (Tab. \ref{tab:lines}) to derive the temperature map of this system (Fig. \ref{fig:specmap358608}~d). The temperature constraints mostly come from the relative line strengths of the O\,VII(f), O\,VIII, and Fe\,XVII lines, and will be most sensitive to trace temperature between 0.15 and 0.45\,keV. 
We note that for the majority of regions the effects of resonant scattering and photoionization are small, and the plasma is close to CIE (see also appendix \ref{ch:deviation_cie}).
The typical statistical uncertainties vary between $\SI{0.03}{keV}$ and $\SI{0.005}{keV}$, so relative uncertainties range between 1\% and 10\% (see Fig. \ref{fig:specmap358608}~d). We note that for regions that are less dense and cooler, the uncertainties will be higher, e.g., at the virial radius and beyond (\citealp{Bogdan2023-qg}). 

Comparing this observed temperature map with the idealized, emission weighted temperature (0.5-1\,keV band) derived from the simulation (Fig \ref{fig:specmap358608}~f), we can reproduce the brighter parts to the north of the core and the south-east. The emission weighted temperature is even higher in these regions, mostly because the lines that we used from our mock observation to probe the gas temperature are not sensitive enough to the hotter gas components. We note that a mass weighted temperature will be biased toward high mass, low temperature gas cells, and therefore will be lower than the emission weighted temperature, which more closely reflects our measured quantity in the mock observations. 

For every fitted region, we leave the redshift free to vary during the spectral fit, which is used to derive the line-of-sight velocity information. We note that all redshifts of CGM redshift components (emission from the various chemical elements) are linked together in the spectral fit, and we therefore derive a single velocity from all the line centroids. 
The velocity map (Fig. \ref{fig:specmap358608}~g ) reveals a split between east and west, with an average difference between the two sides of about $\SI{300}{km\,s^{-1}}$. As this is most pronounced in the central region, it can be explained by the rotation of the disk. The higher velocity gas (red region southeast of the core) is likely an outflow, since it overlaps with the hotter regions. 
The predicted, emission weighted velocity map from the simulation (Fig. \ref{fig:specmap358608}~i) confirms this, as it shows structures that are very consistent: The central rotation of the disk, the higher velocity part to the south-east, and the large scale velocity structure of the hot gas. 
A subsequent paper (\citealp{ZuHone2023-le})  will analyze the velocity structure of simulated galaxy halos in great detail. 
The statistical uncertainty mainly depends on the number of counts in a line. With our adaptive binning described in section \ref{ch:spec_analysis}, we find a velocity uncertainty of about 25 to $\SI{45}{km\,s^{-1}}$ in the brighter central regions, and about $\SI{80}{km\,s^{-1}}$ in the lower surface brightness region north-east of the center. We assumed conservatively a 2\,eV response across the field of view. 

We are not constraining individual elemental abundances, since there is a degeneracy with density. There are cases, where this degeneracy can be broken, e.g., by observing the CGM line emission and the absorption of a sufficiently bright background AGN by the CGM, the former proportional to the square of the density (emission measure), and the latter proportional to the column density. 
However, we are able to reliably determine abundance ratios (with respect to solar), and show the observed oxygen-to-iron ratio in Fig. \ref{fig:specmap358608}~j. 
This abundance ratio is sensitive to the enrichment history, mainly the  SNIa versus core-collapsed supernovae ratio (\citealp{Mernier2020-gc}).
We find typical values in the central region $[\rm O/Fe]=-0.3$ ($Z_{\rm O} / Z_{\rm Fe} = 0.5$), and values closer to $[\rm O/Fe]=0$ ($Z_{\rm O} / Z_{\rm Fe} = 1$) and above, in the outer regions. This is consistent with the predicted O/Fe brightness from the simulation (Fig. \ref{fig:specmap358608}~l), which shows the center and outflows to be more Fe-rich. 
For galaxy clusters and groups, the oxygen abundance distribution has been found to be flat, while iron is centrally peaked \cite{Werner2006-rv,Mernier2017-li,Vogelsberger2018-qb}, which leads to an increasing O/Fe profile. A similar trend can be expected for galaxies (\citealp{Geisler2007-hg,Segers2016-gj,Matthee2018-md}). 
Some regions, especially in the south-east, have very high oxygen abundances, up to 2.5. Typical uncertainties range from 10-20\% in the center to 70\% in the faintest regions.

The second galaxy (467415) that we map in detail is also selected from  TNG50, but with a lower halo mass of $10^{12.32}\,\si{M_\odot}$, which would place it in the medium mass sample. 
\begin{figure*}
    \centering
    \includegraphics[width=0.99\textwidth]{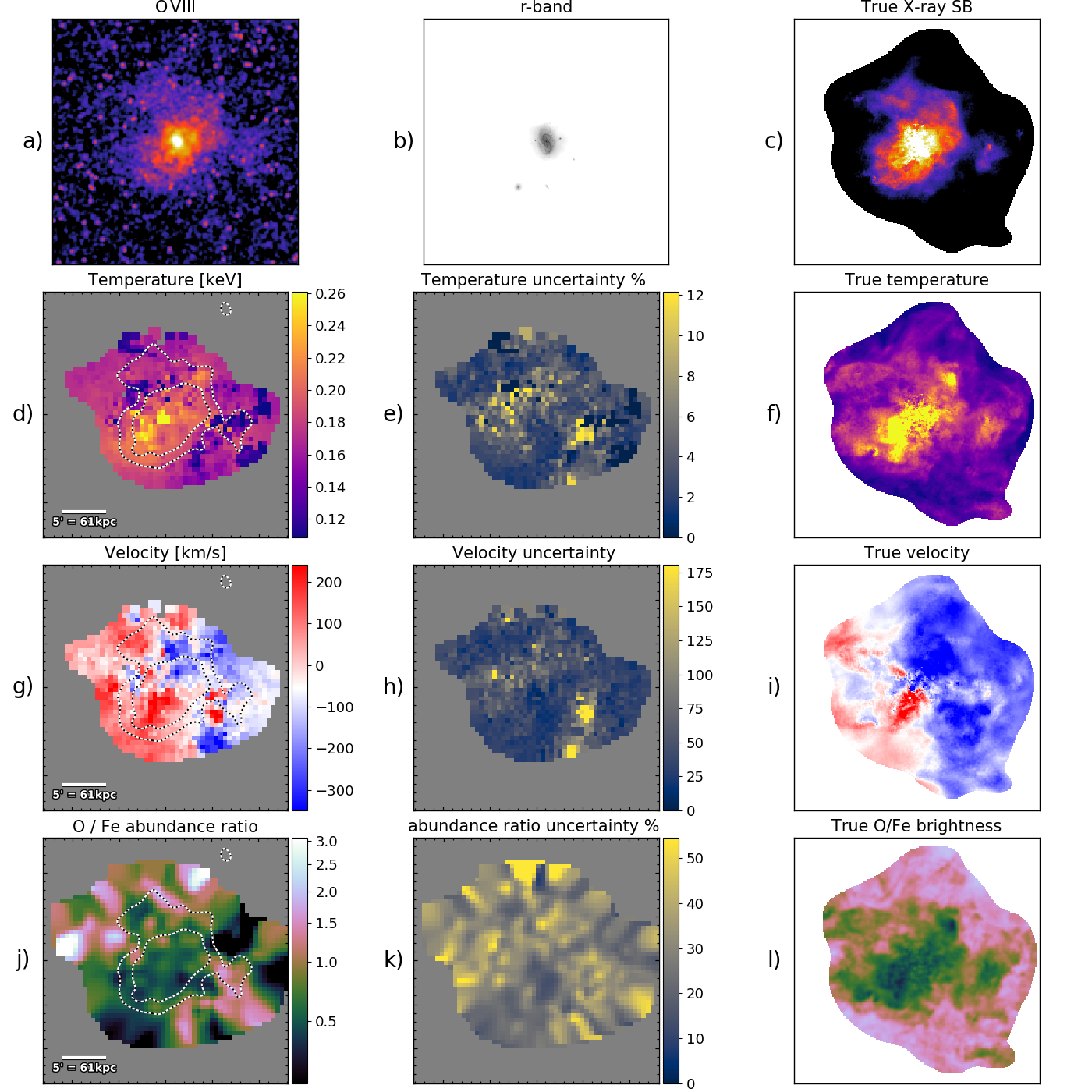}
    \caption{As for Fig. \ref{fig:specmap358608}, but for the lower mass galaxy from TNG50 (ID 467415, details in section \ref{ch:spectral_analysis}).}
    \label{fig:specmap467415}
\end{figure*}
The stellar mass, $10^{10.94}\,\si{M_\odot}$ is relatively high for its size, and the star formation rate of $\SI{8.9}{M_\odot\,yr^{-1}}$ might still dominate its halo environment. Also, its SMBH mass is at the higher end with $10^{8.33}\,\si{M_\odot}$, which makes this another interesting target to study the effects of stellar and AGN feedback. The radius, $R_{500}=\SI{183}{kpc}$, is within the  field of view at $z=0.01$. 
The observed O\,VIII emission (Fig. \ref{fig:specmap467415}~a) is brightest in the center, but is detected out to almost 100\,kpc, and some filaments beyond that. The distribution is not azimuthally uniform, but seems to be aligned along filaments, mainly to the south-east, the north, and a narrow region to the west. 
The r-band contours in Fig. \ref{fig:specmap467415}~b show one single galaxy at the center, and some much smaller and fainter structures to the south, possibly small satellite galaxies. Figure \ref{fig:specmap467415}~c displays the X-ray surface brightness in the simulation. 

The observed temperature map (Fig. \ref{fig:specmap467415}~d, and the error map \ref{fig:specmap467415}~e) shows the hottest emission ($\sim \SI{0.3}{keV}$) near the center of the galaxy, while the temperature in the outer filaments drops to $\SI{0.18}{keV}$. We identify slightly hotter structures, extending from the south-east to the north-west, while the north-east to south-west axis has cooler gas.  
The uncertainties are again in the 1\% to 10\% range. The temperatures are broadly consistent with emission weighted temperature in the simulation (Fig \ref{fig:specmap467415}~f). 

The velocity (Fig. \ref{fig:specmap467415}~g, and the error map \ref{fig:specmap467415}~h) ranges from $\SI{250}{km\,s^{-1}}$ in the south-east to $\SI{-250}{km\,s^{-1}}$ in the north-west, outside the stellar disk. These are even higher velocities than the within first, more massive galaxy. 
This coincides with the hotter regions in the temperature map, and is consistent with the predicted velocity from the simulation (Fig. \ref{fig:specmap467415}~i). 
Along the faint, low-temperature regions in the south-west and north-east, the velocities are lower than the surrounding. Statistical uncertainties are similar to the other galaxy, ranging from 20 to $\SI{80}{km\,s^{-1}}$. 

Lastly, the oxygen-iron map (Fig. \ref{fig:specmap467415}~j, and the error map \ref{fig:specmap467415}~k) has typical values between 0.5 and 1, but also several small regions have ratios well above 1 (statistical uncertainties are between 20\% and 70\%). It is consistent with the predicted O/Fe brightness in the simulation (Fig. \ref{fig:specmap467415}~l), where the iron rich gas is again in the center and along the high velocity, high temperature trajectory from the south-east to slightly north of the core. 

The level of detail that is revealed in these spectral maps is unprecedented for galaxy-sized halos. Typical spectral maps from CCD based detectors (e.g., Chandra or XMM-Newton) can only reach large radii (e.g., R$_{500}$) for galaxy clusters and massive galaxy groups, however, without any line-of-sight velocity information.

\section{Discussion} \label{ch:discussion} 

\subsection{Tracing the CGM with X-ray microcalorimeters}
The extended gaseous halos around MW-like galaxies are predicted by simulations, and have been detected in stacked broad-band images. Constraining their extent, brightness profile, azimuthal distribution, and enrichment with various elements for individual galaxies will allow us to distinguish between simulation models and ultimately enable us to understand the changing feedback processes of galaxies on various scales. The X-ray continuum  emission of these galaxies is very faint, but the bright emission lines, namely O\,VII, O\,VIII, and Fe\,XVII, are clearly detectable over the local background and foreground. 
Selecting a narrow $\sim \SI{2}{eV}$ energy band around these lines results in a high signal to noise detection with LEM-like observatory, in a 1\,Ms exposure of a $z=0.01$ galaxy, and even allows the 2D-mapping of these galaxy halos. Only focusing on emission lines and not having the continuum information will still allow the majority of science questions to be answered: How is the hot gas distributed, and what are the relative metal abundances (e.g., wrt iron). Only the degeneracy between metallicity and density cannot be easily broken. 

The faint CGM line emission around individual, nearby spiral and elliptical galaxies cannot be detected and mapped with current X-ray CCD instruments due to the bright Milky-Way foreground. Even state-of-the-art DEPFET detectors such as the Athena/WFI 
(\citealp{Meidinger2017-kg}) with its $\sim \SI{80}{eV}$ energy resolution will not be able to distinguish the CGM emission lines from the much brighter foreground. A galaxy redshift of at least $0.12$ is necessary to shift the O\,VIII line from the foreground, which will also reduce the apparent size of the galaxy to about $\SI{1}{arcmin}$, and the total flux to about 8\% with respect to a galaxy at redshift $0.01$.  

The development of microcalorimeters marks the start of a new epoch in X-ray astronomy, reaching unprecedented energy resolution, while spatially resolving the source structure. The currently planned Athena/X-IFU instrument  (\citealp{Barret2018-hb}) will have a large effective area and good spatial resolution. However the field-of-view of $\sim 5\times 5 \si{arcmin^2}$ (before reformulation) is clearly not sufficient to observe the extended CGM of nearby galaxies. At $z=0.01$ a Milky-Way-sized galaxy has $R_{500}\approx 15\arcmin$, and therefore requires about 30 pointings. 
Moving to a higher redshift and utilizing Athena's large effective area can reduce the required amount of observing time to a factor of 5-10 times what a LEM-like mission would need. Galaxies at $z\gg 0.01$ will also not offer the same amount of structure that can be resolved. 
Since simulations predict variance between galaxies, also based e.g., on the star formation rate, one would like to observe a medium-sized sample of 10 to 20 galaxies.

The X-Ray Imaging and Spectroscopy Mission (XRISM, \citealp{Tashiro2018-wi} successfully launched in 2023)  also has a microcalorimeter onboard. However, its field-of-view is limited to only $\SI{3}{\arcmin}\times\SI{3}{\arcmin} $, the effective area is about 10-15 times smaller than LEM, and together with the arcmin spatial resolution and only $6\times6$ pixels, it will not be able to map the extended CGM. 

Other mission concepts with a large effective area microcalorimeter include the Line Emission Mapper (\citealp{Kraft2022-vi}), and HUBS (\citealp{Zhang2022-ze}). While HUBS does not have enough spatial resolution to map the structure, and distinguish X-ray point sources in the field, LEM is clearly optimized to the CGM science by having sufficient energy resolution ($\SI{2}{eV}$), a large effective area similar to XMM-Newton, a $\SI{10}{arcsec}$ PSF, and a large field-of-view of $\sim \SI{900}{arcmin^2}$, allowing one to map nearby galaxies in a single pointing. 
In contrast to typical X-ray observations of faint, diffuse sources, the instrumental background level plays only a minor role when using a narrow energy band of a microcalorimeter: The requirement for Athena/X-IFU is to reach an internal particle background level of $\SI{5e-3}{cts\,s^{-1}\,cm^{-2}\,keV^{-1}}$ (\citealp{Lotti2021-wm}). This should be achieved through a graded anti-coincidence shield, while the background is predicted to be about an order of magnitude higher without the shielding. 
For our results, we conservatively assume a constant particle background level in the soft band of $\SI{8.6e-2}{cts\,s^{-1}\,cm^{-2}\,kev^{-1}}$, which is more than 15 times higher than the Athena requirement.  
The foreground emission however, scales with the effective area of the mirror. In the case of LEM, it will be the dominant background component, and even at the foreground continuum around the O\,VIII line, the particle background is still below all other components (foreground and CXB).

\subsection{Model distinction}
We have demonstrated that a LEM-like microcalorimeter will be able to detect the CGM of MW-mass galaxies to large radii $\sim R_{500}$, even in low mass galaxies below the ``transition'' regime (Fig. \ref{fig:stellar_mass}). 
Long exposure times with CCD instruments such as Chandra ACIS or XMM-Newton EPIC spent on individual massive galaxies have revealed only the innermost part of the CGM, and at best give us a vague idea of the temperature structure, especially if they are not in an ongoing starburst phase. 
\cite{Bogdan2013-vw}  used Chandra to image NGC\,266, a massive (M$_{200}\approx \SI{8e12}{M_\odot}$), nearby galaxy, and detected the CGM out to about 60\,kpc, which is about 20\% of R$_{500}$. \cite{Bogdan2017-mm} used XMM-Newton to detect and characterize the CGM around the massive galaxy NGC\,6753, which has a virial mass of $\SI{e13}{M_\odot}$. The authors could reliably make a detection out to 50\,kpc, before background systematics made any conclusions impossible. This is about 17\% of R$_{500}$. These exceptional cases demonstrate, that only with massive efforts, we are currently able to explore up to 1\% of the volume that the CGM fills out to R$_{500}$, and this only for the most massive, hand-picked galaxies, which are at the high mass end. 
Our mock observations show that with a large grasp microcalorimeter we can not only detect these type of galaxies beyond R$_{500}$ (see green profiles in Figs. \ref{fig:profiles_Eagle} and \ref{fig:profiles_TNG}) in individual lines, such as O\,VIII, but also map their dynamical, thermal, and chemical abundance. These galaxies are expected to be dominated by AGN feedback, which we see as outbursts in the velocity map or the O/Fe ratio map. Features in the abundance ratio map, such as high O/Fe ratios maybe indicative of strong early feedback or a recent starburst, whereas a high abundance of metals from AGB winds, such as carbon and nitrogen may be evidence for efficient gas entrainment from the ISM in SNe-driven winds (e.g., \citealp{Nomoto2006-lh,Das2019-uz,Carr2023-mw}). 

A LEM-like instrument will explore unknown territory by also mapping galaxies of much lower mass, down to M$_{200} \approx \SI{3e11}{M_\odot}$, which has not been done so far.
For these lowest mass galaxies, depending on the star formation rate, radial profiles can be derived out  $0.5 R_{500}$.  
In this regime, we do not expect AGNs to be important in the feedback cycle, while stellar winds are expected to enrich and reheat the CGM. 

The details of the transition from stellar to AGN feedback are largely not understood and are implemented ad-hoc in simulations to match some observational constraints, such as stellar scaling relations ($M-\sigma$, $M_\star-M_{\rm halo}$, galaxy morphologies, quiescent fractions as a function of stellar mass, and SFR$-z$ relations). 
Many measurements by a LEM-like observatory can be conducted that will lead to a new understanding of the processes within the CGM: Central regions, where a spectral continuum of the CGM emission can be measured, allowing us to constrain the absolute metal abundance, and at larger radii, the steepening of the X-ray line emission to distinguish the contribution from SN feedback, as seen, e.g., between EAGLE and TNG100, where TNG100 produces centrally peaked profiles with a steeper decrease in surface brightness (see e.g., \citealp{Chadayammuri2022-ei} for a comparison of profiles with simulations that have different feedback mechanisms). 
Measurements of the X-ray luminosities, surface brightness profiles, temperature distributions that relate to the outflow energies, will show if the gas is ejected from the the disk, and allow one to distinguish if feedback is instantaneously stopping a cooling flow, or whether a cumulative feedback effect is preventing gas phases to cool (\citealp{Davies2019-nz,Truong2020-fz,Terrazas2020-hs,Oppenheimer2020-ix}). 
With a few assumptions such as a metallicity profile, a total gas mass can be derived. Supplementary observations such as Sunyaev-Zeldovich (SZ, e.g., \citealp{Wu2020-ws,Bregman2022-tb,Moser2022-uc}), or fast radio bursts (FRB, e.g., \citealp{Ravi2019-zu,Macquart2020-wv,Wu2023-lg})  will also help to derive the gas mass.
The impact of the central AGN will be observed in the range of azimuthal asymmetry observed in the CGM emission. 

\cite{Chadayammuri2022-ei} and \cite{Comparat2022-es} have demonstrated, through stacking of optically detected galaxies in the eROSITA Final Equatorial Depth Survey (eFEDS), that the X-ray bright CGM exists even in low mass galaxies. Based on these results, simulations such as EAGLE and TNG100 are likely underestimating the surface brightness, especially in the low mass regime (see, e.g., the $\SI{2.5e10}{M_\odot}$ stellar mass bin in Fig. 4 of \citealp{Chadayammuri2022-ei}, which is a factor of 3-5 above the simulation predictions). Furthermore, the dichotomy between star forming galaxies being brighter in simulations with respect to quiescent galaxies, might not be true, at least not to the extent that it is predicted. 
Quiescent galaxies with little to no star formation tend to have massive and dominant AGNs, and are well-suited to understand the AGN cycles. 
These results cast doubt on the validity of the Simba CGM profiles, as Simba appears to drive gas to too large radii, making the galaxy halos fainter than observed by \cite{Chadayammuri2022-ei}.  
Simba also appears to be too X-ray faint, compared to low-mass groups \citep{Robson2020-ir}, as the energy output from the bipolar jets evacuates the halos.
Although stacking analyses have provided some distinctions among the simulations, stacking cannot replace detailed analyses of individual galaxies as it might be biased by a few bright objects.

\subsection{Observing strategies}
In the previous sections, we  demonstrated that a LEM-like mission with a large grasp microcalorimeter will be able to map nearby galaxies over a wide range of mass, star formation rate, and black hole mass. 
We argued that an instrument such as the Athena X-IFU will not be able to dedicate enough observing time to this science area. However, even an observatory such as LEM will not be able to spend 1\,Ms on 120 galaxies that we assumed to have observations available (40 in each of the three mass samples). 
\begin{figure}
    \centering
    \includegraphics[width=0.475\textwidth]{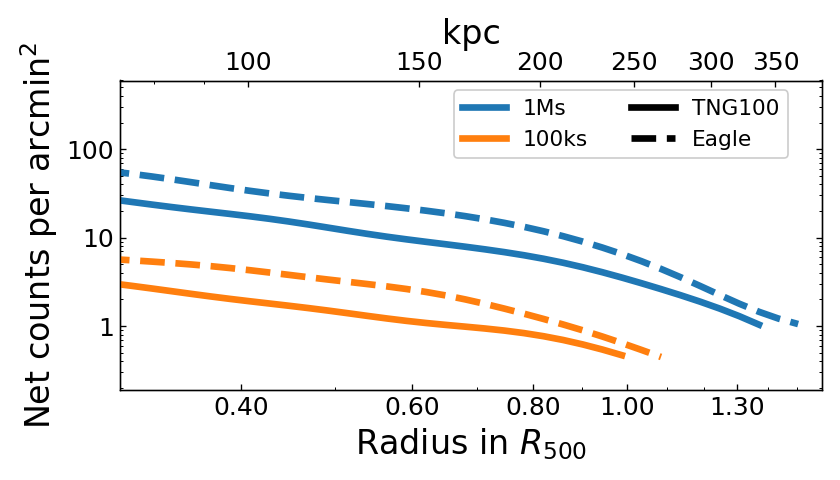}
    \caption{Median O\,VIII surface brightness profile of the total high mass sample of EAGLE (dashed lines) and TNG100 (solid lines) for two simulated exposure times, 1\,Ms (blue), and 100\,ks (orange) for a mission such as LEM. Each profile ends where the criteria for statistical and systematic uncertainties are not fulfilled (see section \ref{ch:method_sbr}).}
    \label{fig:shorter_obs}
\end{figure}
We test the impact on constraining the line surface brightness distribution with shorter observations (Fig. \ref{fig:shorter_obs}), and found that even with 100\,ks per galaxy, a LEM-like mission will be able to map the O\,VIII emission out to R$_{500}$, in both EAGLE and TNG100 simulated galaxies. In an observing plan for a LEM-like mission, fainter galaxies will take up significantly more observing time, especially in the crucial transition regime of Milky Way mass galaxies. 
Having 10 low, 10 medium, and 10 high mass galaxies, where each galaxy is observed for 1\,Ms with the exception of the high mass halos (100\,ks), one can achieve such an ambitious program within about 20\,Ms, which is a typical directed science program for a probe class mission. 
While the trend of the average properties (e.g., O\,VIII brightness) in each mass bin is important, the dispersion around the median, also will be important to understand. Therefore, galaxies should be selected to cover a range of properties that might shape the CGM, such as the stellar mass $M_\star$ at a given halo mass, the star formation rate, and the mass of the supermassive black hole.

We also tested whether dedicated background observations are necessary, or whether surface brightness profiles can be extracted using a model of the foreground and background emission. This model can be fit to the same observation in an outer region, and then constrain the expected background plus foreground counts in each annulus at the CGM line spectral window. This method achieves comparable results, and reduces the overhead, but requires a good model of the foreground spectrum. 

\section{Summary} \label{ch:summary} 
Mapping the X-ray emission of the hot circumgalactic medium (CGM) is one important key to understanding the evolution of galaxies from smaller galaxies with star formation driven feedback, to larger, quiescent galaxies. Milky Way mass galaxies appear to be at the transition point between these regimes. However, the current generation of X-ray instruments are unable to capture the emission from the hot CGM, that is dominant in the soft X-ray band, and distinguish it from the bright Milky Way foreground. We demonstrate that a high spectral resolution microcalorimeter with a large field of view and large effective area cannot only detect the CGM line emission to R$_{500}$, but also can map the individual physical properties such as temperature and velocity. 
A mission designed to study the hot CGM, similar to the Line Emission Mapper probe concept, will transform all fields of astrophysics (\citealp{Kraft2022-vi}). 

We created realistic mock observations, based on hydrodynamical simulations from EAGLE, IllustrisTNG, and Simba, for a large effective area instrument with 2\,eV spectral resolution, and a $\SI{32}{\arcmin}\times\SI{32}{\arcmin}$ FoV. We included all background and foreground components in these mock observations. 
The galaxies span a mass range from $\log M_{200}[M_\odot]=11.5-13$, and have been divided into three samples, based on their halo mass to represent the dominating feedback regime. 
For the mock observations, the low and medium mass galaxies (up to $\log M_{200}[M_\odot] \leq 12.5$) are placed at $z=0.01$, while the high mass galaxies are at $z=0.035$, and an exposure time of $\SI{1}{Ms}$ is used. For each galaxy we constrain the surface brightness profile of the O\,VII(f), O\,VIII, Fe\,XVII (725 and 729\,eV), and C\,VI. For galaxies at $z=0.035$ we also include the O\,VII(r) and the Fe\,XVII (826\,eV) lines, but have to omit the Fe\,XVII (729\,eV) line, since it is blended with the Milky Way foreground. Our findings are summarized as follows:

\begin{itemize}
    \item The median galaxy surface brightness profile for Milky Way sized galaxies at  $z=0.01$ can be traced to R$_{500}$, which is typically $\SI{170}{kpc}$ or $\SI{14}{\arcmin}$. 
    \item The CGM in more massive galaxy halos up to $\log M_{200} = 13$ at  $z=0.035$ can be measured out to R$_{200}$. Even for the lowest mass halos (down to $\log M_{200} = 11.5$ we typically will measure CGM emission  to $\sim 0.5 R_{500}$. 
    \item O\,VIII emission line is brightest in most cases, followed by O\,VII and Fe\,XVII. 
    \item Subdividing the galaxy samples by star formation rate reveals that star forming TNG100 galaxies are brighter in the core. 
    \item There is significant scatter in the CGM brightness due to galaxy-to-galaxy variation. Also the different simulations produce slightly different CGM luminosities at a given mass scale, where EAGLE galaxies are brightest, especially at higher masses, and Simba galaxies are typically the faintest, due to the strong AGN feedback expelling the gas.  
    \item We demonstrate that the O\,VII to O\,VIII line ratio in the mock observations can be used as a temperature tracer out to R$_{500}$ for more massive galaxies at $z=0.035$, and to $0.5 R_{500}$ for the less massive galaxies at $z=0.01$. We find that EAGLE galaxies are hotter in the center compared to TNG100, while having similar line ratios to TNG100 at large radii.
    \item We are able to map the substructure of galaxies out to R$_{500}$ by quantifying the azimuthal asymmetry. Interestingly, we find that TNG100 and Simba galaxies with a smaller SMBH reach high values of substructure beyond $0.4 R_{500}$, while EAGLE galaxies do not show that level of clumping. For massive SMBHs, all simulations predict  lower CGM clumping factors. This observable appears to be crucial to understand the mechanisms of AGN feedback, as it directly points to the efficiency of the AGN to pressurize the CGM gas. 
    \item Finally, we test the 2D properties of the gaseous halos around galaxies with spectral maps of properties, such as the temperature, the line-of-sight velocity, and the O/Fe ratio. Together, these quantities can be used to pin-point signatures of AGN feedback, such as the AGN duty cycle, or energy output. 
\end{itemize}

A LEM-like mission would revolutionize our understanding of the CGM, place dramatic new constraints on the variety of numerical simulations and on key feedback processes.

\begin{acknowledgments}
JAZ, AB, and RPK are funded by the Chandra X-ray Center, which is operated by the Smithsonian Astrophysical Observatory for and on behalf of NASA under contract NAS8-03060. DN acknowledges funding from the Deutsche Forschungsgemeinschaft (DFG) through an Emmy Noether Research Group (grant number NE 2441/1-1). I.K. acknowledges support by the COMPLEX project from the European Research Council (ERC) 
under the European Union’s Horizon 2020 research and innovation program grant agreement ERC-2019-AdG 882679. 

The material is based upon work supported by NASA under award number 80GSFC21M0002. 

The TNG50 simulation was run with compute time granted by the Gauss Centre for Supercomputing (GCS) under Large-Scale Projects GCS-DWAR on the GCS share of the supercomputer Hazel Hen at the High Performance Computing Center Stuttgart (HLRS).

This study made use of high-performance computing facilities at Liverpool John Moores University. 

\end{acknowledgments}

\software{
    AstroPy \citep{The_Astropy_Collaboration2013-lw,The_Astropy_Collaboration2018-gx}
    CIAO \citep{Fruscione2006-wt},
    Matplotlib \citep{Hunter2007-uf},
    NumPy \citep{Harris2020-ed},
    pyXSIM \citep{ZuHone2016-vy},
    Sherpa \citep{Freeman2001-hz,Burke2020-vd},
    SOXS \citep{ZuHone2023-fx}
}

\bibliographystyle{aasjournal}
\bibliography{Paperpile_remote.bib}

\appendix

\section{Analysis details}
\label{ch:analysis_appendix}
The analysis of simulated microcalorimeter observations of nearby galaxies resembles the traditional X-ray analysis. However, a few details have been altered to take advantage of the high spectral resolution. We describe this here in more detail. 

\subsection{Point source removal}
\label{ch:appendix_ps}
\begin{figure*}[h]
    \centering
    \includegraphics[width=0.99\textwidth]{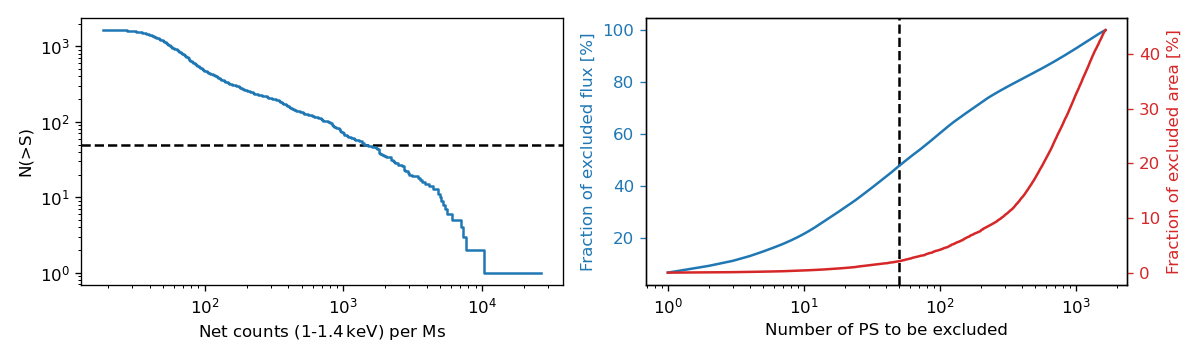}
    \caption{Removal of point sources in the observation. \textit{Left:} Cumulative number counts as a function of flux (in net counts). \textit{Right:} The blue line represents the fraction of excised flux in point sources as a function of the number of point sources (highest flux sources first), and the red line shows, for the same number of excised point sources, the area fraction that is masked. }
    \label{fig:logNlogS}
\end{figure*}

The point sources in the simulated  event files are detected through \verb|wavdetect| in the $\SIrange{1}{1.4}{keV}$ band, where only CXB and NXB dominate. At softer energies the Galactic emission will dominate, and at higher energies the particle background will dilute the CXB signal due to the decrease in effective area. We show in Fig. \ref{fig:logNlogS} (left) the distribution of cumulative number counts as a function of the threshold flux $S$ (in net source counts within a 1\,Ms observation). The distribution is consistent with the expected trend based on the Chandra observations by \cite{Lehmer2012-th}, where a broken powerlaw is found for distant AGNs and galaxies. 
If we excise the 50 brightest sources (Fig. \ref{fig:logNlogS} right) we remove about 50\% of the total flux in point sources, while only removing about 2\% of the detector area. For the 100 brightest sources we would remove more than 4\% of the area and 60\% of the flux, and if we removed 250 sources, we would excise 75\% of the flux and almost 10\% of the total detector area. 

\subsection{Surface brightness extraction}\label{ch:appendix_sbr}
In order to trace the emission lines (such as O\,VIII) in a narrow band out to large radii, it is important to quantify the background precisely. While the MW foreground should be greatly reduced, due to the redshift, the exact level of the background is still crucial, when we want to make a detection with a signal that is only 10\% of the total background level. 
We separate the background components into MW foreground (Local Hot Bubble - LHB, Galactic Halo Emission - GHE, and North Polar Spur - NPS), the unresolved/un-removed point source contribution from distant AGNs and galaxies (Cosmic X-ray Background - CXB), and the interaction of charged particles (galactic cosmic rays) with the detector and/or the satellite (producing secondary particles, such as fluorescent X-rays or electrons), which is not focused by the mirrors (Non X-ray Background - NXB). The modeling of these components in our simulated X-ray observations has been described in section \ref{ch:mocks}. 

To estimate the background counts in a narrow band at the redshifted CGM line, we take a blank field observation without a science target, and spatially close to the observation of each nearby galaxy of interest. We assume that the MW foreground emission does not vary within the FoV of our observation, and is also consistent with the MW foreground in the blank field observation. The same assumption is made for the NXB, although its contribution is less important at the emission lines of interest. 
Our focus is set on the CXB contribution to the total background, as it is different not only between the blank field and the galaxy observation, but also varies slightly from each extraction region (annulus) of the surface brightness profile. 
Therefore, we include a CXB correction factor to the blank field background counts, that is determined from the hard band (e.g., $\SIrange{1}{1.4}{keV}$), where the MW component is insignificant. 

\begin{figure*}[t]
    \centering
    \includegraphics[width=0.85\textwidth]{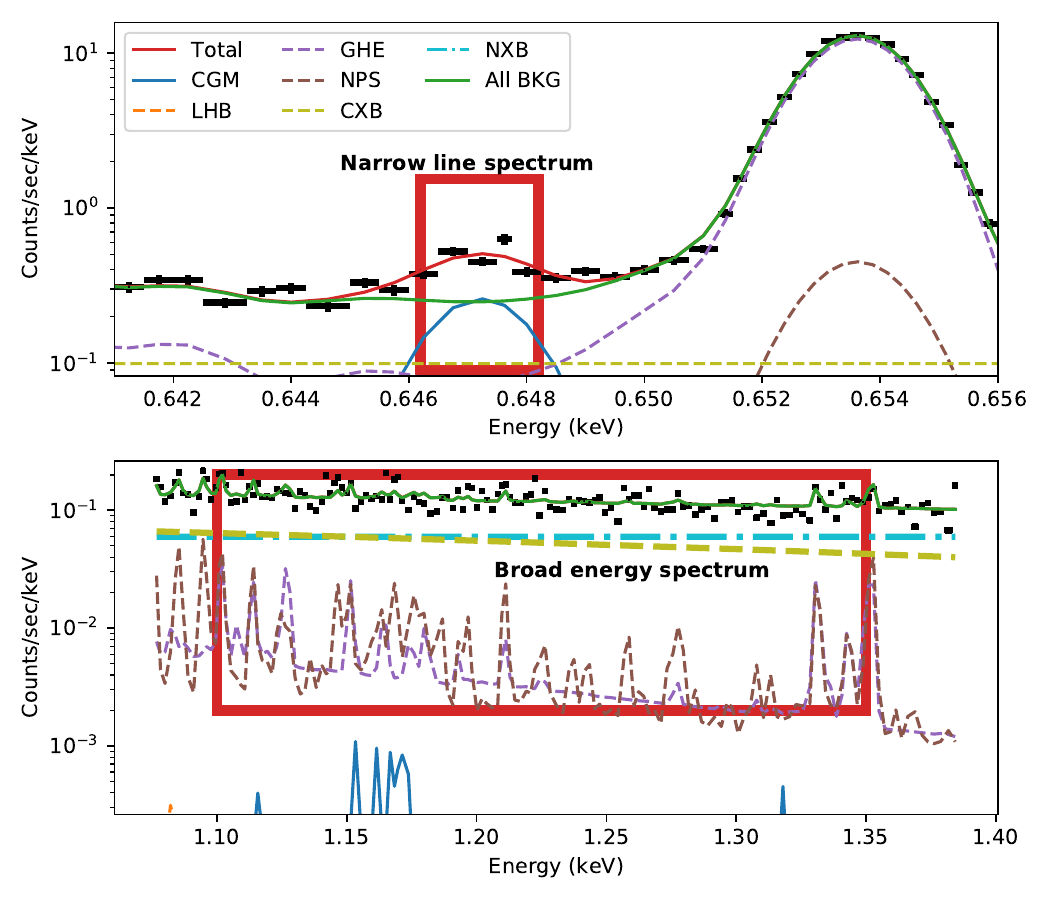}
    \caption{Narrow and broad bands to estimate the CXB scaling of the total background.}
    \label{fig:narrow_broad}
\end{figure*}

We label the extracted total counts (CGM and total background) in a narrow line band and small extraction region within our observation as ${\rm cts}^{\rm Obs}_{\rm line}$. We define the counts in the broad band and the blank field observation (Bkg) accordingly,
\begin{eqnarray}
    {\rm cts}^{\rm Obs}_{\rm line} &=& {\rm CGM}^{\rm Obs}_{\rm line} + {\rm MW}^{\rm Obs}_{\rm line} + {\rm CXB}^{\rm Obs}_{\rm line} + {\rm NXB}^{\rm Obs}_{\rm line} = {\rm CGM}^{\rm Obs}_{\rm line} + \mathcal{B}^{\rm Obs}_{\rm line} \\
    {\rm cts}^{\rm Obs}_{\rm broad} &=& {\rm CGM}^{\rm Obs}_{\rm broad} + {\rm MW}^{\rm Obs}_{\rm broad} + {\rm CXB}^{\rm Obs}_{\rm broad} + {\rm NXB}^{\rm Obs}_{\rm broad} = {\rm CGM}^{\rm Obs}_{\rm broad} + \mathcal{B}^{\rm Obs}_{\rm broad} \\
    {\rm cts}^{\rm Bkg}_{\rm line} &=&  {\rm MW}^{\rm Bkg}_{\rm line} + {\rm CXB}^{\rm Bkg}_{\rm line} + {\rm NXB}^{\rm Bkg}_{\rm line} = \mathcal{B}^{\rm Bkg}_{\rm line} \\
    {\rm cts}^{\rm Bkg}_{\rm broad} &=& {\rm MW}^{\rm Bkg}_{\rm broad} + {\rm CXB}^{\rm Bkg}_{\rm broad} + {\rm NXB}^{\rm Bkg}_{\rm broad} = \mathcal{B}^{\rm Bkg}_{\rm broad} ~.
\end{eqnarray}
Note that the MW and NXB components scale between the observation and blank field region only with the area (the observing time has already been taken into account), and in the broad band, the CGM component has no real contribution. 
We utilize a broad band above 1\,keV (e.g.,  $\SIrange{1}{1.4}{keV}$, see Fig. \ref{fig:narrow_broad}), to estimate the difference in the CXB between our extraction region in the observation and the total blank field (whole FoV). 

We derive a scaling factor,
\begin{equation}
    f_{\rm broad} = \frac{{\rm cts}^{\rm Obs}_{\rm broad}}{f_A \times \mathcal{B}^{\rm Bkg}_{\rm broad}}~,
\end{equation}
where $f_A$ is the ratio of the extraction area of the observation and the background. $f_{\rm broad}$ is typically close to one, especially if the extraction region in the observation is large (outer annuli). 
In order to minimize Poisson noise, we estimate the narrow line band background from the entire blank field observation and scale it by the area and $f_{\rm broad}$. We tested this method with a randomly chosen galaxy halo from TNG100 (ID 419061), and define 7 radial bins of $0.15 R_{500}$ width to reach an outer radius of $1.05 R_{500}$. In each of these regions, we extracted the spectrum from the observation, and fitted it with the model components (see, e.g., Fig. \ref{fig:narrow_broad}), so we have a precise knowledge of background components and the CGM emission. From the fitted model components, we are able to calculate the precise total background in each annulus $\mathcal{B^{\rm Obs}_{\rm line}}$, as well as the CGM counts ${\rm CGM}^{\rm Obs}_{\rm line}$, and make a comparison with the corrected blank field background estimate, $\mathcal{B}^{\rm Bkg}_{\rm line} \times f_{\rm broad}$. We find that the difference in counts between the actual background ,$\mathcal{B^{\rm Obs}_{\rm line}}$ , and the blank field predicted background is always much less than 1\% of the CGM counts in each annulus (reaching 0.5\% around R$_{500}$). Therefore, we employ this method of scaling the blank field by the broad band CXB contribution to estimate the background. Deriving the background from the spectral fit is computationally expensive and not feasible for each galaxy and extraction region. 

In order to define statistically significant radial bins of the surface brightness profile, we require a minimum signal to noise ratio of 3. The signal is directly calculated from the measured ${\rm cts}^{\rm Obs}_{\rm line}$ minus the background estimate $\mathcal{B}^{\rm Bkg}_{\rm line} \times f_{\rm broad}$. The noise is  assumed to be 
\begin{equation}
    {\rm Noise} = \sqrt{{\rm cts}^{\rm Obs}_{\rm line} + \mathcal{B}^{\rm Obs}_{\rm line} \times (1-f_{\rm broad}) + 0.1\times \mathcal{B}^{\rm Obs}_{\rm line}}~,
\end{equation}
where we account for systematic uncertainties in the background, as well as uncertainties in the CXB contribution. 

We also tested the impact of MW foreground uncertainties on the narrow line counts. With the typical uncertainties of the foreground model fitting (using only the narrow lines of C\,VI, O\,VII, O\,VIII, and Fe\,XVII) in the temperature, e.g., of the LHB, we find a difference in background counts of less than 1\% of the CGM counts.  

\section{Deviation from CIE}
\label{ch:deviation_cie}
\begin{figure}
    \centering
    \includegraphics[width=0.99\textwidth]{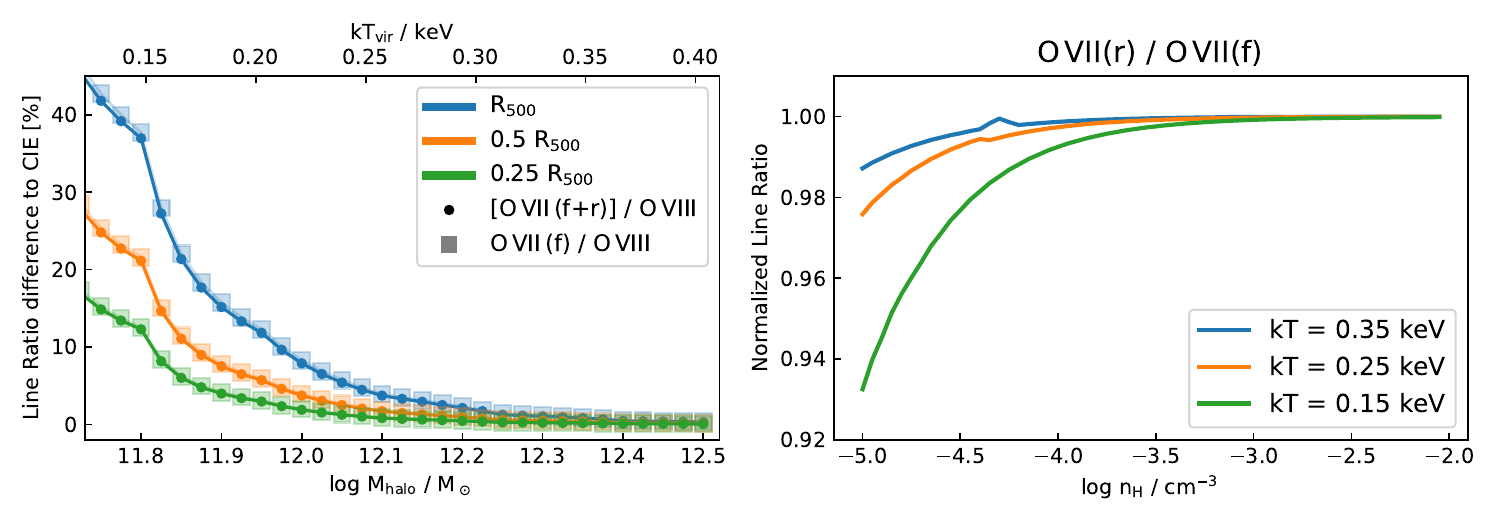}
    \caption{\textit{Left panel:} Difference in line ratio of O\,VII to O\,VIII with respect to the CIE line ratio for a set of halo masses. The dependence on the galaxy halo mass and the measurement of the line ratios is performed at specific radii (blue $R_{500}$, orange $0.5 R_{500}$, green $0.25 R_{500}$), assuming that the density profile follows a simple $\beta$-model, and the temperature is assumed to be the virial temperature. The black circles indicate the O\,VII line flux from the sum of the forbidden and resonant lines, gray circles use only the forbidden line as O\,VII flux. 
    \textit{Right panel:} CIE-normalized line ratios of the resonant to forbidden line of O\,VII for a given density illustrating the fractional difference between between CIE and photoionized plasma that includes CXB resonant scattering. The three colors represent different temperatures. }
    \label{fig:line_ratio_cie}
\end{figure}
In low density plasmas the photoexcitation and ionization rates can become high relative to the electron collisional excitation and ionization rates, which affects emission lines. Therefore, the ratio of emission lines (e.g., O\,VII and O\,VIII) is no longer independent of the gas density, as in CIE (\citealp{Churazov2001-ap,Khabibullin2019-ls}). To estimate the effect for typical galaxy halos we simulate the hot gas distribution with a simple $\beta$-model ($\beta = 0.4$, $r_{c} = \SI{40}{kpc}$, see \citealp{Li2017-bk,Zhang2024-ge}). 
The hot CGM mass is assumed to be the baryonic mass (7.5\% halo mass) minus stellar mass (4\% halo mass). For a given halo mass we can estimate the density at a given radius, and we assume the temperature to be close to the virial temperature (\citealp{Ponti2023-sg}),
\begin{equation}
    k_{\rm B} T_{\rm vir} = \frac{2}{3} G \mu m_{\rm p} \frac{M_{\rm vir}}{r_{\rm vir}} ~,
\end{equation}
where $\mu = 1.32$ is the mean molecular weight of the gas (\citealp{Ponti2023-sg}), $G$ is the gravitational constant, $m_{\rm p}$ is the proton mass. 
We derive the line ratios of both, O\,VII(f) to O\,VIII, and the co-added O\,VII(f+r) to O\,VIII, and quantify the difference to the simple case of CIE. The results are shown in Fig. \ref{fig:line_ratio_cie} (left panel): At a radial distance of R$_{500}$ (blue line) we see that for halos with a mass above $\SI{e12}{M_\odot}$ density dependent effects are less than 8\%. Closer to the center (orange and green lines) the densities are higher and differences to CIE even smaller. At higher masses, above 2 times Milky Way ($\log M/M_\odot = 12.4$), the differences are negligible even at R$_{500}$. 
We also find that the difference between the resonant and forbidden line becomes less than 1\% for higher temperature (mass) halos, above densities of $\SI{e-4}{cm^{-3}}$ (Fig. \ref{fig:line_ratio_cie} right). 

\section{List of simulated galaxy halos}
\begin{figure}
    \centering
    \includegraphics[width=0.95\textwidth]{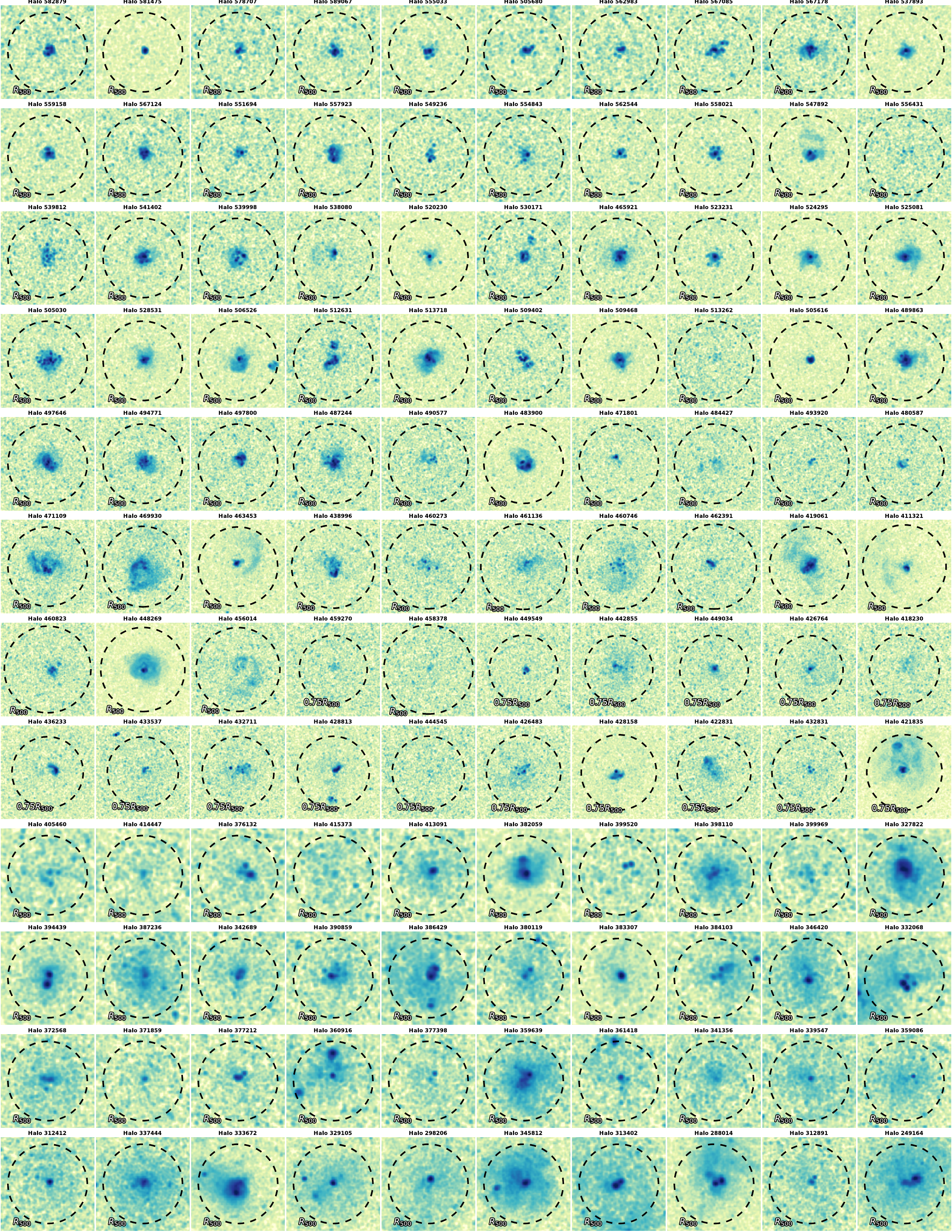}
    \caption{Stacked O\,VII(f)  and O\,VIII images from mock observations of the TNG100 galaxy halos, shown in log-scale with a fraction of the characteristic radius $R_{500}$ indicated.}
    \label{fig:tng100}
\end{figure}

\begin{figure}
    \centering
    \includegraphics[width=0.95\textwidth]{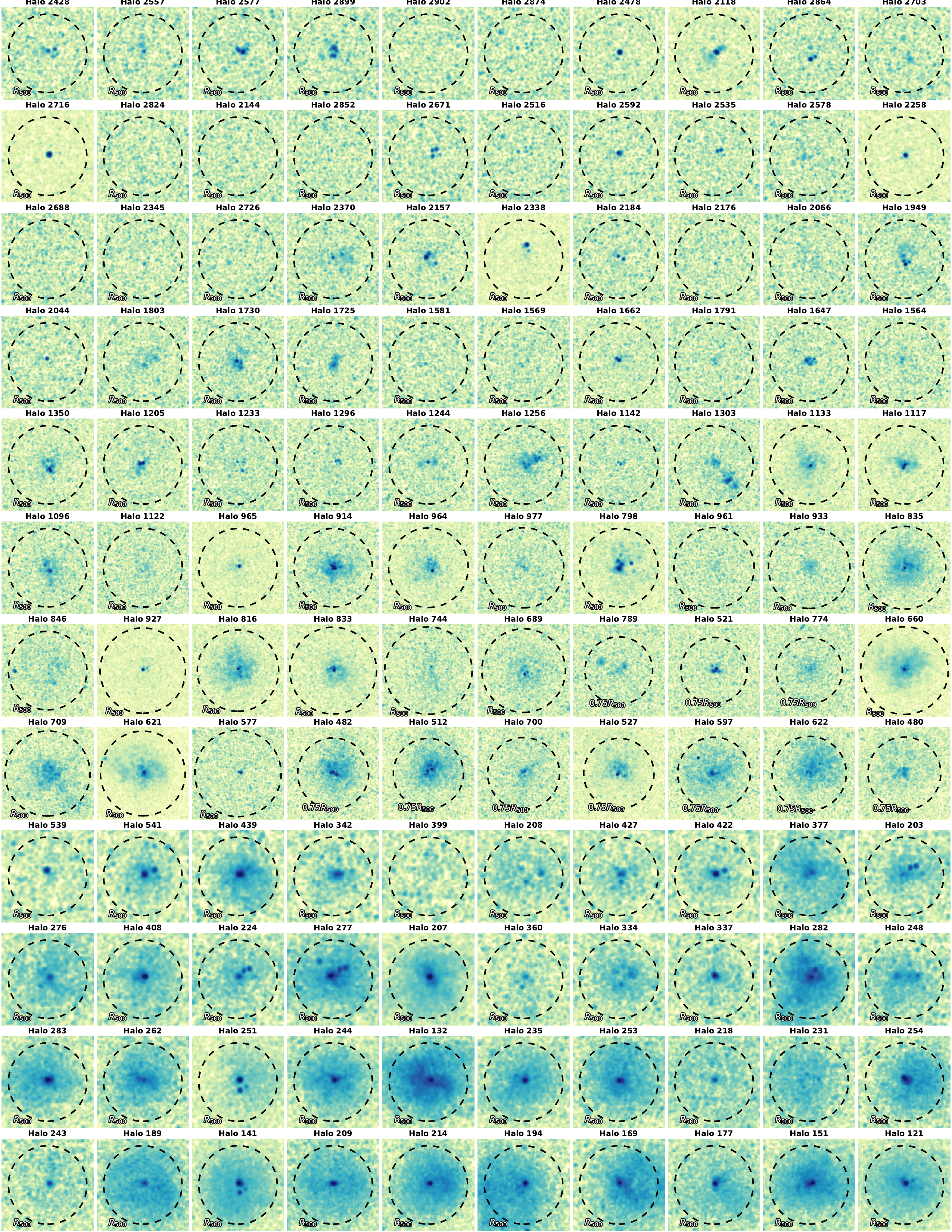}
    \caption{As Fig. \ref{fig:tng100} but for the EAGLE galaxy halos.}
    \label{fig:eagle}
\end{figure}

\begin{figure}
    \centering
    \includegraphics[width=0.95\textwidth]{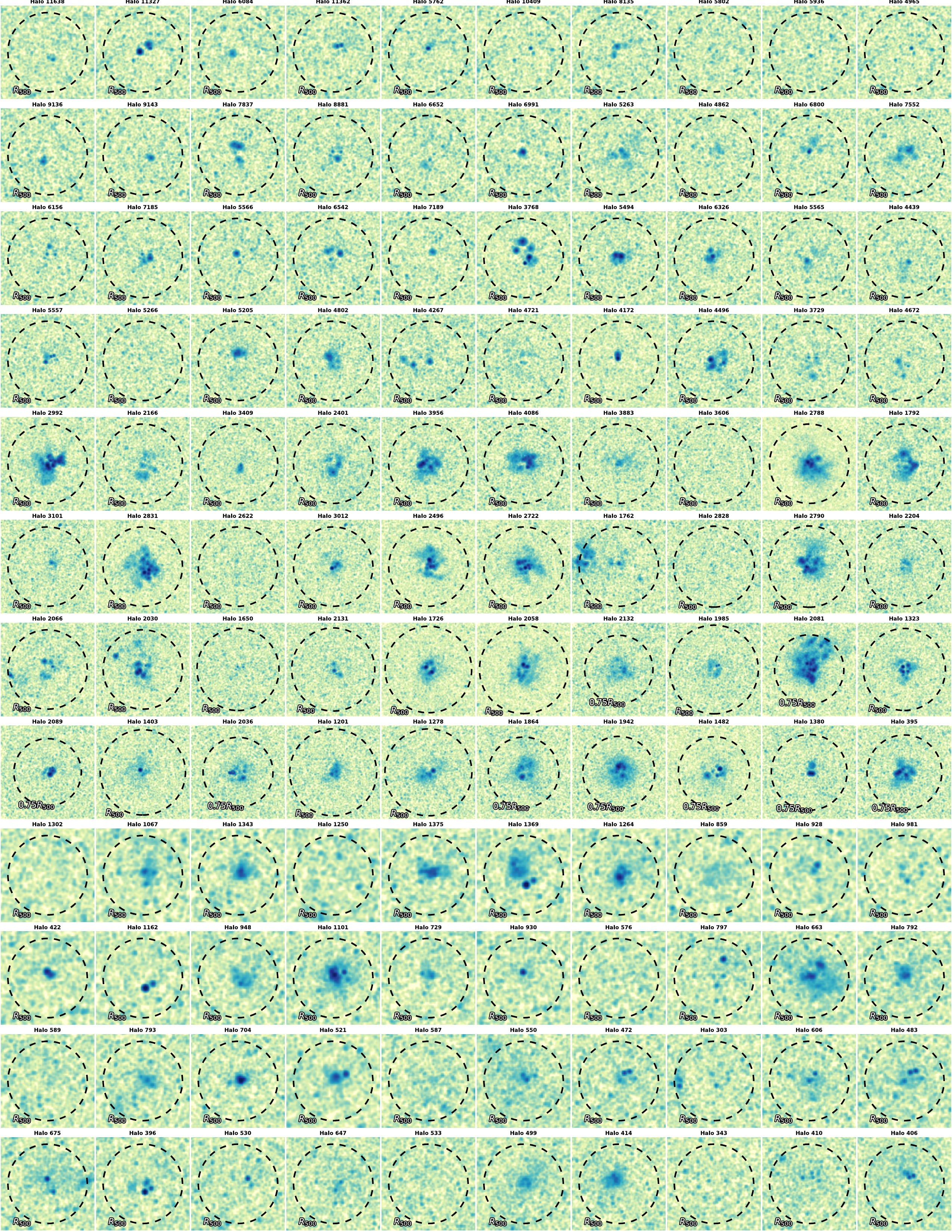}
    \caption{As Fig. \ref{fig:tng100} but for the Simba galaxy halos.}
    \label{fig:simba}
\end{figure}

Tables \ref{tab:TNG}, \ref{tab:Eagle}, and \ref{tab:Simba} list all the simulated galaxy halos that are used for the line surface brightness profiles, the line ratios, and the azimuthal substructure test. 
The columns denote the Halo ID/subID, the halo stellar mass within 30\,kpc, the total halo mass $M_{200}$, the central black hole mass, the star formation rate, and $R_{500}$.
As described in section \ref{ch:mocks}, we use three simulation suits, IllustrisTNG, EAGLE (100\,Mpc box) and Simba (100\,Mpc box), and subselect from each 120 galaxies. These galaxies are equally divided into a low mass sample $\log_{10} M_{200}/M_\odot = 11.5-12$,  a medium mass sample $\log_{10} M_{200}/M_\odot = 12-12.5$, and  a high mass sample $\log_{10} M_{200}/M_\odot = 12.5-13$. In Figs. \ref{fig:tng100}, \ref{fig:eagle}, and \ref{fig:simba}, we show stacked O\,VII and \,OVIII images for each of the galaxies, including instrumental background, foreground emission, and CXB (see section \ref{ch:analysis}). 
\startlongtable 
\begin{deluxetable*}{rrrrrrr|rrrrrr}
	\tablecaption{Basic properties of the selected IllustrisTNG halos \label{tab:TNG}}
	\tablehead{
		\colhead{Halo} & \colhead{M$_\star$} & \colhead{M$_{200}$} & \colhead{M$_{\rm BH}$} & \colhead{SFR} &  \colhead{R$_{500}$} & \colhead{   }  & \colhead{Halo} & \colhead{M$_\star$} & \colhead{M$_{200}$} & \colhead{M$_{\rm BH}$} & \colhead{SFR}  &  \colhead{R$_{500}$}\\
		\colhead{}  &  \colhead{$\si{10^{10}M_\odot}$} & \colhead{$\si{10^{12}M_\odot}$} & \colhead{$\si{10^8 M_\odot}$} & \colhead{$\si{M_\odot\,yr^{-1}}$} & \colhead{kpc}     & \colhead{}  & \colhead{}  &    \colhead{$\si{10^{10}M_\odot}$} & \colhead{$\si{10^{12}M_\odot}$} & \colhead{$\si{10^8 M_\odot}$} & \colhead{$\si{M_\odot\,yr^{-1}}$} & \colhead{kpc} 
	}
	\startdata
\multicolumn{13}{l}{TNG100 low mass sample} \\
582879 & 0.63 & 0.33 & 0.32 & 0.487 & 99 & &  541402 & 1.27 & 0.64 & 0.54 & 1.189 & 123 \\
539812 & 1.56 & 0.63 & 0.62 & 0.418 & 125 & &  581475 & 1.71 & 0.34 & 0.86 & 0.076 & 102 \\
513718 & 3.71 & 0.95 & 1.25 & 2.564 & 142 & &  530171 & 0.98 & 0.75 & 0.22 & 1.217 & 125 \\
509402 & 3.21 & 0.96 & 1.08 & 0.507 & 142 & &  489863 & 1.99 & 0.99 & 0.79 & 4.143 & 130 \\
505030 & 2.49 & 0.84 & 0.54 & 2.334 & 138 & &  513262 & 3.98 & 0.97 & 1.89 & 0.000 & 142 \\
557923 & 1.31 & 0.49 & 0.34 & 1.372 & 114 & &  512631 & 0.44 & 0.93 & 0.19 & 0.667 & 143 \\
562983 & 0.29 & 0.38 & 0.09 & 0.466 & 94 & &  509468 & 4.90 & 0.96 & 1.01 & 4.559 & 144 \\
551694 & 1.04 & 0.49 & 0.41 & 0.967 & 111 & &  567124 & 1.01 & 0.47 & 0.57 & 1.141 & 113 \\
537893 & 1.24 & 0.44 & 0.53 & 1.357 & 110 & &  525081 & 1.77 & 0.84 & 0.88 & 6.793 & 135 \\
524295 & 2.99 & 0.82 & 0.98 & 2.155 & 135 & &  539998 & 2.16 & 0.66 & 1.01 & 0.534 & 127 \\
562544 & 0.52 & 0.51 & 0.23 & 0.479 & 116 & &  528531 & 1.43 & 0.86 & 0.81 & 1.569 & 141 \\
555033 & 0.77 & 0.37 & 0.36 & 0.623 & 103 & &  554843 & 0.41 & 0.50 & 0.13 & 0.402 & 109 \\
556431 & 2.02 & 0.61 & 1.35 & 0.000 & 122 & &  505616 & 4.03 & 0.98 & 1.66 & 0.000 & 145 \\
538080 & 3.26 & 0.70 & 1.35 & 0.000 & 129 & &  523231 & 2.35 & 0.81 & 0.78 & 1.281 & 135 \\
559158 & 1.43 & 0.45 & 0.50 & 1.081 & 106 & &  558021 & 1.10 & 0.52 & 0.32 & 1.231 & 116 \\
520230 & 1.68 & 0.73 & 0.53 & 0.885 & 130 & &  465921 & 0.91 & 0.78 & 0.39 & 1.842 & 127 \\
506526 & 2.07 & 0.89 & 1.17 & 1.163 & 134 & &  505680 & 0.34 & 0.38 & 0.08 & 0.403 & 87 \\
578707 & 0.43 & 0.34 & 0.30 & 0.449 & 99 & &  567178 & 1.42 & 0.44 & 0.29 & 1.512 & 112 \\
589067 & 1.00 & 0.35 & 0.59 & 0.306 & 103 & &  567085 & 0.63 & 0.41 & 0.22 & 0.970 & 107 \\
547892 & 1.76 & 0.58 & 1.06 & 1.285 & 121 & &  549236 & 0.67 & 0.50 & 0.31 & 0.774 & 110 \\
&  &  &  &  && &        &      &      &      &      &       \\
\hline
\hline
\multicolumn{13}{l}{TNG100 medium mass halos} \\
497646 & 4.90 & 1.09 & 0.91 & 2.781 & 149 & &  497800 & 4.45 & 1.20 & 1.56 & 0.000 & 156 \\
487244 & 6.27 & 1.20 & 1.56 & 2.622 & 155 & &  490577 & 6.15 & 1.30 & 1.50 & 0.005 & 160 \\
421835 & 7.19 & 3.10 & 1.69 & 0.177 & 208 & &  419061 & 3.76 & 1.92 & 1.01 & 5.303 & 164 \\
460273 & 6.27 & 1.79 & 2.39 & 0.005 & 175 & &  484427 & 5.20 & 1.38 & 1.90 & 0.000 & 157 \\
422831 & 6.59 & 2.97 & 2.70 & 0.000 & 203 & &  449034 & 8.66 & 2.34 & 2.75 & 0.000 & 189 \\
411321 & 6.72 & 1.93 & 1.80 & 0.479 & 172 & &  449549 & 7.59 & 2.24 & 2.28 & 0.000 & 189 \\
428813 & 6.57 & 2.70 & 1.75 & 0.002 & 199 & &  442855 & 7.61 & 2.24 & 2.47 & 0.001 & 188 \\
483900 & 5.75 & 1.31 & 1.61 & 0.718 & 159 & &  460746 & 8.07 & 1.86 & 2.49 & 0.015 & 174 \\
444545 & 5.25 & 2.79 & 3.56 & 0.000 & 200 & &  438996 & 5.76 & 1.78 & 2.60 & 0.000 & 172 \\
463453 & 5.26 & 1.59 & 1.89 & 0.000 & 163 & &  494771 & 4.13 & 1.11 & 1.04 & 3.695 & 149 \\
461136 & 6.04 & 1.80 & 1.81 & 0.117 & 177 & &  493920 & 5.39 & 1.40 & 2.56 & 0.000 & 162 \\
432711 & 8.37 & 2.67 & 2.62 & 0.649 & 198 & &  471801 & 4.27 & 1.32 & 1.83 & 0.000 & 155 \\
418230 & 8.11 & 2.42 & 1.61 & 0.019 & 192 & &  428158 & 7.14 & 2.95 & 3.57 & 0.000 & 207 \\
417281 & 5.60 & 2.39 & 1.55 & 5.086 & 185 & &  433537 & 4.76 & 2.65 & 3.24 & 0.000 & 196 \\
459270 & 4.32 & 2.14 & 2.95 & 0.000 & 187 & &  426764 & 8.91 & 2.39 & 2.65 & 0.248 & 187 \\
462391 & 4.59 & 1.90 & 2.11 & 0.000 & 175 & &  460823 & 5.29 & 1.98 & 2.19 & 0.000 & 179 \\
436233 & 5.18 & 2.63 & 2.75 & 0.000 & 196 & &  456014 & 4.82 & 2.04 & 1.70 & 0.000 & 173 \\
432831 & 5.17 & 2.97 & 3.41 & 0.000 & 206 & &  480587 & 5.02 & 1.43 & 2.23 & 0.000 & 161 \\
426483 & 10.13 & 2.88 & 3.15 & 0.002 & 205 & &  471109 & 6.03 & 1.50 & 1.47 & 0.021 & 164 \\
458378 & 6.63 & 2.19 & 2.72 & 0.000 & 184 & &  469930 & 7.65 & 1.51 & 1.58 & 1.236 & 166 \\
&  &  &  &  && &        &      &      &      &      &       \\
\hline
\hline
\multicolumn{13}{l}{TNG100 high mass halos} \\
390859 & 8.48 & 4.75 & 4.49 & 0.000 & 237 & &  415373 & 9.43 & 3.66 & 4.95 & 0.000 & 219 \\
383307 & 10.31 & 4.87 & 4.52 & 0.365 & 232 & &  332068 & 8.15 & 5.34 & 2.78 & 0.034 & 217 \\
387236 & 12.15 & 4.46 & 3.65 & 0.865 & 230 & &  341356 & 11.92 & 7.18 & 5.27 & 0.064 & 273 \\
377398 & 17.08 & 6.09 & 5.98 & 0.003 & 260 & &  414447 & 8.42 & 3.44 & 3.35 & 0.000 & 211 \\
333672 & 10.32 & 8.70 & 2.34 & 0.472 & 288 & &  312412 & 13.09 & 7.79 & 4.97 & 0.000 & 282 \\
394439 & 12.03 & 4.29 & 3.39 & 1.286 & 230 & &  399520 & 8.74 & 3.93 & 4.83 & 0.000 & 220 \\
249164 & 13.77 & 9.95 & 6.87 & 0.069 & 298 & &  372568 & 13.64 & 5.53 & 5.60 & 0.104 & 254 \\
413091 & 12.81 & 3.67 & 2.39 & 0.021 & 218 & &  342689 & 13.01 & 4.74 & 3.69 & 0.201 & 238 \\
346420 & 11.97 & 5.31 & 5.37 & 0.056 & 244 & &  359639 & 14.99 & 6.41 & 7.83 & 0.456 & 261 \\
360916 & 5.75 & 5.98 & 4.77 & 0.000 & 238 & &  405460 & 12.18 & 3.23 & 5.59 & 0.000 & 206 \\
345812 & 10.51 & 8.82 & 3.41 & 5.729 & 295 & &  288014 & 7.10 & 9.55 & 2.24 & 0.246 & 292 \\
384103 & 6.69 & 4.91 & 2.94 & 0.010 & 218 & &  339547 & 11.91 & 7.57 & 4.71 & 0.000 & 271 \\
359086 & 8.14 & 7.75 & 4.70 & 0.000 & 283 & &  382059 & 11.56 & 3.80 & 2.87 & 2.505 & 221 \\
377212 & 14.52 & 5.98 & 7.86 & 0.079 & 258 & &  376132 & 7.35 & 3.61 & 4.06 & 0.000 & 214 \\
337444 & 10.97 & 7.94 & 6.50 & 0.000 & 278 & &  329105 & 14.29 & 8.71 & 8.22 & 1.058 & 294 \\
361418 & 6.22 & 6.46 & 4.64 & 0.000 & 252 & &  399969 & 11.73 & 4.23 & 4.43 & 0.000 & 227 \\
313402 & 10.00 & 9.02 & 6.59 & 0.002 & 245 & &  386429 & 9.24 & 4.78 & 1.69 & 4.350 & 208 \\
327822 & 10.97 & 4.27 & 2.22 & 4.068 & 222 & &  371859 & 8.90 & 5.62 & 7.07 & 0.000 & 253 \\
298206 & 12.79 & 8.75 & 5.82 & 1.032 & 282 & &  398110 & 11.38 & 4.01 & 4.98 & 0.033 & 230 \\
312891 & 15.21 & 9.78 & 11.30 & 0.088 & 303 & &  380119 & 12.16 & 4.85 & 5.02 & 0.290 & 239 \\
\hline
\hline
\multicolumn{13}{l}{TNG50 halos} \\
358608 & 15.06 & 5.04 & 4.38 & 4.018 & 243 & &  467415 & 8.68 & 2.08 & 2.13 & 9.224 & 183 \\
\enddata
\end{deluxetable*}
\startlongtable 
\begin{deluxetable*}{rrrrrrr|rrrrrr}
	\tablecaption{Basic properties of the selected EAGLE halos \label{tab:Eagle}}
	\tablehead{
		\colhead{Halo} & \colhead{M$_\star$} & \colhead{M$_{200}$} & \colhead{M$_{\rm BH}$} & \colhead{SFR} &  \colhead{R$_{500}$} & \colhead{   }  & \colhead{Halo} & \colhead{M$_\star$} & \colhead{M$_{200}$} & \colhead{M$_{\rm BH}$} & \colhead{SFR}  &  \colhead{R$_{500}$}\\
		\colhead{}  &  \colhead{$\si{10^{10}M_\odot}$} & \colhead{$\si{10^{12}M_\odot}$} & \colhead{$\si{10^8 M_\odot}$} & \colhead{$\si{M_\odot\,yr^{-1}}$} & \colhead{kpc}     & \colhead{}  & \colhead{}  &    \colhead{$\si{10^{10}M_\odot}$} & \colhead{$\si{10^{12}M_\odot}$} & \colhead{$\si{10^8 M_\odot}$} & \colhead{$\si{M_\odot\,yr^{-1}}$} & \colhead{kpc} 
	}
	\startdata
\multicolumn{13}{l}{low mass sample} \\
2703 & 0.45 & 0.42 & 0.02 & 0.691 & 104 & &  2144 & 0.79 & 0.44 & 0.01 & 0.292 & 110 \\
2258 & 0.27 & 0.52 & 0.02 & 0.477 & 109 & &  2824 & 0.66 & 0.44 & 0.01 & 0.431 & 109 \\
2577 & 0.21 & 0.37 & 0.00 & 0.498 & 95 & &  1949 & 1.15 & 0.69 & 0.02 & 1.131 & 125 \\
2578 & 1.15 & 0.52 & 0.01 & 0.480 & 117 & &  1791 & 0.99 & 0.86 & 0.05 & 0.606 & 139 \\
2044 & 1.07 & 0.71 & 0.01 & 0.190 & 117 & &  2478 & 0.53 & 0.39 & 0.02 & 0.306 & 104 \\
2535 & 0.71 & 0.51 & 0.02 & 0.614 & 114 & &  2370 & 0.92 & 0.55 & 0.02 & 0.992 & 117 \\
1662 & 1.32 & 0.85 & 0.07 & 1.099 & 133 & &  2428 & 0.41 & 0.33 & 0.01 & 0.411 & 98 \\
2899 & 0.40 & 0.38 & 0.01 & 0.750 & 98 & &  2726 & 0.71 & 0.53 & 0.08 & 0.000 & 119 \\
2592 & 1.14 & 0.50 & 0.05 & 0.205 & 115 & &  1730 & 2.86 & 0.78 & 0.05 & 1.402 & 132 \\
2345 & 0.36 & 0.53 & 0.01 & 0.230 & 117 & &  2864 & 0.56 & 0.41 & 0.01 & 0.355 & 105 \\
2557 & 0.28 & 0.37 & 0.02 & 0.326 & 98 & &  2516 & 0.83 & 0.48 & 0.03 & 0.434 & 113 \\
2671 & 0.25 & 0.48 & 0.01 & 0.408 & 100 & &  1725 & 1.93 & 0.81 & 0.11 & 0.003 & 131 \\
2118 & 0.85 & 0.39 & 0.01 & 0.823 & 105 & &  2902 & 0.36 & 0.38 & 0.01 & 0.241 & 97 \\
1647 & 2.13 & 0.87 & 0.08 & 0.577 & 138 & &  2874 & 0.31 & 0.39 & 0.02 & 0.147 & 105 \\
2184 & 1.47 & 0.65 & 0.03 & 0.690 & 125 & &  1803 & 1.64 & 0.73 & 0.03 & 1.379 & 128 \\
1581 & 1.21 & 0.82 & 0.12 & 0.018 & 123 & &  2688 & 0.73 & 0.52 & 0.04 & 0.338 & 115 \\
2176 & 1.09 & 0.68 & 0.01 & 0.547 & 124 & &  2338 & 0.29 & 0.63 & 0.03 & 0.304 & 125 \\
1564 & 2.37 & 0.92 & 0.07 & 0.885 & 140 & &  2852 & 0.63 & 0.46 & 0.04 & 0.578 & 112 \\
2716 & 0.56 & 0.42 & 0.01 & 0.478 & 100 & &  1569 & 1.50 & 0.84 & 0.03 & 0.983 & 132 \\
2066 & 1.59 & 0.69 & 0.05 & 0.768 & 126 & &  2157 & 1.43 & 0.59 & 0.05 & 1.220 & 121 \\
&  &  &  &  && &        &      &      &      &      &       \\
\hline
\hline
\multicolumn{13}{l}{medium mass halos} \\
1096 & 2.81 & 1.45 & 0.28 & 0.111 & 160 & &  798 & 4.79 & 1.67 & 0.31 & 0.010 & 162 \\
700 & 4.45 & 2.70 & 0.81 & 1.100 & 201 & &  816 & 4.28 & 1.86 & 0.07 & 2.487 & 170 \\
521 & 2.76 & 2.19 & 0.73 & 0.255 & 185 & &  1256 & 1.61 & 1.20 & 0.04 & 0.776 & 155 \\
480 & 6.90 & 3.12 & 0.76 & 0.218 & 203 & &  1303 & 1.44 & 1.27 & 0.06 & 0.632 & 156 \\
964 & 4.46 & 1.55 & 0.04 & 2.218 & 166 & &  846 & 2.02 & 1.82 & 0.23 & 0.041 & 164 \\
789 & 2.47 & 2.19 & 0.47 & 0.000 & 189 & &  1233 & 2.02 & 1.06 & 0.10 & 0.548 & 144 \\
1244 & 2.56 & 1.10 & 0.27 & 1.034 & 147 & &  774 & 4.40 & 2.20 & 0.53 & 0.021 & 185 \\
965 & 2.84 & 1.53 & 0.28 & 0.577 & 158 & &  1133 & 3.28 & 1.28 & 0.16 & 1.896 & 153 \\
1350 & 2.59 & 1.05 & 0.15 & 1.869 & 147 & &  1296 & 3.15 & 1.09 & 0.26 & 0.910 & 149 \\
927 & 5.02 & 1.83 & 0.49 & 1.128 & 179 & &  833 & 5.81 & 2.03 & 0.79 & 2.467 & 182 \\
482 & 4.98 & 2.67 & 0.58 & 1.394 & 198 & &  660 & 5.16 & 2.23 & 0.14 & 7.143 & 183 \\
744 & 3.43 & 2.06 & 0.72 & 0.053 & 183 & &  1122 & 3.16 & 1.51 & 0.31 & 0.500 & 165 \\
1205 & 2.47 & 1.06 & 0.19 & 1.044 & 144 & &  597 & 4.57 & 3.01 & 0.48 & 2.220 & 203 \\
933 & 3.30 & 1.74 & 0.52 & 0.000 & 170 & &  977 & 3.68 & 1.63 & 0.30 & 1.095 & 168 \\
527 & 4.90 & 2.92 & 0.80 & 2.603 & 196 & &  914 & 2.22 & 1.53 & 0.02 & 3.448 & 164 \\
1142 & 2.33 & 1.25 & 0.49 & 0.024 & 155 & &  961 & 1.82 & 1.69 & 0.39 & 0.000 & 168 \\
1117 & 3.30 & 1.32 & 0.13 & 2.053 & 155 & &  835 & 3.91 & 1.79 & 0.08 & 2.951 & 173 \\
512 & 4.32 & 2.70 & 0.39 & 2.959 & 196 & &  622 & 3.81 & 3.03 & 0.41 & 1.113 & 207 \\
621 & 4.63 & 2.41 & 0.10 & 2.462 & 177 & &  689 & 2.65 & 2.14 & 0.27 & 0.412 & 175 \\
709 & 3.29 & 2.35 & 0.24 & 1.544 & 178 & &  577 & 4.82 & 2.42 & 0.45 & 1.189 & 181 \\
&  &  &  &  && &        &      &      &      &      &       \\
\hline
\hline
\multicolumn{13}{l}{high mass halos} \\
207 & 8.57 & 4.86 & 0.34 & 6.738 & 223 & &  218 & 6.27 & 7.19 & 1.27 & 0.533 & 261 \\
214 & 12.05 & 8.57 & 2.52 & 0.101 & 294 & &  262 & 8.67 & 6.05 & 1.67 & 0.094 & 255 \\
209 & 9.18 & 8.53 & 2.77 & 2.621 & 287 & &  541 & 8.47 & 3.29 & 1.22 & 1.588 & 210 \\
235 & 7.45 & 6.75 & 1.71 & 3.997 & 264 & &  399 & 5.15 & 3.89 & 2.18 & 0.000 & 226 \\
251 & 9.04 & 6.17 & 2.02 & 3.160 & 252 & &  342 & 4.51 & 3.89 & 1.20 & 0.199 & 224 \\
253 & 9.42 & 6.85 & 1.34 & 4.110 & 262 & &  132 & 11.83 & 6.70 & 1.35 & 1.723 & 255 \\
244 & 8.89 & 6.58 & 0.89 & 4.195 & 255 & &  169 & 7.16 & 9.12 & 0.68 & 4.680 & 256 \\
377 & 9.35 & 4.19 & 0.85 & 0.538 & 222 & &  224 & 5.07 & 4.59 & 1.44 & 2.278 & 236 \\
283 & 6.32 & 5.87 & 1.17 & 3.746 & 252 & &  231 & 13.12 & 7.43 & 2.13 & 0.000 & 271 \\
208 & 8.55 & 3.93 & 1.21 & 0.000 & 226 & &  422 & 6.10 & 4.14 & 0.91 & 1.463 & 230 \\
282 & 3.92 & 5.62 & 0.27 & 2.259 & 219 & &  141 & 12.53 & 8.53 & 2.14 & 4.794 & 285 \\
189 & 12.05 & 7.94 & 4.10 & 0.124 & 275 & &  121 & 9.44 & 9.98 & 2.75 & 1.904 & 298 \\
248 & 4.91 & 5.62 & 2.34 & 0.081 & 247 & &  439 & 6.53 & 3.58 & 0.66 & 3.127 & 212 \\
151 & 13.37 & 9.33 & 2.03 & 4.601 & 285 & &  203 & 9.59 & 4.32 & 1.53 & 1.344 & 229 \\
194 & 11.61 & 8.99 & 1.71 & 3.159 & 277 & &  360 & 8.57 & 4.95 & 1.52 & 0.008 & 240 \\
539 & 6.25 & 3.24 & 1.01 & 0.019 & 210 & &  177 & 8.83 & 9.25 & 2.88 & 2.211 & 292 \\
427 & 6.19 & 4.05 & 0.79 & 0.765 & 223 & &  254 & 11.07 & 7.55 & 1.64 & 0.385 & 274 \\
334 & 8.81 & 5.09 & 1.80 & 0.000 & 240 & &  243 & 7.60 & 7.78 & 0.21 & 0.124 & 277 \\
408 & 6.70 & 4.39 & 0.30 & 3.627 & 227 & &  277 & 3.39 & 4.65 & 0.43 & 1.054 & 212 \\
276 & 8.18 & 4.37 & 1.13 & 0.104 & 221 & &  337 & 8.00 & 5.22 & 1.60 & 0.793 & 235 \\
\enddata
\end{deluxetable*}
\startlongtable 
\begin{deluxetable*}{rrrrrrr|rrrrrr}
	\tablecaption{Basic properties of the selected Simba halos \label{tab:Simba}}
	\tablehead{
		\colhead{Halo} & \colhead{M$_\star$} & \colhead{M$_{200}$} & \colhead{M$_{\rm BH}$} & \colhead{SFR} &  \colhead{R$_{500}$} & \colhead{   }  & \colhead{Halo} & \colhead{M$_\star$} & \colhead{M$_{200}$} & \colhead{M$_{\rm BH}$} & \colhead{SFR}  &  \colhead{R$_{500}$}\\
		\colhead{}  &  \colhead{$\si{10^{10}M_\odot}$} & \colhead{$\si{10^{12}M_\odot}$} & \colhead{$\si{10^8 M_\odot}$} & \colhead{$\si{M_\odot\,yr^{-1}}$} & \colhead{kpc}     & \colhead{}  & \colhead{}  &    \colhead{$\si{10^{10}M_\odot}$} & \colhead{$\si{10^{12}M_\odot}$} & \colhead{$\si{10^8 M_\odot}$} & \colhead{$\si{M_\odot\,yr^{-1}}$} & \colhead{kpc} 
	}
	\startdata
\multicolumn{13}{l}{low mass sample} \\
4721 & 1.94 & 0.93 & 0.51 & 1.702 & 140 & &  6326 & 1.38 & 0.68 & 0.28 & 2.072 & 126 \\
5266 & 0.63 & 0.75 & 0.04 & 1.432 & 128 & &  4496 & 1.43 & 0.94 & 0.16 & 4.941 & 133 \\
7185 & 1.60 & 0.58 & 0.24 & 1.664 & 121 & &  6991 & 1.10 & 0.50 & 0.20 & 1.610 & 109 \\
3729 & 1.68 & 0.97 & 0.43 & 2.214 & 124 & &  5205 & 2.40 & 0.83 & 0.12 & 2.447 & 136 \\
7837 & 0.91 & 0.49 & 0.17 & 2.523 & 113 & &  4172 & 1.34 & 0.93 & 0.38 & 1.880 & 135 \\
5566 & 1.74 & 0.60 & 0.49 & 1.522 & 119 & &  11638 & 0.83 & 0.32 & 0.05 & 1.070 & 98 \\
6542 & 1.42 & 0.60 & 0.37 & 1.558 & 122 & &  4862 & 1.65 & 0.53 & 0.38 & 1.122 & 119 \\
6652 & 1.18 & 0.50 & 0.26 & 2.086 & 111 & &  5494 & 2.18 & 0.65 & 0.57 & 1.301 & 123 \\
7552 & 1.63 & 0.58 & 0.27 & 1.385 & 119 & &  4267 & 0.92 & 0.92 & 0.34 & 2.029 & 132 \\
9143 & 0.87 & 0.47 & 0.09 & 1.036 & 111 & &  6156 & 0.87 & 0.58 & 0.04 & 5.991 & 116 \\
7189 & 0.85 & 0.62 & 0.14 & 1.226 & 115 & &  3768 & 1.89 & 0.64 & 0.46 & 3.384 & 119 \\
5263 & 2.38 & 0.51 & 0.56 & 3.635 & 113 & &  5762 & 1.15 & 0.36 & 0.33 & 2.065 & 104 \\
8135 & 0.83 & 0.38 & 0.16 & 1.581 & 103 & &  4965 & 0.30 & 0.40 & 0.29 & 2.137 & 107 \\
4672 & 0.95 & 0.97 & 0.24 & 1.679 & 140 & &  5557 & 1.31 & 0.75 & 0.28 & 1.001 & 130 \\
6084 & 1.18 & 0.35 & 0.18 & 2.453 & 100 & &  9136 & 1.11 & 0.43 & 0.22 & 1.063 & 107 \\
11327 & 1.04 & 0.33 & 0.24 & 1.031 & 101 & &  6800 & 1.06 & 0.54 & 0.14 & 1.076 & 116 \\
5936 & 0.93 & 0.39 & 0.21 & 2.142 & 102 & &  8881 & 0.82 & 0.50 & 0.08 & 1.207 & 116 \\
4439 & 1.65 & 0.73 & 0.47 & 1.755 & 126 & &  11362 & 0.69 & 0.35 & 0.10 & 1.166 & 100 \\
10409 & 0.70 & 0.37 & 0.11 & 1.507 & 103 & &  5802 & 1.31 & 0.38 & 0.37 & 1.842 & 104 \\
4802 & 3.35 & 0.91 & 0.44 & 4.681 & 141 & &  5565 & 1.33 & 0.70 & 0.21 & 1.498 & 124 \\
&  &  &  &  && &        &      &      &      &      &       \\
\hline
\hline
\multicolumn{13}{l}{medium mass halos} \\
2788 & 7.33 & 1.34 & 0.87 & 5.462 & 158 & &  2992 & 5.23 & 1.06 & 0.40 & 15.196 & 148 \\
2722 & 4.66 & 1.55 & 0.93 & 10.480 & 159 & &  2401 & 4.21 & 1.12 & 1.10 & 1.439 & 151 \\
1323 & 6.02 & 2.22 & 0.96 & 3.030 & 170 & &  1792 & 3.99 & 1.38 & 0.21 & 11.855 & 147 \\
3956 & 4.32 & 1.13 & 0.40 & 5.601 & 150 & &  1864 & 11.30 & 2.48 & 1.17 & 3.974 & 195 \\
1278 & 7.51 & 2.37 & 1.30 & 2.963 & 180 & &  2496 & 2.55 & 1.49 & 0.23 & 9.434 & 156 \\
2132 & 11.38 & 2.12 & 0.77 & 4.576 & 188 & &  1650 & 2.30 & 1.84 & 1.38 & 1.849 & 170 \\
2030 & 4.23 & 1.79 & 0.49 & 4.188 & 160 & &  2166 & 2.76 & 1.08 & 0.42 & 5.268 & 146 \\
395 & 8.61 & 3.13 & 3.03 & 3.697 & 206 & &  1201 & 6.09 & 2.33 & 1.25 & 1.188 & 180 \\
2828 & 4.97 & 1.57 & 1.19 & 5.388 & 167 & &  2622 & 2.52 & 1.47 & 0.99 & 1.495 & 157 \\
2058 & 6.82 & 2.03 & 0.99 & 5.198 & 183 & &  1380 & 6.64 & 3.13 & 0.61 & 1.297 & 209 \\
2131 & 3.20 & 1.92 & 0.88 & 1.257 & 172 & &  2089 & 6.11 & 2.23 & 1.74 & 1.040 & 186 \\
4086 & 4.22 & 1.15 & 0.15 & 6.362 & 152 & &  3606 & 5.71 & 1.23 & 1.43 & 2.548 & 154 \\
1403 & 3.92 & 2.24 & 0.90 & 3.556 & 176 & &  1726 & 7.23 & 1.99 & 2.01 & 6.853 & 179 \\
1985 & 7.44 & 2.13 & 0.67 & 1.127 & 183 & &  2081 & 5.61 & 2.19 & 0.40 & 9.351 & 191 \\
1482 & 6.83 & 2.86 & 1.91 & 3.100 & 197 & &  1942 & 15.99 & 2.56 & 0.63 & 5.588 & 199 \\
2204 & 3.86 & 1.59 & 0.48 & 1.382 & 164 & &  2790 & 5.33 & 1.58 & 0.29 & 14.680 & 168 \\
3101 & 6.34 & 1.45 & 0.36 & 9.689 & 163 & &  3409 & 3.16 & 1.12 & 1.04 & 2.281 & 148 \\
1762 & 2.39 & 1.57 & 0.79 & 4.764 & 154 & &  3012 & 4.14 & 1.49 & 1.02 & 3.209 & 163 \\
2831 & 3.98 & 1.45 & 0.16 & 13.514 & 152 & &  2036 & 10.33 & 2.31 & 1.31 & 5.877 & 193 \\
2066 & 1.14 & 1.65 & 0.24 & 3.519 & 144 & &  3883 & 8.74 & 1.22 & 0.55 & 1.652 & 158 \\
&  &  &  &  && &        &      &      &      &      &       \\
\hline
\hline
\multicolumn{13}{l}{high mass halos} \\
793 & 11.00 & 5.82 & 1.25 & 5.868 & 236 & &  483 & 7.07 & 7.41 & 2.06 & 1.367 & 241 \\
414 & 11.41 & 9.12 & 9.96 & 7.806 & 296 & &  928 & 5.38 & 3.69 & 1.77 & 2.034 & 201 \\
343 & 17.38 & 9.31 & 8.13 & 1.533 & 292 & &  406 & 17.12 & 9.79 & 4.20 & 6.775 & 300 \\
704 & 15.05 & 6.15 & 2.46 & 1.044 & 259 & &  859 & 19.38 & 3.63 & 2.00 & 8.954 & 209 \\
550 & 5.75 & 6.84 & 2.71 & 2.130 & 256 & &  729 & 11.17 & 4.49 & 3.22 & 2.392 & 236 \\
589 & 8.08 & 5.37 & 4.71 & 1.272 & 234 & &  499 & 9.01 & 8.77 & 2.22 & 2.046 & 280 \\
1101 & 13.65 & 4.24 & 1.20 & 5.909 & 231 & &  533 & 9.96 & 8.53 & 4.88 & 1.674 & 283 \\
606 & 8.68 & 7.33 & 1.81 & 1.538 & 276 & &  576 & 6.72 & 4.69 & 4.46 & 3.734 & 233 \\
797 & 5.25 & 5.10 & 3.96 & 2.142 & 242 & &  981 & 19.67 & 3.73 & 2.51 & 1.825 & 220 \\
587 & 6.22 & 6.79 & 3.92 & 5.185 & 244 & &  530 & 10.34 & 7.74 & 7.76 & 1.340 & 259 \\
422 & 6.25 & 3.88 & 3.30 & 7.754 & 204 & &  792 & 4.40 & 5.33 & 2.31 & 1.513 & 242 \\
1162 & 9.05 & 3.98 & 1.35 & 1.839 & 214 & &  1250 & 5.21 & 3.48 & 2.66 & 1.492 & 195 \\
675 & 6.11 & 7.43 & 3.49 & 1.278 & 269 & &  303 & 10.63 & 7.19 & 4.63 & 4.744 & 267 \\
663 & 5.61 & 5.19 & 2.86 & 3.604 & 239 & &  521 & 9.28 & 6.21 & 1.21 & 4.473 & 200 \\
647 & 10.31 & 7.93 & 4.45 & 1.167 & 283 & &  1369 & 3.15 & 3.58 & 1.66 & 9.063 & 217 \\
1302 & 4.14 & 3.16 & 1.78 & 2.846 & 202 & &  1375 & 4.02 & 3.56 & 0.83 & 4.062 & 219 \\
1264 & 11.70 & 3.60 & 0.96 & 4.972 & 218 & &  930 & 5.35 & 4.52 & 0.83 & 3.742 & 226 \\
410 & 9.51 & 9.42 & 4.18 & 1.048 & 303 & &  1067 & 5.37 & 3.26 & 2.46 & 2.147 & 201 \\
948 & 4.69 & 4.21 & 3.68 & 3.268 & 229 & &  396 & 13.91 & 7.52 & 5.18 & 1.300 & 263 \\
472 & 7.78 & 7.03 & 3.55 & 2.297 & 257 & &  1343 & 11.42 & 3.41 & 0.34 & 1.601 & 216 \\
\enddata
\end{deluxetable*}

\end{document}